 \documentclass[aps,prb,eqsecnum,amsmath,floatfix,showpacs,amssymb,superscriptaddress,twocolumn,10pt]{revtex4-1}
 \usepackage{graphicx}
 \usepackage{float,rotating}
 \usepackage[caption=false]{subfig}
 \usepackage{subeqnarray}
 \usepackage{bbold}
 \usepackage{amsmath}
 \usepackage{verbatim}
 \usepackage{amssymb}
 \usepackage{mathrsfs}
 \usepackage{float}
 \usepackage{color}
 \usepackage [english]{babel}
 \usepackage [autostyle, english = american]{csquotes}\MakeOuterQuote{"}
 \usepackage[breaklinks=true,colorlinks=true,linkcolor=blue,urlcolor=blue,citecolor=magenta]{hyperref}
 \usepackage{mathtools}
\DeclarePairedDelimiter\bra{\langle}{\rvert}
\DeclarePairedDelimiter\ket{\lvert}{\rangle}

\begin{document}
\title{SU(3) Landau-Zener interferometry with a transverse periodic drive}
\author{M. B. Kenmoe}
\affiliation{Mesoscopic and Multilayer Structures Laboratory, Faculty of Science, Department of Physics, University of Dschang, Cameroon}
\author{A. B. Tchapda}
\affiliation{Mesoscopic and Multilayer Structures Laboratory, Faculty of Science, Department of Physics, University of Dschang, Cameroon}
\author{L. C. Fai}
\affiliation{Mesoscopic and Multilayer Structures Laboratory, Faculty of Science, Department of Physics, University of Dschang, Cameroon}
\date{\today}
\begin{abstract}
Quantum triangles can work as interferometers. Depending on their geometric size and interactions between paths, "beats" {\it and/or} "steps" patterns are observed.  We show that when inter-level distances between level positions  in quantum triangles periodically change with time, formation of beats {\it and/or} steps no longer depends only on the geometric size of the triangles but also on the characteristic frequency of the transverse signal. For large-size triangles, we observe the coexistence of beats {\it and} steps when the frequency of the signal matches that of non-adiabatic oscillations and for large frequencies, a maximum of four steps  instead of two as in the case with constant interactions  is observed.  Small-size triangles also revealed counter-intuitive interesting dynamics for large frequencies of the field: unexpected two-step patterns are observed. When the frequency is large and tuned such that it matches the uniaxial anisotropy, three-step patterns are observed. We have equally observed that when the transverse signal possesses a static part, steps maximize to six.  These  effects are semi-classically explained in terms of Fresnel integrals and quantum mechanically in terms of quantized fields with a photon-induced tunneling process. Our expressions for populations are in excellent agreement with the gross temporal profiles of exact numerical solutions. We compare the semi-classical and quantum dynamics in the triangle and establish the conditions for their equivalence. 
 
\end{abstract}
\maketitle

\section{Introduction}\label{Sec1}
The possibility of crossing three energy levels at more than one point has opened a new avenue for exploring Landau\cite{lan}-Zener\cite{zen}-St\"uckelberg\cite{stu}-Majorana\cite{Majorana} (LZSM) interferometry in three-level systems (ThLSs)\cite{Gallego, Kiselev, KenmoePRB}.  When three diabatic levels cross and form a triangular geometry, this hides the dynamical symmetry of the $SU(3)$ group (spanned in the Lie space by Gell-Mann matrices\cite{Georgi}) and may be exploited as a quantum interferometer\cite{Kiselev, KenmoePRB} or used as the building block for qutrits (the unit of ternary quantum computing\cite{KenmoePRB}). If in addition the inter-level distance between level positions is maintained constant throughout the course of variation of a control parameter (time, chemical potential, flux, magnetic field, pressure, temperature etc), the relevant model leads to the so called $SU(3)$ LZSM interferometry\cite{Kiselev, KenmoePRB}. Such a system has stimulated  active theoretical researches\cite{Gallego, Kiselev, KenmoePRB, Henriet} and is currently attracting tremendous interests from both fundamental and experimental physics due to versatile applications in Bose-Josephson junctions\cite{Lee, Bouk} (BJJ), quantum spectroscopy\cite{Hamley}, quantum metrology\cite{Huang11}, quantum information processing\cite{Patton} etc. 

 Quantum triangles are observed in various experimental protocols\cite{Broers, Ivanov2008, Dieter, Raymond, Jose} (see also the triangle model in Ref.\onlinecite{Ashab}).  A typical example is achieved when in a spin-1 $SU(2)$ LZSM model (three levels crossing at a single point) with constant coupling between levels, one vertically shifts the  zero-energy level downwards or upwards. This can technically be done by adding the zero-energy
 splitting term $D(S^{z})^{2}$  to the  $SU(2)$ Hamiltonian and where $D$ is the single ion easy-axis anisotropy whilst $S^{z}$ is the projection of the spin vector onto the quantization direction. This action creates two additional crossings making a total of three and consequently increases the order of the symmetry group by a unit. The resulting configuration is a quantum triangle which  works as an interferometer and may be exploited in spectroscopy analysis to harvest information about  the complex dynamics of a three-level atom (or a qutrit) bathing in its environment or for achieving high-precision measurements. Other examples are triplet states  energy levels of a linearly driven two-spin-$1/2$ system\cite{Dieter, Raymond, Jose, Henriet, Garanin2003}. Indeed, if the triplets are coupled through Ising interactions and   bath in a boson sea consisting of harmonic oscillators at room temperature, various types of quantum triangles   form depending on the value of the Ising coupling\cite{Henriet}.
  Ultracold atoms in optical lattices with lattice sites converted into biased double-well also depict a triangular geometry\cite{Blochs}. A linear triple quantum dots geometry connected to leads also works as an $SU(3)$ interferometer providing two additional {\it ac} gate voltages that are tuned such that the phase difference between them achieves $\pi$ (see  Ref.\onlinecite{Gallego} for ample discussions). Bose-Einstein condensates (BEC) also hide the non trivial dynamical symmetry of the $SU(3)$ group and $D$ here acts as the atom-atom interaction term\cite{Huang}.
 
 It is formally established that interferences between paths in triangles through level crossings result into formation of quantum beats and quantum steps (multiple LZSM transitions)\cite{Kiselev, Leuch, Racheal, Garanin2013}. For small-size triangles, constant interactions for an initialization of the ThLS in the middle diabatic state leads to the formation of beats patterns when the dwell time $\delta t$ (time between two non-adiabatic transitions or the time spent in between two vertices of the triangle) in the triangle is shorter than the characteristic time $t_{\rm relax}$ of relaxation  (and the time $t_{\rm LZ}$ of non-adiabatic oscillations\cite{Vitanov1998, comment3}) i.e. ($\delta t<t_{\rm LZ}<t_{\rm relax}$) (see Fig.\ref{Figure1}(b) and also Ref.\onlinecite{comment4}). In contrast, for large-size triangles, these lead to step  patterns when $\delta t>t_{\rm relax}>t_{\rm LZ}$ (see Fig.\ref{Figure1}(c)). These are unquestionably due to the $SU(3)$ deformation and also likely to couplings between diabatic states that are all equal and maintained constant (not evolving with time or any other control parameter). These patterns, predicted to be observed in ultracold atoms\cite{Kiselev},  double and triple quantum dots\cite{Gallego, Gallego2015}, double-well trap potentials\cite{Blochs}, render the  $SU(3)$ protocol yet another interesting means for exploring the strong coupling dynamics of the cited set-ups.  
 
 For the aforementioned reasons, deciphering the complex dynamics of $SU(3)$ LZSM interferometers when the coupling between levels changes with time is at the heart of discussions ongoing in this piece of work. We investigate three versions of the $SU(3)$ LZSM model obtained by allowing levels spacing to  periodically change with time while upholding the linear time-dependence of diabatic energies  as detuned by a uni-axis anisotropy $D$. Our concern is to  enquire how these changes affect the beats and steps patterns observed in the original $SU(3)$ LZSM interferometry.

The remaining part of the paper is organized as follows: in Sections \ref{Sec2} and \ref{Sec3} respectively, the model is presented and evolutions in the non-adiabatic limit are investigated. Section \ref{Sec4} concentrates on adiabatic evolutions while in Section \ref{Sec5} some experimental relevances are pointed out. Finally in Section \ref{Sec6}, we conclude with our main achievements.

\section{Model}\label{Sec2}

We couple the inter-level distance between level positions in the $SU(3)$ LZSM Hamiltonian to a periodic signal. Interactions between the dipole moment of the ThLS and the classical radiation reads $-\hat{\mathbf{D}}\cdot \mathbf{E}(t)$ where $\hat{\mathbf{D}}=\hat{d}e_{x}$ and $\mathbf{E}(t)=E(t)e_{x}$ are respectively the dipole moment operator and the electric field vector (of amplitude $A$, frequency $\omega$ and phase shift $\phi$) oriented along the direction of the polarization vector $e_{x}$. In the dipole moment and  rotative-wave approximations,  the governing  model reads ($\hbar=1$)
\begin{eqnarray} \label{equ1} 
\mathcal{H}(t)=
\alpha tS^{z}+ f(t)S^{x}+D(S^{z})^{2},
\end{eqnarray}
where
\begin{eqnarray} \label{equ2} 
f(t)=
A\cos (\omega t+\phi).
\end{eqnarray}
Here, $\alpha>0$ represents the constant sweep velocity of the external magnetic field and $D$ the uni-axial anisotropy.  $S^{\nu}$ ($\nu=x, y, z$) are spin operators generators of the $su(2)$ algebra $[S^{\mu}, S^{\nu}]=i\epsilon^{\mu\nu}_{\gamma}S^{\gamma}$ where $\epsilon^{\mu\nu}_{\gamma}$ are structure constants on $SU(2)$ (see Ref.\onlinecite{Georgi}). For our case, the total spin $S=1$  and this maps Eq.(\ref{equ1}) to a three-level model. Thus, the last term in Eq.(\ref{equ1}) is the zero-energy splitting term added to the spin-$1$ $SU(2)$ LZSM model. As a consequence of the addition of this term, the order of the symmetry is increased by a unit. It is clearly demonstrated in Ref.\onlinecite{Kiselev} that Eq.(\ref{equ1}) has an $SU(3)$ symmetry. Indeed, in the space of the $SU(2)$ group generators, the model (\ref{equ1}) depicts a non-linearity which is removed in the $SU(3)$ space by reformulating the model in terms of Gell-Mann matrices\cite{Georgi} and neglecting an Abelian term without affecting the quantities of central interest\cite{Kiselev} (see next paragraph). It should also be noted that when the $SU(3)$ symmetry breaks down and the transverse drive is switched off ($D=\omega=0$) the model Eq.(\ref{equ1}) is integrable and has  exact solutions\cite{Patra, Kenmoe2013}.

\begin{figure}[]
	\includegraphics[width=9cm, height=7cm]{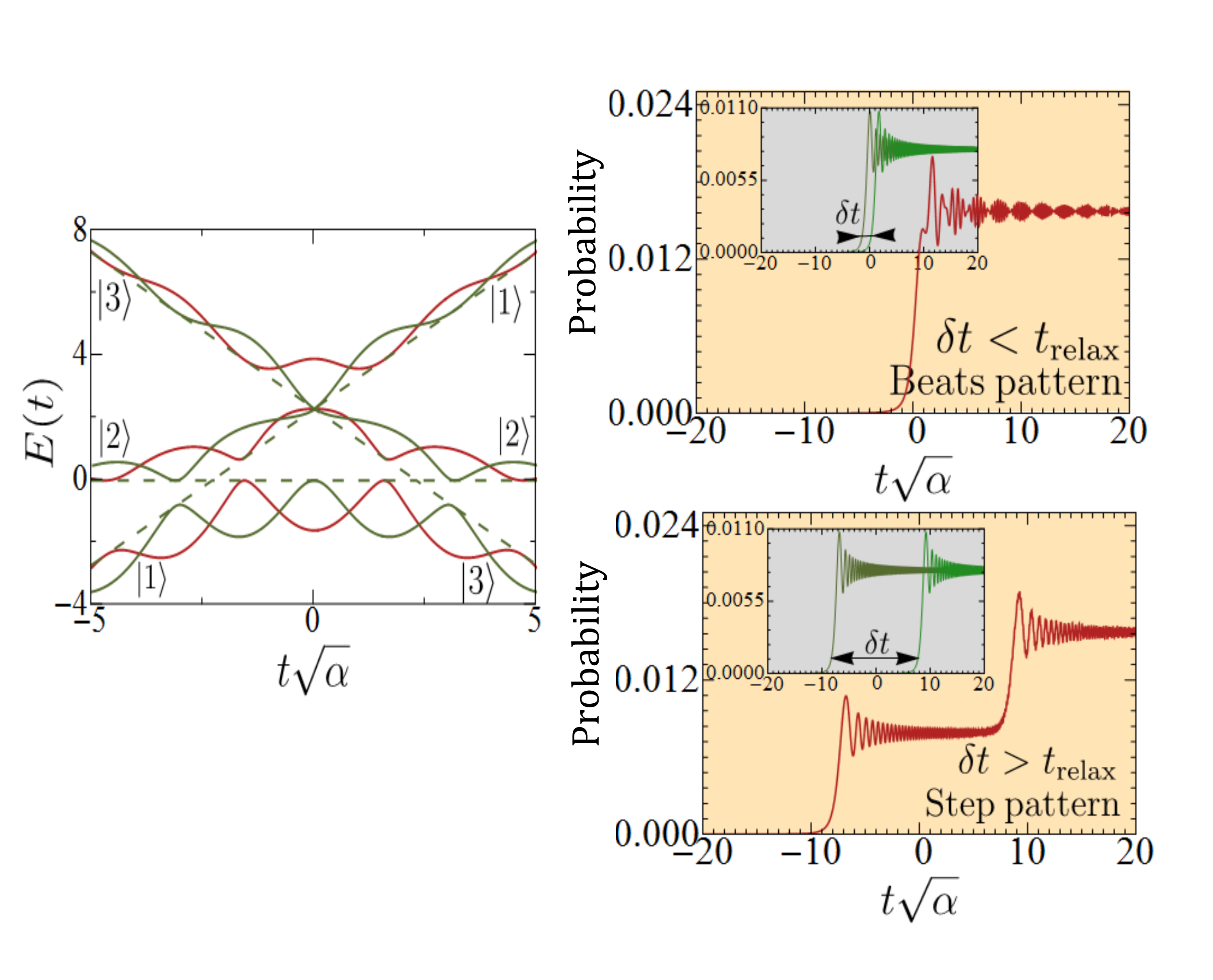}
	\vspace{-0.9cm}
	\caption{(Color Online) (a): Time-evolution of adiabatic eigenenergies Eq.(\ref{equ2a}) of the Hamiltonian (\ref{equ1}). Gray solid lines correspond to $\phi=\pi/2$ and  red solid lines to $\phi=0$. The gray dashed lines correspond to the limit $A=0$.   On (b) and (c), gray and green curves respectively correspond to the transition probabilities $\pi\hat{\delta} F(t+\mathbf{Q})$ and $\pi\hat{\delta} F(t-\mathbf{Q})$ while red curves are $\pi\hat{\delta}(F(t+\mathbf{Q})+F(t-\mathbf{Q}))$ with $\mathbf{Q}=0.5\sqrt{\alpha}$ in (b) and $\mathbf{Q}=8\sqrt{\alpha}$ in (d). The time is in the unit of $t_{\rm LZ}=1/\sqrt{\alpha}$ and $\hat{\delta}=0.0025$. Let us recall that in the non-adiabatic limit, the finite-time LZSM transition probability for a two-level system is given by $\pi\hat{\delta} F(t)$ where $\hat{\delta}$ is the LZSM parameter\cite{Biao, Kenmoe2013}. Up to a factor of $\pi\hat{\delta}$, the function $F(t)$ presented in appendix \ref{App1} represents the population that non-adiabatically traverses a crossing in the fast sweep limit\cite{Kenmoe2013}. On (b) and (c) $\delta t$ is the time between two consecutive LZSM transitions and $t_{\rm relax}$ is the relaxation time\cite{comment3}. Beats form (Panel (b)) when within a time interval $\delta t<t_{\rm relax}$, returning to its original diabatic state, the system traverses several crossings. If in contrast (Panel (c)) $\delta t>t_{\rm relax}$ and the system goes through several crossings and returns to its diabatic state, oscillations in the final population show steps\cite{comment4}. 
		} \label{Figure1}
\end{figure}

For further relevant purposes, it might be interesting to proceed  to a geometrical description of the model (\ref{equ1}). An adapted space for this task is formed by a rotated $\mu$-basis (a set of eight matrices alternate to Gell-Mann matrices\cite{Georgi} preserving all Casimir operators and the $su(3)$ algebra) constructed in Refs.[\onlinecite{Kiselev,Davidson}] through suitable combination of Gell-Mann matrices (see Ref.\onlinecite{Davidson} for matrix representations). Thus, the magnetic field vector  $\vec{B}(t)=[f(t), \alpha t, -D/\sqrt{3}]$ and the spinor $\vec{\Lambda}=[\mu_1, \mu_3, \mu_8]^T$ with $T$ standing for the transposed vector, help in rewriting the model in the $\mu$-basis topologically as a trajectory\cite{Kiselev}  
\begin{eqnarray} \label{equ2aa} 
\mathcal{H}(t)=\vec{B}(t)\cdot\vec{\Lambda},
\end{eqnarray}
providing an Abelian term $2D\hat{\mathbb{1}}/3$ (where $\hat{\mathbb{1}}$ is a $(3\times3)$ unit matrix). This term is neglected as it merely induces a trivial exponential phase factor of modulus $1$ into the expressions for transition probabilities. The representation (\ref{equ2aa}) is yet an indication that (\ref{equ1})  possesses the symmetry operations of the $SU(3)$ group.
 The population dynamics in the $8$-dimensional Bloch's hypersphere corresponds to precession of the Bloch's vector onto the hypersurface of the sphere at a rate determined by the magnetic field. The fact that a minimal number of three $\mu$ matrices out of the eight is sufficient to describe the model has a nontrivial consequence on the  description of the population dynamics.  We demonstrate in what follows that only three components of the Bloch's vector respectively associated with $\mu_3$, $\mu_6$ and $\mu_8$ are also enough to describe the precession (population dynamics).

The eigen-energies of the model (\ref{equ1}) are given by
\begin{eqnarray}\label{equ2a}
E_{n}(t)=\frac{2D}{3}+2\sqrt{\frac{p(t)}{3}}\cos\Big[\frac{\vartheta_{n}(t)}{3}\Big],
\end{eqnarray}
where
\begin{eqnarray}\label{equ2b}
\vartheta_{n}(t)=\arccos\Big[\frac{3q(t)}{2p(t)}\sqrt{\frac{3}{p(t)}}\Big]-\delta_{n},
\end{eqnarray}
with $\delta_{1}=4\pi$, $\delta_{2}=2\pi$ and $\delta_{3}=0$. Here, $p(t)=(\alpha^{2}t^{2}+D^{2}/3+f^{2})$ and $q(t)=2D(\alpha^{2}t^{2}-D^{2}/9-f^{2}/2)/3$. The time evolution of $E_{n}(t)$ is plotted in Fig.\ref{Figure1}(a). It appears that when the linear sweep  quickly changes with time and the transverse field is tuned such that $A^{2}/\alpha\ll 1$, the ThLS evolves {\it on the triangle}. When in contrast it slowly changes and $A^{2}/\alpha\gg 1$, the ThLS follows adiabatic trajectories and evolves {\it out of the triangle}. We consider, explore and discuss these two limits in the remaining part of the paper.

\section{Non-adiabatic evolution}\label{Sec3}
We probe the model (\ref{equ1}) in the non-adiabatic limit $A^{2}/\alpha\ll 1$ by numerically and analytically solving the von Neuman equation $i\dot{\boldsymbol{\mathrm{\rho}}}(t)=[\mathcal{H}(t), \boldsymbol{\mathrm{\rho}}(t)]$  for the density matrix $\boldsymbol{\mathrm{\rho}}(t)=\sum_{n,m=1}^{3}\rho_{nm}(t)|n\rangle\langle m|$ whose nine elements $\rho_{nm}(t)$ with $n,m =(1,2,3)$ satisfy
\begin{equation} \label{equ3}
i\dot{\rho}_{nm}(t)=\sum_{\kappa=1}^{3}\Big(\mathcal{H}_{n\kappa}(t)\rho_{\kappa m}(t)-\rho_{n\kappa}(t)\mathcal{H}_{\kappa m}(t)\Big),
\end{equation} 
 and where $\mathcal{H}_{n\kappa}(t)$ are matrix elements of $\mathcal{H}(t)$. The indices $1$, $2$ and $3$ respectively match the diabatic states $|1\rangle$, $|2\rangle$ and $|3\rangle$. Therefore, Eq.(\ref{equ3}) contains nine equations: three (diagonal elements) for populations and six (off-diagonal elements) for coherence factors between diabatic states. We are interested in seeing how the gross temporal profile of populations changes with time for an initialization of the ThLS in the state $|\kappa'\rangle$. Thus, 
  \begin{equation} \label{equ3a}
\rho_{\kappa'\kappa}(t_{0})=\delta_{\kappa'\kappa},
\end{equation}
stand for initial conditions. This task is numerically performed. We construct analytical expressions describing the time-evolution of  populations in the non-adiabatic limit. Furthering our goal, we define three supplementary variables: two for population differences, $ R(t)=\rho_{11}(t)-2\rho_{22}(t)+\rho_{33}(t)=-\sqrt{3}{\rm Tr}(\boldsymbol{\mathrm{\rho}}(t)\mu_{8})$, $ Q(t)=\rho_{11}(t)-\rho_{33}(t)={\rm Tr}(\boldsymbol{\mathrm{\rho}}(t)\mu_{3})$, and one for coherence factors $\hat{W}(t)=2\sqrt{2}{\rm Re}(\rho_{12}(t)-\rho_{23}(t))=-2{\rm Tr}(\boldsymbol{\mathrm{\rho}}(t)\mu_{6})$ (where $\text{Re(...)}$ indicates the real part) and notice that they are components of the Bloch's vector. These three functions are sufficient to fully parametrize the density matrix (see Ref.\onlinecite{Kiselev} for further explanations). Thus, the precession of the Bloch's vector is completely ruled by three of its components.  This fact directly stems from the geometrical representation of $\mathcal{H}(t)$ in Eq.(\ref{equ2aa}). The $\mu$-basis helps in minimizing the number of basis vectors for the trajectory $\mathcal{H}(t)$ in the space of $SU(3)$ group generators, consequently minimizing the number of Bloch's vector components necessary for describing precession (spin dynamics) on the hypersphere.  Thus, after eliminating all coherence factors from the equations for population differences, we obtain the set of coupled integral-differential equations 
\begin{widetext}

\begin{subeqnarray}
&\dfrac{dQ}{dt}=-\frac{1}{2}\int_{-\infty}^{t} f(t) f(t_{1})\Big[Kr^{-}(t, t_{1})R(t_{1})+Kr^{+}(t,t_{1})Q(t_{1})\Big]dt_{1}+\frac{f(t)}{2}\hat{\Phi}_{-}(t), \slabel{equ4}
\\\nonumber\\
&\dfrac{dR}{dt}=-\frac{3}{2}\int_{-\infty}^{t} f(t) f(t_{1})\Big[Kr^{+}(t, t_{1})R(t_{1})+Kr^{-}(t,t_{1})Q(t_{1})\Big]dt_{1}+\frac{3f(t)}{2}\hat{\Phi}_{+}(t), \slabel{equ5}
\\\nonumber\\
&\hat{W}(t)=\int_{-\infty}^{t} f(t_{1})\Big[Ki^{+}(t, t_{1})R(t_{1})+Ki^{-}(t,t_{1})Q(t_{1})\Big]dt_{1}+\hat{\Phi}_{0}(t). \slabel{equ6}
\end{subeqnarray}

\end{widetext}
Here, $\hat{\Phi}_{\pm}(t)$ and $\hat{\Phi}_{0}(t)$ (different from the ones in Ref.\onlinecite{Kiselev}) are functions of $R(t)$ and $\hat{W}(t)$. In addition, $\hat{\Phi}_{\pm}(t)$ are functions of $f(t)f(t_{1})$ and do not contribute to population differences in the non-adiabatic limit as they lead to terms of higher orders of  $A^2/\alpha$ that are discarded/overlooked. On the other hand, we have defined
$ K\mu^{\pm}(t, t_{1})=K\mu^{\Omega^{+}}(t, t_{1})\pm K\mu^{\Omega^{-}}(t, t_{1})$
 where $K\mu^{\xi}(t, t_{1})=\text{L}\mu[\exp[i(\xi(t)-\xi(t_{1}))]]$ ($\mu=r,i$ and $\text{L}r=\text{Re}$, $\text{L}i=\text{Im}$) 
with $\xi(t)=(\Omega^{+}(t),\Omega^{-}(t))$ and where
$\Omega^{\pm}(t)=\pm\frac{\alpha}{2}(t\pm \frac{D}{\alpha})^{2}\mp\frac{D^2}{2\alpha}$, is the phase picked up by the components of the wave function during the sweep of the external field. ${\rm Im}(...)$ designates the imaginary part of the term inside the brackets. For initial conditions,  let us define the projectors
\begin{eqnarray}\label{equ6a}
\nonumber\mathcal{P}^{R}=|\kappa'\rangle R(t_{0})\langle\kappa'|=-\sqrt{3}\mu_{8}, \quad \mathcal{P}^{Q}=|\kappa'\rangle Q(t_{0})\langle\kappa'|=\mu_{3},\\
\end{eqnarray}
such that when the ThLS starts off in the diabatic state $|\kappa'\rangle$, then the matrix elements of $\mathcal{P}^{R/Q}$ yield $\mathcal{P}_{\kappa'\kappa'}^{R}=R(t_{0})=-\sqrt{3}\langle\kappa'|\mu_{8}|\kappa'\rangle$ and $\mathcal{P}_{\kappa'\kappa'}^{Q}=Q(t_{0})=\langle\kappa'|\mu_{3}|\kappa'\rangle$. After solving Eqs.(\ref{equ4})-(\ref{equ6}) with appropriate initial conditions, the desired populations are extracted as $\rho_{11}(t)=\frac{1}{3}(1+\frac{R(t)}{2}+\frac{3Q(t)}{2})$, $\rho_{22}(t)=\frac{1}{3}(1-R(t))$ and $\rho_{33}(t)=\frac{1}{3}(1+\frac{R(t)}{2}-\frac{3Q(t)}{2})$. Therefore, for each preparation, the probability of transition  $|\kappa'\rangle\to|\kappa\rangle$ reads
\begin{eqnarray}\label{equ6b}
P_{\kappa'\to\kappa}(t)=\rho_{\kappa\kappa}(t).
\end{eqnarray}
Interestingly, Eqs.(\ref{equ4})-(\ref{equ6}) are useful to tackle (quantify and qualify) hyperfine interactions effects on the $SU(3)$ LZSM interferometry when the latter is set in quantum dots where such effects prevail\cite{comment2}. Several interesting problems with/without periodic drives may also be tackled as well. For the case of noise and especially a fast colored noise (i.e. $f(t)$ turns to a zero-mean noise field), the terms $\hat{\Phi}_{\pm}(t)$ and $\hat{\Phi}_{0}(t)$ are irrelevant and vanish as results of the Bloch's averaging procedure over the noise realizations\cite{Comment6, Kenmoe2013}.

\subsection{Numerical results}\label{Sec3.1}
\begin{widetext}
	
	\begin{figure}[]
		\centering
		\vspace{-0.5cm}
		\includegraphics[width=8.5cm, height=6.5cm]{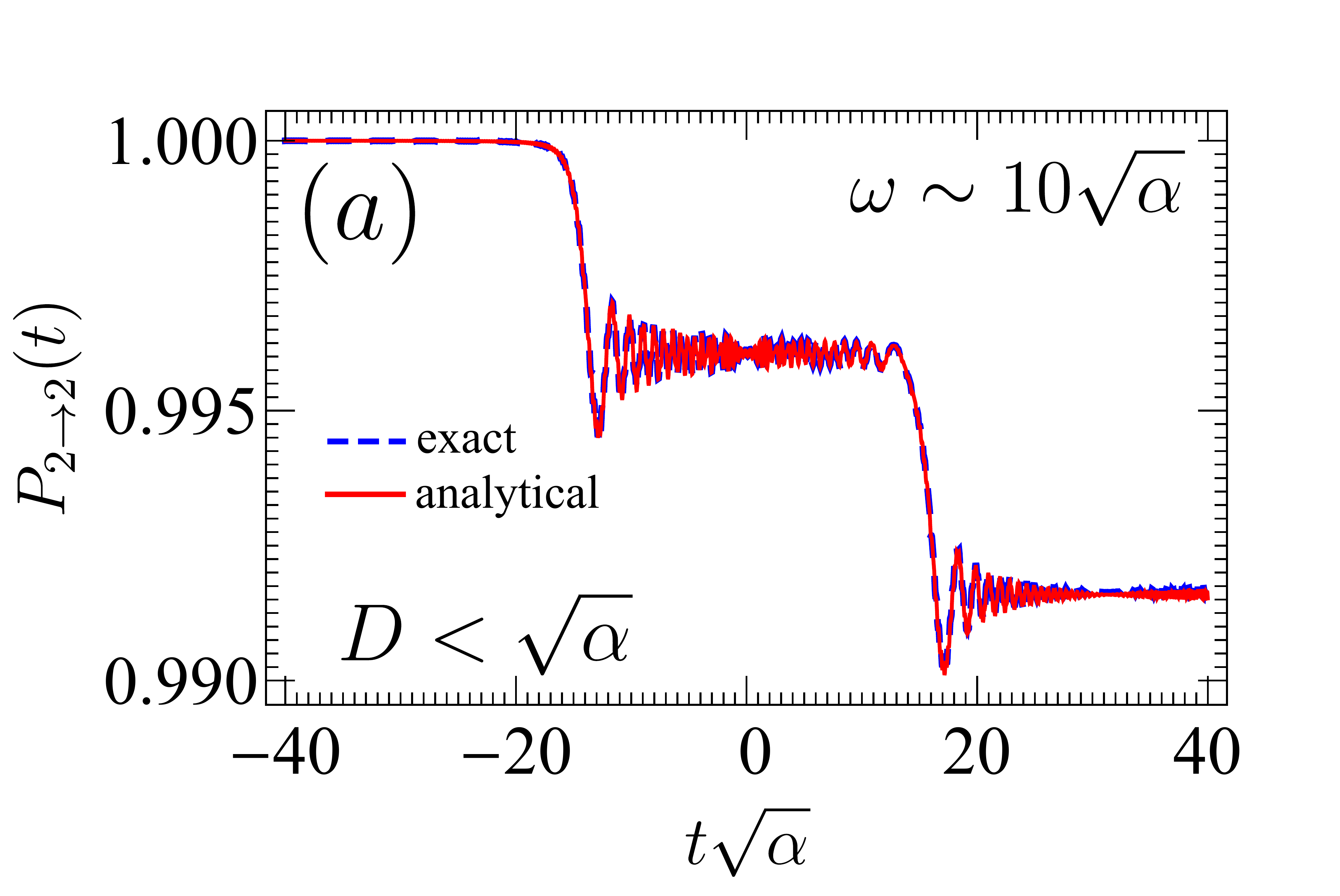}\hspace{-0.6cm}
		\includegraphics[width=8.5cm, height=6.5cm]{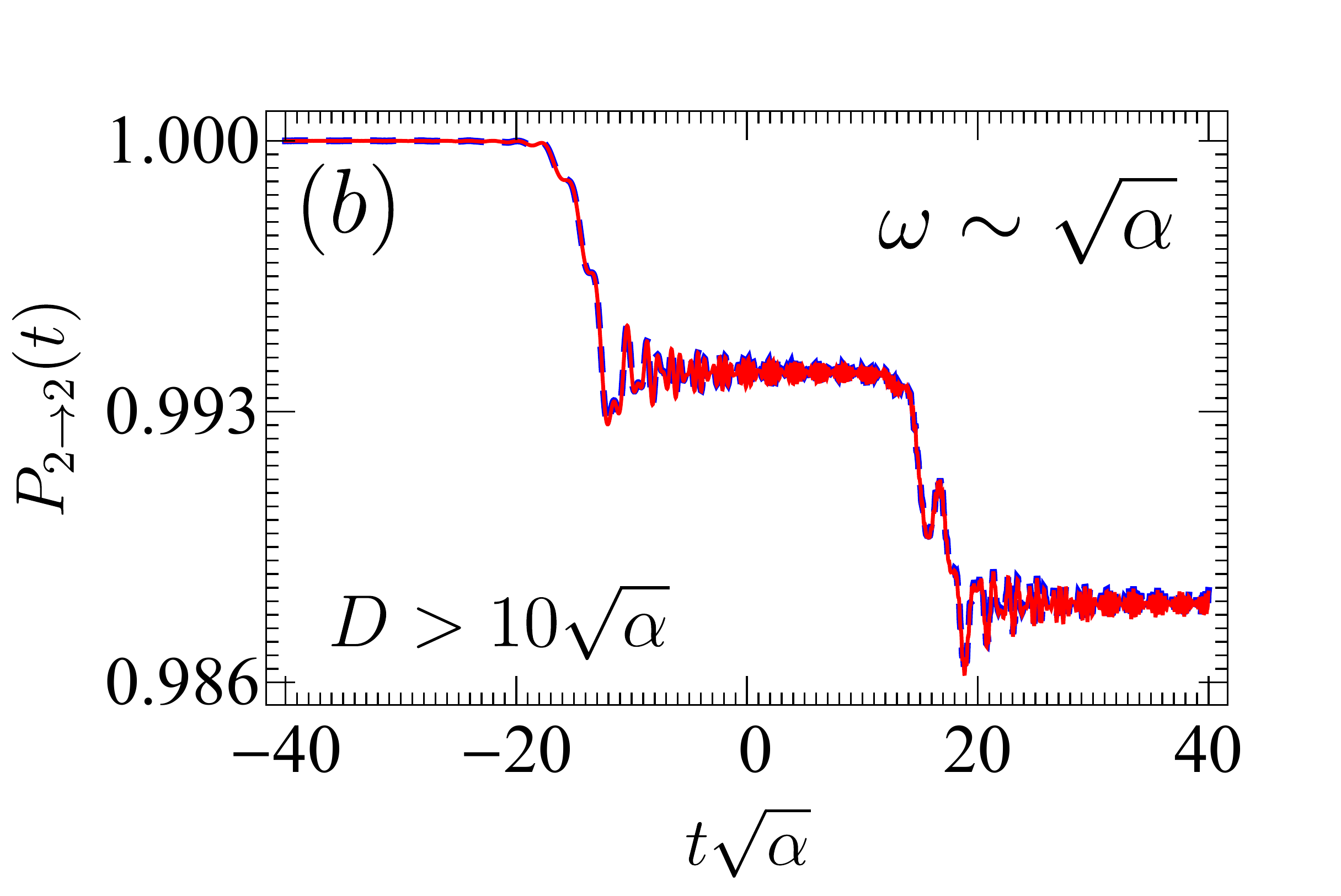}
		\vspace{-0.5cm}
		\caption{(Color Online)  Typical time-evolution of population remaining on the diabatic state $|2\rangle$ in the non-adiabatic limit. On panel $(a)$ we present a two-step pattern corresponding to the case when $D/\sqrt{\alpha}<1$ and $\omega/\sqrt{\alpha}\sim10$. In the circumstance, $D/\sqrt{\alpha}=0.05$ and $\omega/\sqrt{\alpha}=15$. This pattern also forms when $D/\sqrt{\alpha}>1$ and $\omega/\sqrt{\alpha}\ll1$ (see Ref.\onlinecite{Kiselev}). The panel $(b)$ shows coexistence of beats and steps patterns corresponding to  $D/\sqrt{\alpha}>10$ and $\omega/\sqrt{\alpha}\sim1$. Specifically,  $D/\sqrt{\alpha}=15$ and $\omega/\sqrt{\alpha}=1.5$. The time is in the unit of $1/\sqrt{\alpha}$ and $A/\sqrt{\alpha}=0.005$. } \label{Figure2}
	\end{figure}
\end{widetext}

\begin{widetext}
 	
 	\begin{figure}[]
 		\vspace{-0.5cm}
 		\includegraphics[width=8.5cm, height=6.5cm]{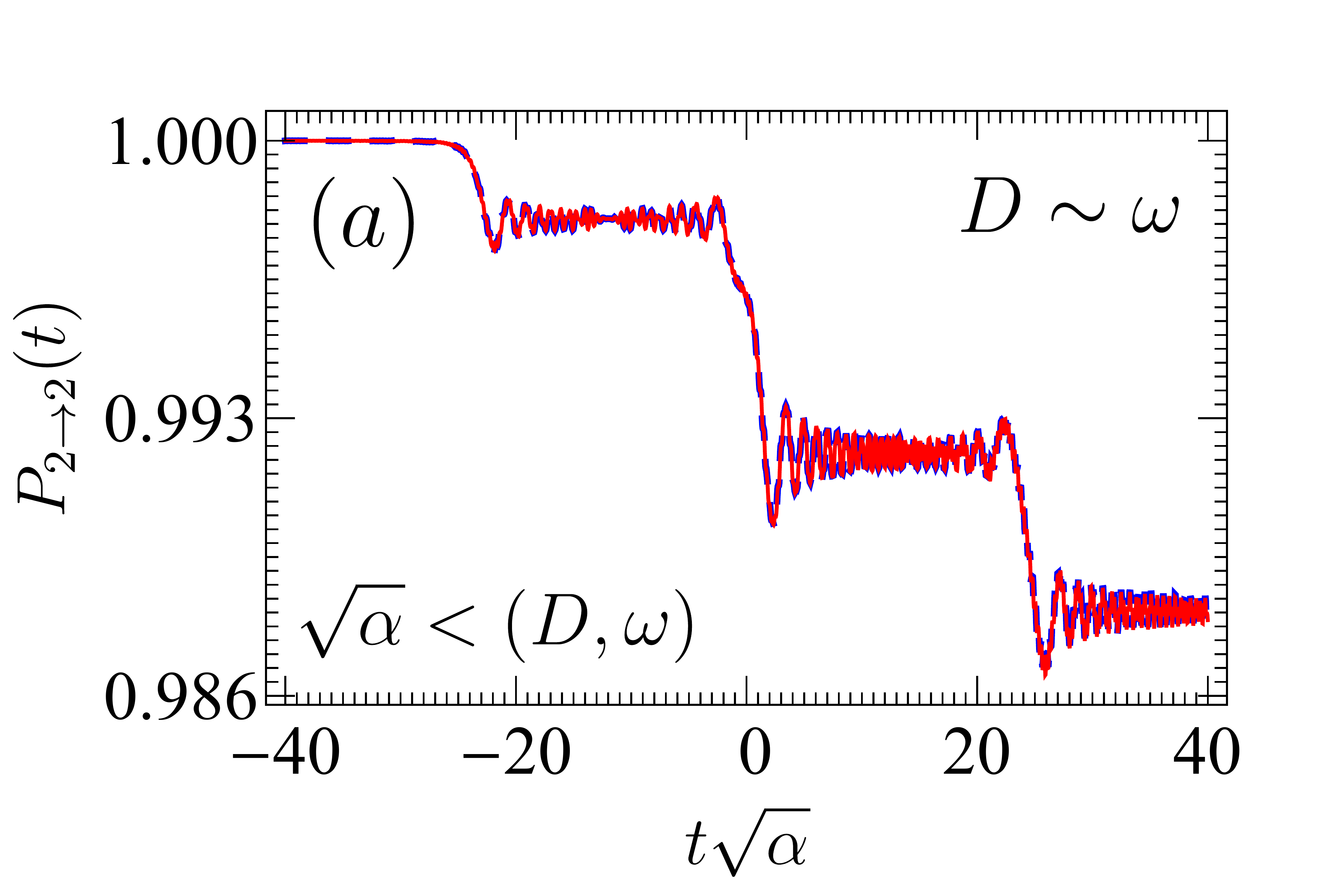}\hspace{-0.8cm}
 		\includegraphics[width=8.5cm, height=6.5cm]{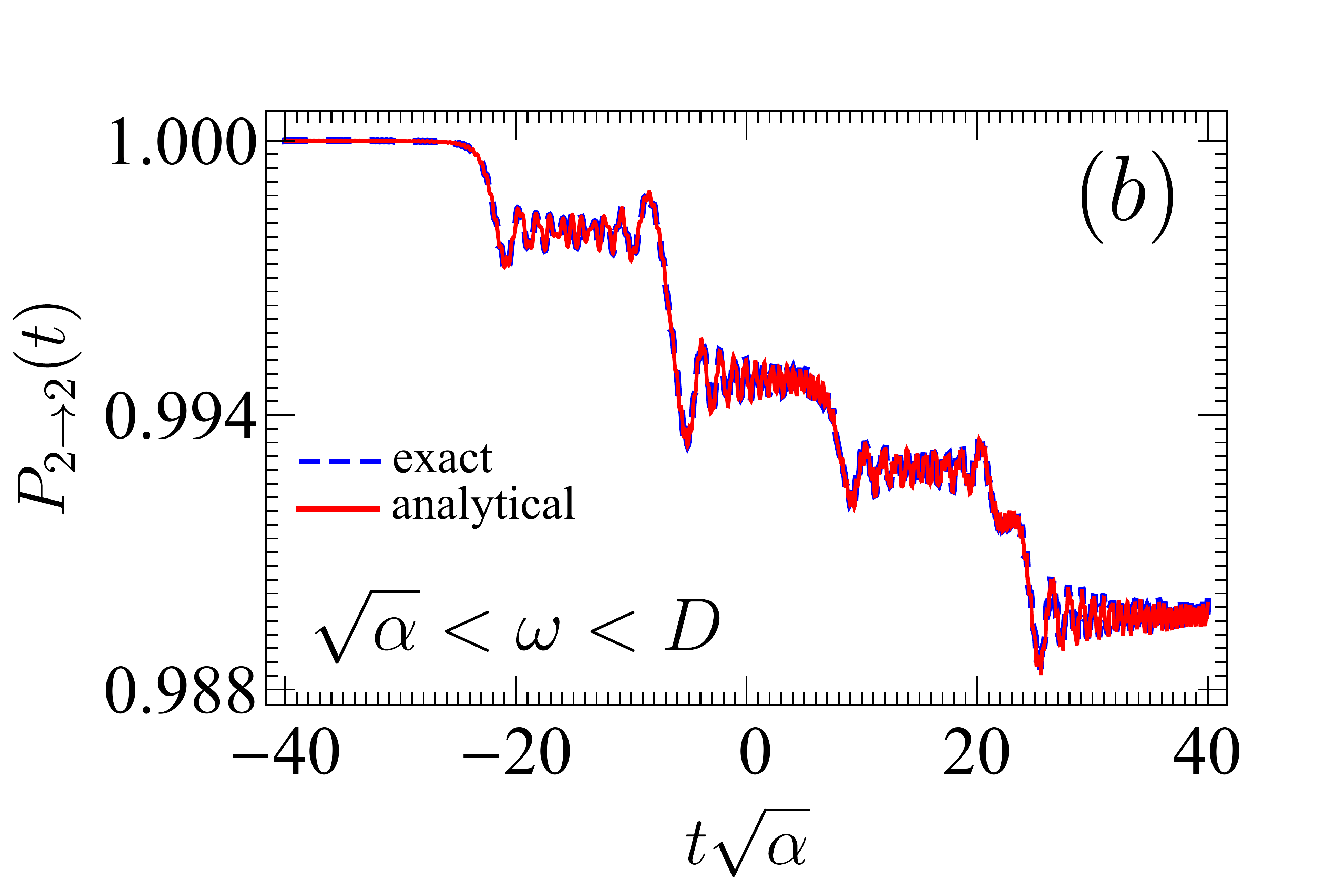}
 		\vspace{-0.5cm}
 		\caption{(Color Online)  Typical time-evolution of population remaining on the diabatic state $|2\rangle$ in the non-adiabatic limit.  On panel $(a)$ we present a three-step pattern achieved when $D/\sqrt{\alpha}$ and $\omega/\sqrt{\alpha}$ are large and of the same order of magnitude ($D=\omega=12\sqrt{\alpha}$). Panel $(b)$ displays a four-step pattern  corresponding to  $1< \omega/\sqrt{\alpha}< D/\sqrt{\alpha}$ or $1< D/\sqrt{\alpha}< \omega/\sqrt{\alpha}$. Specifically,  $D/\sqrt{\alpha}=15$ and $\omega/\sqrt{\alpha}=8$. Remark: $D/\sqrt{\alpha}=8$ and $\omega/\sqrt{\alpha}=15$ produces the same result.  The time is in the unit of $1/\sqrt{\alpha}$ and $A/\sqrt{\alpha}=0.005$ for all plots. } \label{Figure3}
 	\end{figure}
 	
 \end{widetext}

The von Neumann equation (\ref{equ3}) is  numerically solved in the non-adiabatic limit $A^{2}/\alpha\ll 1$.  The three specific cases of interest in this paper are listed below and the relevant observations are presented. The case $\omega/\sqrt{\alpha}\ll1$ is not considered as in this limit, the coupling $f(t)$ weakly changes with time and the study reduces to Ref.\onlinecite{Kiselev}.
For a small-size triangle $D/\sqrt{\alpha}\ll1$ and: $\omega/\sqrt{\alpha}\sim1$, beats are observed (not shown); $\omega/\sqrt{\alpha}\sim10$,  a two-step is observed (see Fig.\ref{Figure2}(a)). For a moderately large-size triangle $D/\sqrt{\alpha}>1$ and:
  $\omega/\sqrt{\alpha}\sim 1$,  beats  are  observed (not shown);  $\omega/\sqrt{\alpha}\sim 10$, a three-step pattern is observed (see Fig.\ref{Figure3}(a)).
For a large-size triangle $D/\sqrt{\alpha}\gg1$ and: $\omega/\sqrt{\alpha}\sim 1$, coexistence of beats and steps (see Fig.\ref{Figure2}(b)); $\omega/\sqrt{\alpha}\sim 10$, a four-step pattern is observed (see Fig.\ref{Figure3}(b)).

 Essential of our numerical results are depicted on Figs.\ref{Figure2} and \ref{Figure3}. The remarkable fact is that  by varying the detuning from negative to positive values ($-40\le t\sqrt{\alpha}\le40$) the system mainly remains in $|2\rangle$ as only a few percentage ($\sim 0.1\%$) of the population is non-adiabatically transferred to other diabatic states. The reason for this is that in the limit $A/\sqrt{\alpha}=0.005\ll1$ as considered, the detuning rapidly changes and the system quickly traverses the crossings without feeling the gaps.  On the panel $(a)$ of Fig.\ref{Figure2}, we show the formation of an unexpected two-step pattern  associated with a regime where $D/\sqrt{\alpha}<1$ and $\omega$ of the order of $10\sqrt{\alpha}$. On panel $(b)$ of the same figure, we have a coexistence of beats and steps when $D/\sqrt{\alpha}>10$ and $\omega$ is of the same order of magnitude as $\sqrt{\alpha}$    (frequency of non-adiabatic oscillations). On Fig.\ref{Figure3}$(a)$, $D/\sqrt{\alpha}$ and $\omega/\sqrt{\alpha}$ are both large and of the same order of magnitude. For this case, we have a three-step pattern while having four-step pattern on Fig.\ref{Figure3}$(b)$ corresponding to $1< \omega/\sqrt{\alpha} < D/\sqrt{\alpha}$ with $\omega$ of the order of $10\sqrt{\alpha}$. It should be noted that satisfying the inequality $1< D/\sqrt{\alpha} < \omega/\sqrt{\alpha}$ reproduces the four-step pattern in Fig.\ref{Figure3}$(b)$.   
 
Beats and/or steps form depending on how the characteristic time $\delta t$ is with respect to the relaxation time\cite{Vitanov1998} $t_{\rm relax}$.
Prepare the system in the diabatic state $\ket{2}$ far from the left of the first crossing. If it goes through two vertices of the triangle within a time interval $\delta t$ shorter than $t_{\rm relax}$ and finally returns to $\ket{2}$ far from the right of the last crossing, then beats are observed in the population of the level. In other words, if within a time interval shorter than the time of relaxation the ThLS undergoes several LZSM transitions, then its final population depicts beats\cite{comment3, comment4}. These observations corroborates the theory of quantum beats, which suggests that if for instance an electron receives several impulses within a time interval shorter than its relaxation time and gets to the excited state, it returns to its original state by emitting beats\cite{Wal}. Steps form in the opposite situation when $\delta t>t_{\rm relax}$ as the system relaxes enough between two crossings. To understand the coexistence of beats and steps as well as three- and four-step patterns one should envisage additional crossings likely induced by the transverse signal.  

 Note that the tendencies observed here are different from those observed in Ref.\onlinecite{Kiselev} with constant couplings ($f(t)={\rm const}$). It was reported in the reference therein that for small-size triangles, interactions between paths on the triangle lead to the formation of beats while for large-size triangles they lead to steps. Here, we observed that such a conclusion is significantly altered when the coupling between levels is no longer constant but periodically changes with time. 
 Remarkably, we have also observed (but not shown) that the populations transferred from $|2\rangle\to |1\rangle$ and  from $|2\rangle\to |3\rangle$ also oscillate in time and lead to the formation of beats and steps which is not the case with constant interactions (see Ref.\onlinecite{Kiselev}). These interesting behavior are no longer fully attributed to the $SU(3)$ dynamics but also to the periodic drive. They are explained in terms of Fresnel integrals. 
 
\subsection{Analytical results}\label{Sec3.2}

We construct analytical expressions that describe the gross temporal profile of populations and help in explaining the behavior numerically observed. Thus, in the non-adiabatic limit $A^{2}/\alpha\ll 1$, we perturbatively solve Eqs.(\ref{equ4})-(\ref{equ6}) and obtain the transition matrix in Eq.(\ref{equ9}). However, as far as this paper is concerned, we are only interested in 
$P_{2\to2}(t)\approx 1-p_{+}(t)-p_{-}(t)+\mathcal{O}(\delta^{2})$ mainly because it describes oscillations of population in the only diabatic state which does not undergo a splitting by Stark effects as $|1\rangle$ and $|3\rangle$.
 Defining the level crossing parameter $\delta=A^2/4\alpha$ and the angles $\vartheta^{\mp}=\phi\mp D\omega/\alpha$, we  have obtained
\begin{widetext}

\begin{eqnarray}\label{equ14}
\nonumber p_{\pm}(t)=\pi\delta\Big[F\Big(t\pm\frac{D\mp\omega}{\alpha},t\pm\frac{D\mp\omega}{\alpha}\Big)+F\Big(t\pm\frac{D\pm\omega}{\alpha},t\pm\frac{D\pm\omega}{\alpha}\Big)\hspace{0.9cm}\\
+2F\Big(t\pm\frac{D\pm\omega}{\alpha},t\pm\frac{D\mp\omega}{\alpha}\Big)\cos2\vartheta^{\mp}+2G\Big(t\pm\frac{D\pm\omega}{\alpha},t\pm\frac{D\mp\omega}{\alpha}\Big)\sin2\vartheta^{\mp}\Big],
\end{eqnarray}

\end{widetext}
where
 \begin{eqnarray}\label{equ14a}
 \nonumber F(x,y)=\frac{1}{2}\Big[\Big(\frac{1}{2}+C\Big(\sqrt{\frac{\alpha}{\pi}}x\Big)\Big)\Big(\frac{1}{2}+C\Big(\sqrt{\frac{\alpha}{\pi}}y\Big)\Big)\\+
 \Big(\frac{1}{2}+S\Big(\sqrt{\frac{\alpha}{\pi}}x\Big)\Big)\Big(\frac{1}{2}+S\Big(\sqrt{\frac{\alpha}{\pi}}y\Big)\Big)\Big],
 \end{eqnarray}
 and
 \begin{eqnarray}\label{equ14b}
 \nonumber G(x,y)=\frac{1}{2}\Big[\Big(\frac{1}{2}+C\Big(\sqrt{\frac{\alpha}{\pi}}x\Big)\Big)\Big(\frac{1}{2}+S\Big(\sqrt{\frac{\alpha}{\pi}}y\Big)\Big)\\-
 \Big(\frac{1}{2}+S\Big(\sqrt{\frac{\alpha}{\pi}}x\Big)\Big)\Big(\frac{1}{2}+C\Big(\sqrt{\frac{\alpha}{\pi}}y\Big)\Big)\Big].
 \end{eqnarray}
 Here, $C(...)$ and $S(...)$ are cosine and sine Fresnel integrals\cite{Abramo}. 
The function $F(x,y)$ describes two "waves" coming/emerging from two vertices respectively located at points $t'$ and $t''$ (where $x(t')=0$ and $y(t'')=0$) and interacting constructively at $x(t)=y(t)$. The function $G(x,y)$ typically describes the same waves in the scenario where they rather interact destructively at $x(t)=y(t)$ and thus, $F(x,x)$ describes a single wave coming/emerging from the vertex $x(t)=0$. 

 Let us have a close look at expression (\ref{equ14}). At first glance, it reproduces the result in Ref.\onlinecite{Kiselev} for $\omega=\phi=0$ (constant coupling) and allows for investigating the effects of the phase shift $\phi$ on transitions between bare states and is then suitable for analyzing interference processes. Indeed, $p_{+}(t)$ and $p_{-}(t)$ describe a  maximum of two consecutive LZSM transitions that take place respectively within the regions $]-\infty,0]$ (negative time domain) and $[0,+\infty[$ (positive time domain). $p_{+}(t)$ depicts two transitions, one that occurs at $t=-(D+\omega)/\alpha$ immediately followed by another occurring at $t=-(D-\omega)/\alpha$. $p_{-}(t)$ does typically the same in the positive domain; the first transition occurs at $t=+(D-\omega)/\alpha$  and is followed at $t=+(D+\omega)/\alpha$ by the second one.  Thus for instance, if the signal is tuned such that $\phi=0$ or $\phi=\pi/4$ and $D^{2}/\alpha=\omega^{2}/\alpha=\pi N$ where $N=1,2,3,...,$ one LZSM transition is suppressed and expression (\ref{equ14}) leads to three consecutive steps. The reason for this is that in each of these cases, contribution from one of the terms (third or fourth) in (\ref{equ14}) vanishes. 

Our analytic results are compared with exact numerical results of the von Neumann  equation and are displayed on Figs. \ref{Figure2} and \ref{Figure3} (solid red lines). We clearly see that both results are barely discernible and hold  in the limit $A/\sqrt{\alpha}\ll1$ for all $D$, $\omega$ and $\phi$. Another critical look at (\ref{equ14})  surprisingly reveals that formation of beats and steps occur even at $D=0$ when the $SU(3)$ symmetry breaks down   allowing us to actually assert that these patterns resulting from interferometry processes appear not necessary as results of $SU(3)$ deformation (addition of the zero-splitting energy $D(S^{z})^{2}$) but unavoidably as consequences of the renormalization of inter-level distance between level positions by a transverse periodic drive.  

Returning to Figs.\ref{Figure2} and \ref{Figure3} provided the analytical results (\ref{equ14}), we confirm that in the weak driving limit ($\omega\ll\sqrt{\alpha}$), the formation of beats and steps mostly depends on $D/\sqrt{\alpha}$. In the  strong driving limit ($\omega\gg\sqrt{\alpha}$), the periodic signal creates additional paths in the ThLS dynamics (photon-induced tunneling). This appears as a dynamical Stark effect (associated with the electric field $f(t)$) accompanying the linear Zeemann effect caused by the magnetic field $\vec{B}_{0}(t)=[0,0,\alpha t]^{T}$ (in the Cartesian coordinates system). Diabatic energies $\alpha t+D$ and $-\alpha t+D$ of  states with extremal spin projections in the original $SU(3)$ Hamiltonian doubly degenerate each. Level $|1\rangle$ splits into two sub-levels $|1,\pm\omega\rangle$ with energies $\alpha t+(D\pm\omega)$ and $|3\rangle$ typically does the same leading to sub-levels $|3,\pm\omega\rangle$ with energies $-\alpha t+(D\pm\omega)$ (see upper panels in Figs.\ref{Figure4} and \ref{Figure4k}). A photon with energy $+\omega$ is absorbed and  another is emitted after transition with an energy $-\omega$ (multi-photon process).

\subsection{Quantized fields description}\label{Sec3c}
For better understanding, it might be relevant to describe the $SU(3)$ protocol in terms of quantized fields. We quantize the electric field in a cavity and show that in the corresponding scenario, the ThLS interacts with a two-mode harmonic oscillator.  Thus,   $\mathcal{H}(t)=\alpha tS^{z}+\mathcal{H}_{\rm bath}+\mathcal{H}_{\rm ThLS-bath}+D(S^{z})^{2}$ 
where the bath Hamiltonian $\mathcal{H}_{\rm bath}=\omega(\hat{a}^{\dagger}\hat{a}-\hat{b}^{\dagger}\hat{b})$  and the system-bath interaction Hamiltonian reads $\mathcal{H}_{\rm ThLS-bath}=[ g_{a}(\hat{a}^{\dagger}+\hat{a})+g_{b}(\hat{b}^{\dagger}+\hat{b})]S^{x}$ and where $g_{a,b}$ is the coupling strength between the ThLS and the mode $a$ or $b$. $\hat{a}^{\dagger}(\hat{a})$ and $\hat{b}^{\dagger}(\hat{b})$ are the creation(annihilation) operators of photons in each mode. $\langle \hat{a}^{\dagger}\hat{a}\rangle_{0}=n_{a}$ and $\langle \hat{b}^{\dagger}\hat{b}\rangle_{0}=n_{b}$ where $n_{a,b}$ is the number of oscillators of $a,b$ type in the cavity; $\langle...\rangle_{0}$ is the thermal disordered average.  The  minus sign in $\mathcal{H}_{\rm bath}$  is chosen  such that in the classical limit its contribution vanishes. Indeed, in the classical limit, the bosonic operators $\hat{a}$ and $\hat{b}$ are replaced by $\sqrt{n_{a}}e^{i(\omega t+\phi_{q})}$ and $\sqrt{n_{b}}e^{i(\omega t+\phi_{q})}$ respectively and $\hat{a}^{\dagger}$ and $\hat{b}^{\dagger}$ are replaced by $\sqrt{n_{a}}e^{-i(\omega t+\phi_{q})}$ and $\sqrt{n_{b}}e^{-i(\omega t+\phi_{q})}$ respectively\cite{Ashab}. Here, $\phi_{q}$ is the quantum analogue of the classical phase shift $\phi$. It clearly appears that $\mathcal{H}_{\rm bath}$ vanishes and $\mathcal{H}_{\rm ThLS-bath}$ leads us to the periodic term in Eq.(\ref{equ1}). The classical and the quantum limits are equivalent when $\phi_{q}$ and $\phi$ are all set to zero and
\begin{eqnarray} \label{equ15a} 
A=2(\sqrt{n_{a}}g_{a}+\sqrt{n_{b}}g_{b}).
\end{eqnarray}
To reproduce the results of the semi-classical treatment, let us consider the $SU(2)$ coherent states and  assume that no more than four photons coexist in the cavity at the same time i.e. $n_{a,b}=(1,2)$ (possible number of excitations). Hence, if $n_{a}=1$ photon is observed in $a$-mode, then a maximum of $n_{b}=2$ is observed in $b$-mode;  similarly, if $n_{a}=2$ photons are found in  $a$-mode, then a maximum of $n_{b}=2$ is observed in $b$-mode, photon-photon interactions being overlooked. We have for instance observed that when $n_{a,b}=2$ i.e. four photons in the cavity, the number of steps reduces to two. Therefore, given that diabatic energies are $E_{m,\{n_{a},n_{b}\}}(t)=m\alpha t+Dm^{2}+\omega(n_{a}-n_{b})$ (where $m$ is now the eigenvalue of $S^{z}$) thus, $E_{+1,\{2,1\}}(t)=\alpha t+(D+\omega)$ and $E_{+1,\{1,2\}}(t)=\alpha t+(D-\omega)$ are respectively diabatic energies of $|1,\omega\rangle$ and $|1,-\omega\rangle$ while $E_{-1,\{1,2\}}(t)=-\alpha t+(D-\omega)$ and $E_{-1,\{2,1\}}(t)=-\alpha t+(D+\omega)$ are those of $|3,-\omega\rangle$ and $|3,\omega\rangle$ respectively. The middle diabatic state $|2\rangle$ is a dark state with energy $E_{0,\{1,1\}}(t)=0$.  In the cavity, we also assume two optical pumps ${\bf p_{a}}$ and ${\bf p_{b}}$ that generate the modes $a$ and $b$ respectively.  They are tuned such that the transitions $|2\rangle \leftrightarrow|1,\pm\omega\rangle$ are ensured by ${\bf p_{a}}$ while $|2\rangle \leftrightarrow|3,\pm\omega\rangle$ are guaranteed by ${\bf p_{b}}$.  Thus, the coupling $g_{a}$ depends on the number $n_a$ of photons injected by ${\bf p_{a}}$ and similarly $g_{b}$ is a function of $n_b$ created by ${\bf p_{b}}$.  Direct transitions $|1,\pm\omega\rangle \leftrightarrow|3,\pm\omega\rangle$ are forbidden. Instead, the middle diabatic state $|2\rangle$ acts as a shuttle mediating population transfer between these states of extremal spin projections. Also, the  transitions $|1,\omega\rangle \leftrightarrow|1,-\omega\rangle$ and $|3,\omega\rangle \leftrightarrow|3,-\omega\rangle$ between the states issued from the same splitting are entirely forbidden. Truncating the Hamiltonian in a space spanned by the basis $\{|1,\omega\rangle, |1,-\omega\rangle, |2\rangle, |3,-\omega\rangle,|3,\omega\rangle\}$, we have (see Appendix \ref{App2})

\begin{widetext}
	
	\begin{eqnarray} \label{equ15} 
	\mathcal{H}(t)=
	\left(
	{\begin{array}{*{20}c}
		\alpha t+(D+\omega) &  0 & \lambda_{1,2}^{\bf p_{a}}/\sqrt{2} & 0 & 0\\
		0 & \alpha t+(D-\omega) & \lambda_{\bar{1},2}^{\bf p_{a}}/\sqrt{2} & 0 & 0\\
		\lambda_{2,1}^{\bf p_{a}}/\sqrt{2} & \lambda_{2,\bar{1}}^{\bf p_{a}}/\sqrt{2}  & 0 &
		\lambda_{2,\bar{3}}^{\bf p_{b}}/\sqrt{2} & \lambda_{2,3}^{\bf p_{b}}/\sqrt{2}\\
		0 & 0 & \lambda_{\bar{3},2}^{\bf p_{b}}/\sqrt{2} & -\alpha t+(D-\omega) & 0\\
		0 & 0 & \lambda_{3,2}^{\bf p_{b}}/\sqrt{2} & 0 &-\alpha t+(D+\omega)\\
		\end{array} } \right). 
	\end{eqnarray} 
	
\end{widetext} 
Here,  $\lambda_{j,2}^{\bf p_{a}/p_{b}}=\langle \omega,j|\mathcal{H}|2\rangle$, $\lambda_{2,j}^{\bf p_{a}/p_{b}}=\langle 2 |\mathcal{H}|j,\omega\rangle$, $\lambda_{\bar{j},2}^{\bf p_{a}/p_{b}}=\langle -\omega,j|\mathcal{H}|2\rangle$ and $\lambda_{2,\bar{j}}^{\bf p_{a}/p_{b}}=\langle 2|\mathcal{H} |j,-\omega\rangle$ with $j=1,3$. The model (\ref{equ15}) as a class of models investigated in Refs.[\onlinecite{Sun, Sinitsyn2, Sinitsyn3, Patra}] is integrable and its large positive times asymptotic solutions are written as by-product of the spin-$1/2$ LZSM formula\cite{lan, zen, stu, Majorana} (see Appendix \ref{App2}).

\begin{figure}[]
	\centering
	\vspace{-0.5cm}
	\includegraphics[width=8.5cm, height=6.5cm]{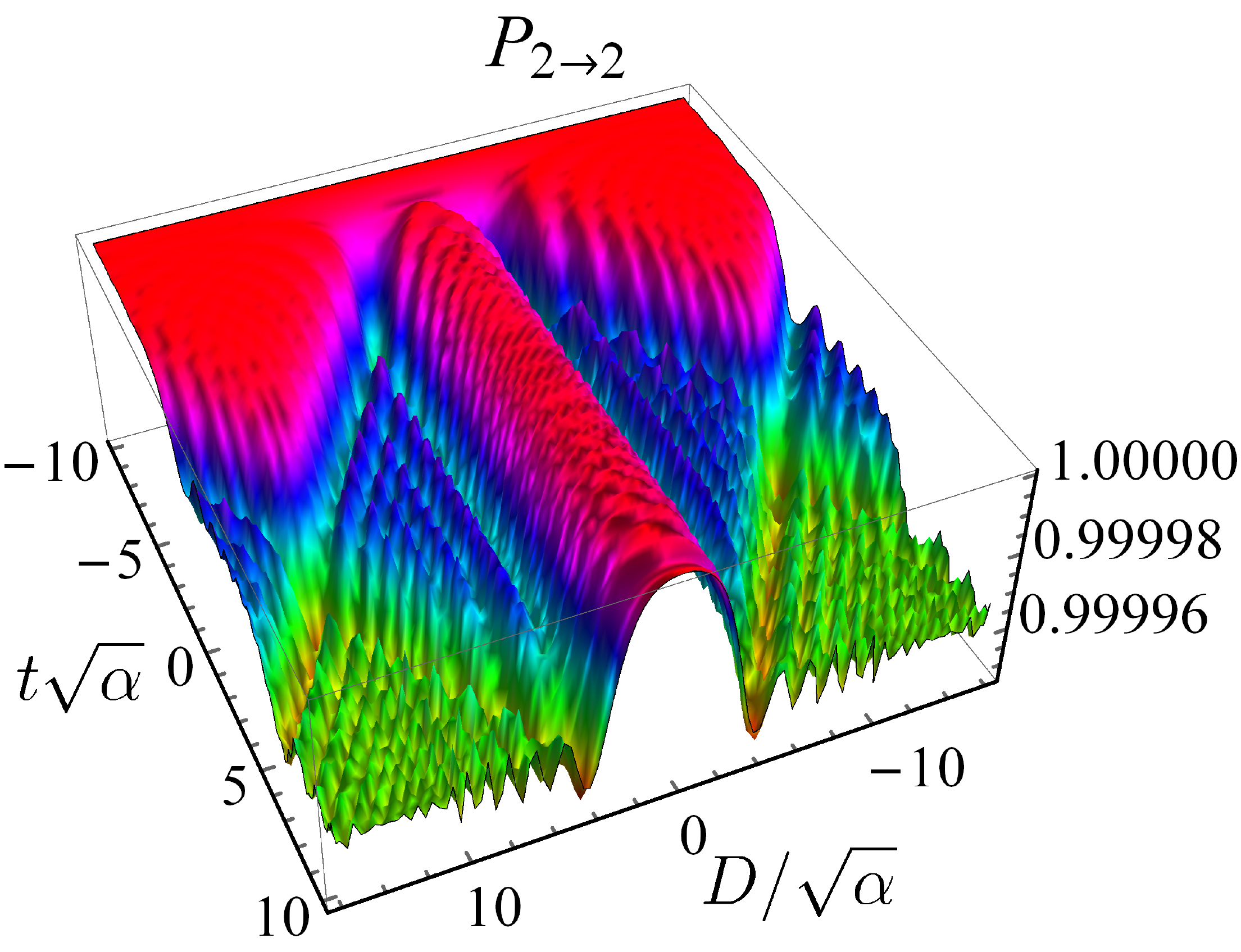}
	\vspace{-0.1cm}
	\caption{(Color Online) Interference pattern corresponding to population $P_{2\to2}(t)$ interpreted as a function of $t\sqrt{\alpha}$ and $D/\sqrt{\alpha}$ for $A/\sqrt{\alpha}= 0.005$, $\phi=0$, $t_{0}=-10/\sqrt{\alpha}$ (initial time),  and $\omega/\sqrt{\alpha}=12$. The graph corresponding to $-10\le D/\sqrt{\alpha}\le 0$ or $0\le D/\sqrt{\alpha}\le 10$ is obtained in Ref.\onlinecite{Kiselev} for constant couplings and  $\omega=0$. The fact that such a figure now appears twice is not only the consequence of the periodic field but the major fact according to which $\omega/\sqrt{\alpha}$ is large. } \label{Figure5}
\end{figure}

\begin{widetext}
	
	\begin{figure}[]
		\centering
		\vspace{-0.5cm}\hspace{0.6cm}
		\includegraphics[width=7.8cm, height=6cm]{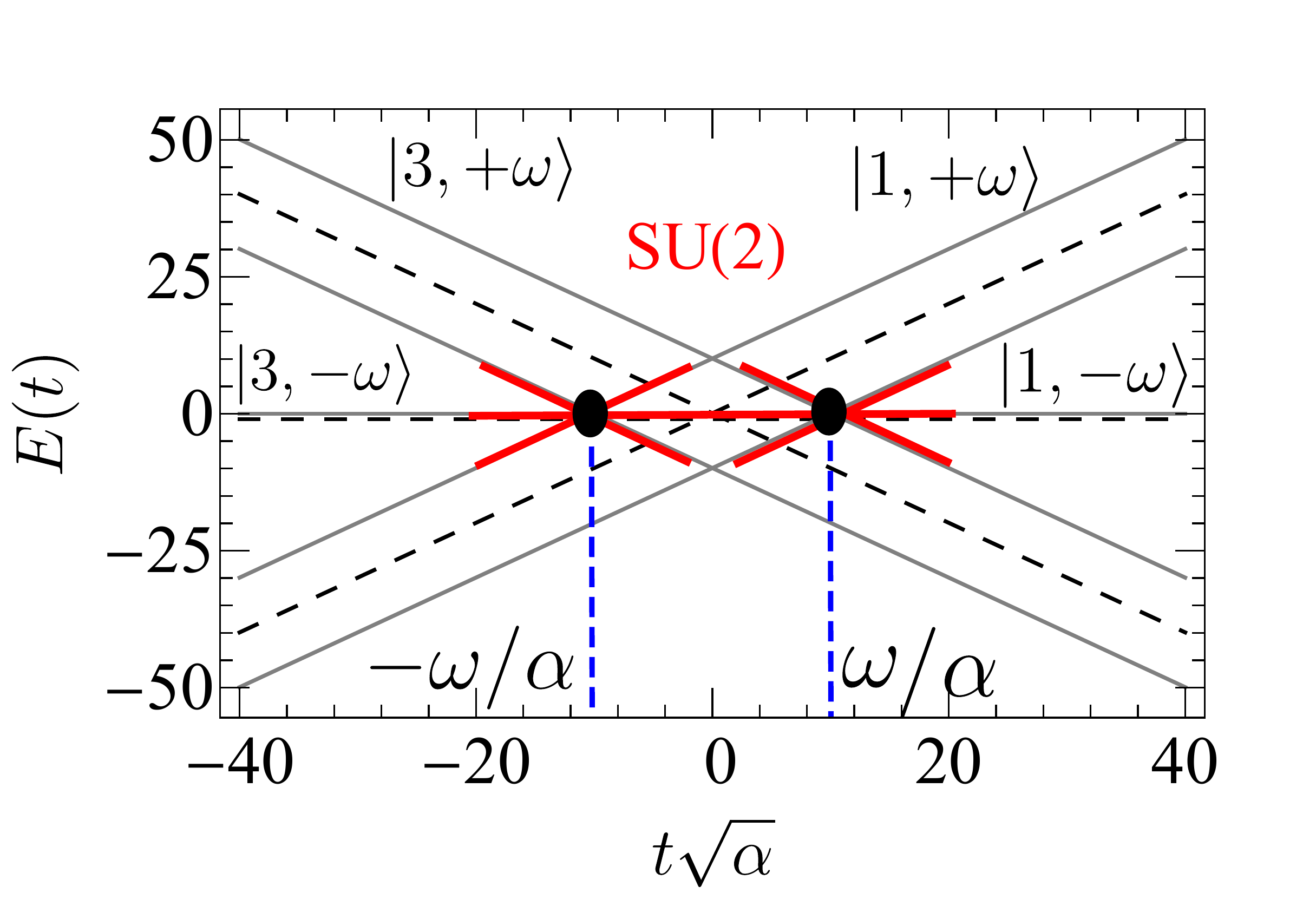}\hspace{0.3cm}
		\includegraphics[width=8.cm, height=6cm]{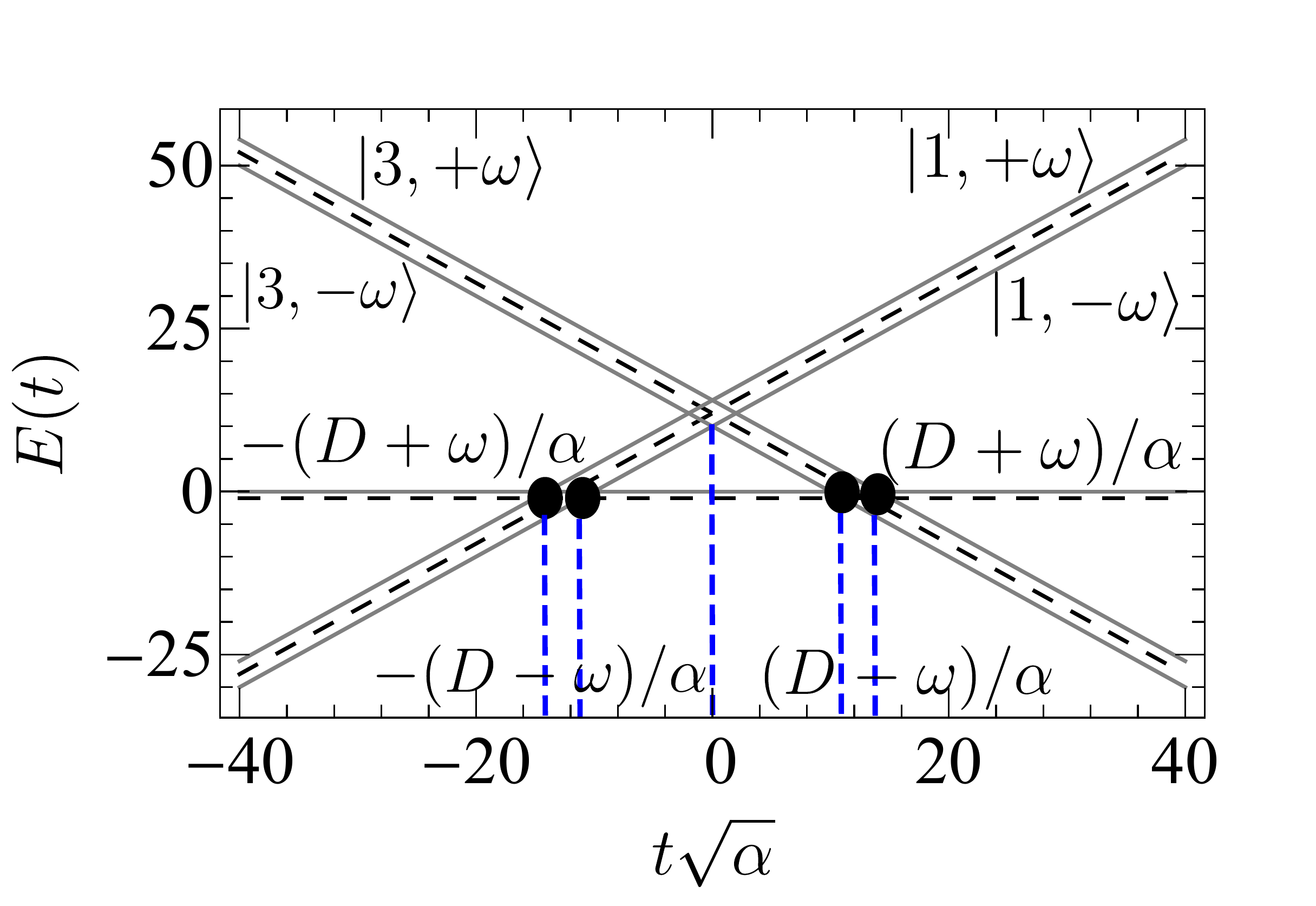}\vspace{-0.7cm}
		\includegraphics[width=9cm, height=6cm]{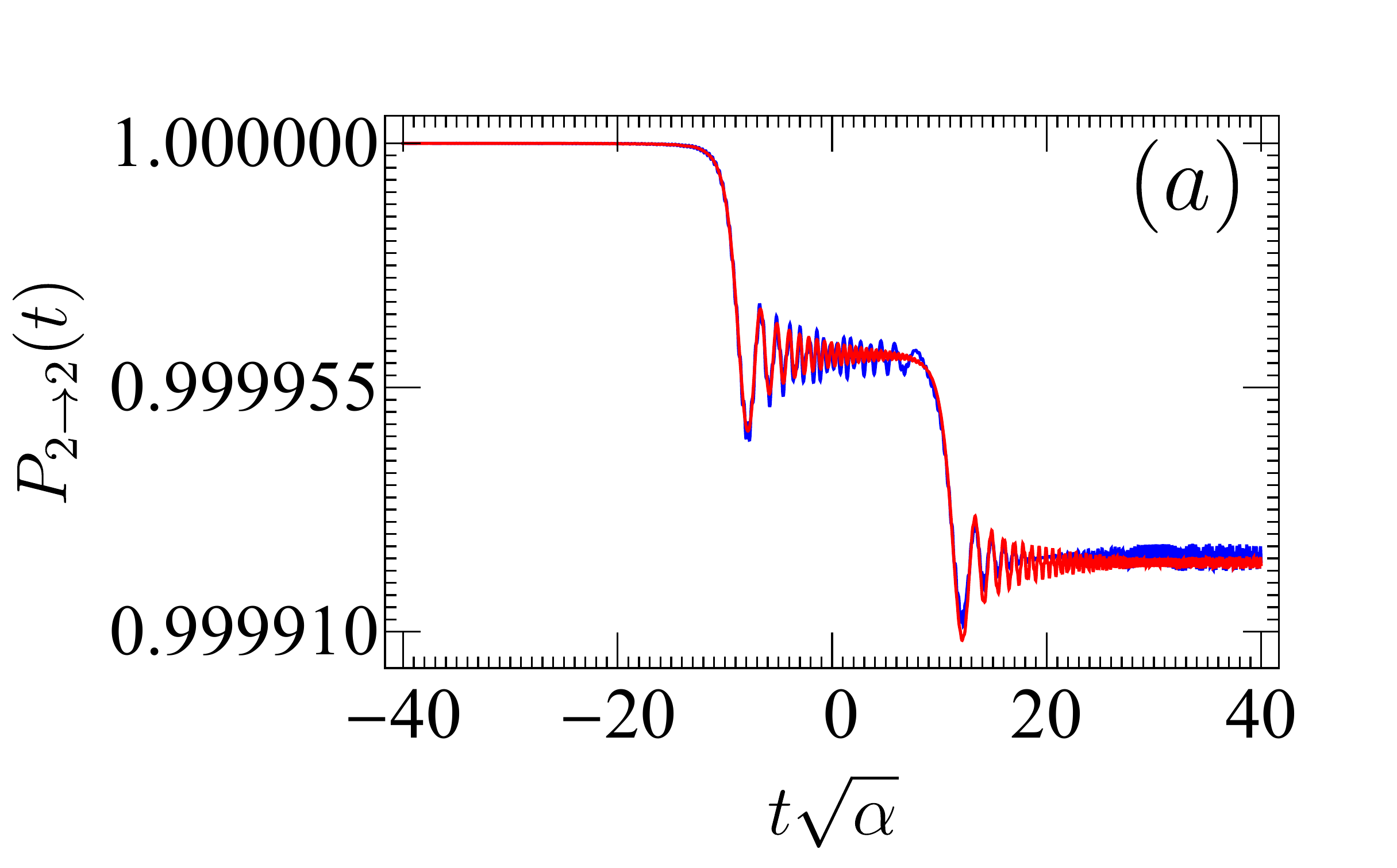}\hspace{-0.7cm}
		\includegraphics[width=9cm, height=6cm]{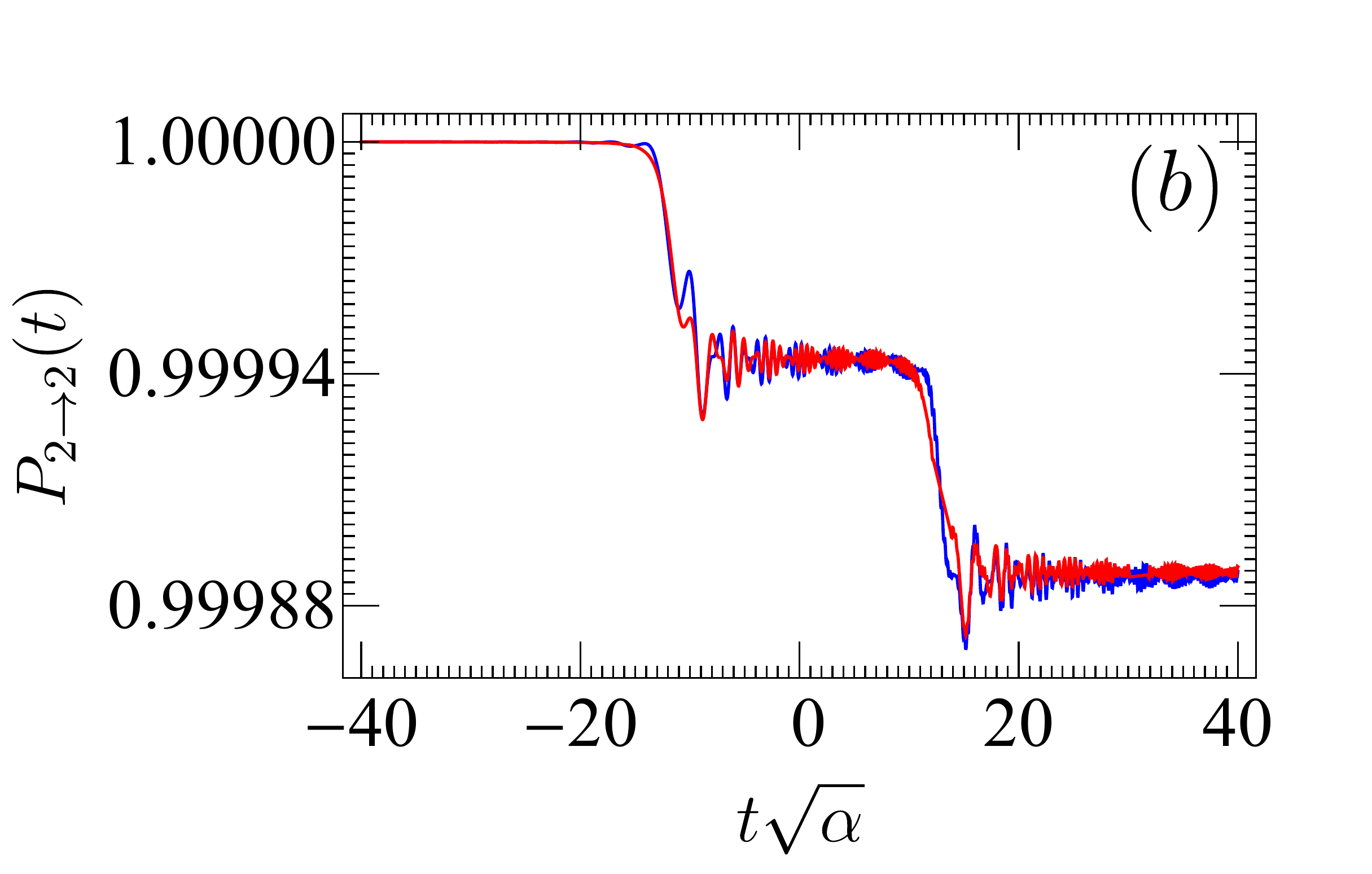}
		\vspace{-0.4cm}
		\caption{(Color Online) 
			{\bf Upper panel:} Energy diagrams of diabatic states of the model Eq.(\ref{equ1}) written in Eq.(\ref{equ15}) in terms of quantized fields. Gray dashed lines correspond to diabatic energies in the absence of periodic signal while black solid lines show their splitting in the presence of the periodic signal (Stark effects). Black balls indicate crossings. {\bf Lower panel:}
			Correspondence between the $SU(3)$ LZSM model with a transverse periodic drive and the $SU(5)$ model Eq.(\ref{equ15}) with time-independent couplings. We have compared our analytical solution $P_{2\to2}(t)$ (blue solid lines) in Eq.(\ref{equ14}) with the numerical solution (red solid lines) of the model Eq.(\ref{equ15}).  Fig.\ref{Figure4}$(a)$ (two-step): $D/\sqrt{\alpha}=0.08$, $\omega/\sqrt{\alpha}=10$ with $0.00176\sqrt{2\alpha}$ for all couplings.
		    	Fig.\ref{Figure4}$(b)$ (two-step coexisting with beats): $D/\sqrt{\alpha}=12$, $\omega/\sqrt{\alpha}=1$ with $0.0021\sqrt{2\alpha}$ for  all couplings.
			  For all plots, $A/\sqrt{\alpha}=0.005$. Panel $(a)$ shows two coupled spin-1 $SU(2)$ LZSM processes. Thus, the two-step pattern can also be viewed as results of interferences between two consecutive LZSM processes occurring within the time $\delta t>t_{\rm LZ}$.} \label{Figure4}
	\end{figure}
	
\end{widetext}

\begin{widetext}
	
	\begin{figure}[]
		\centering
		\vspace{-0.5cm}\hspace{0.5cm}
		\includegraphics[width=8.cm, height=6cm]{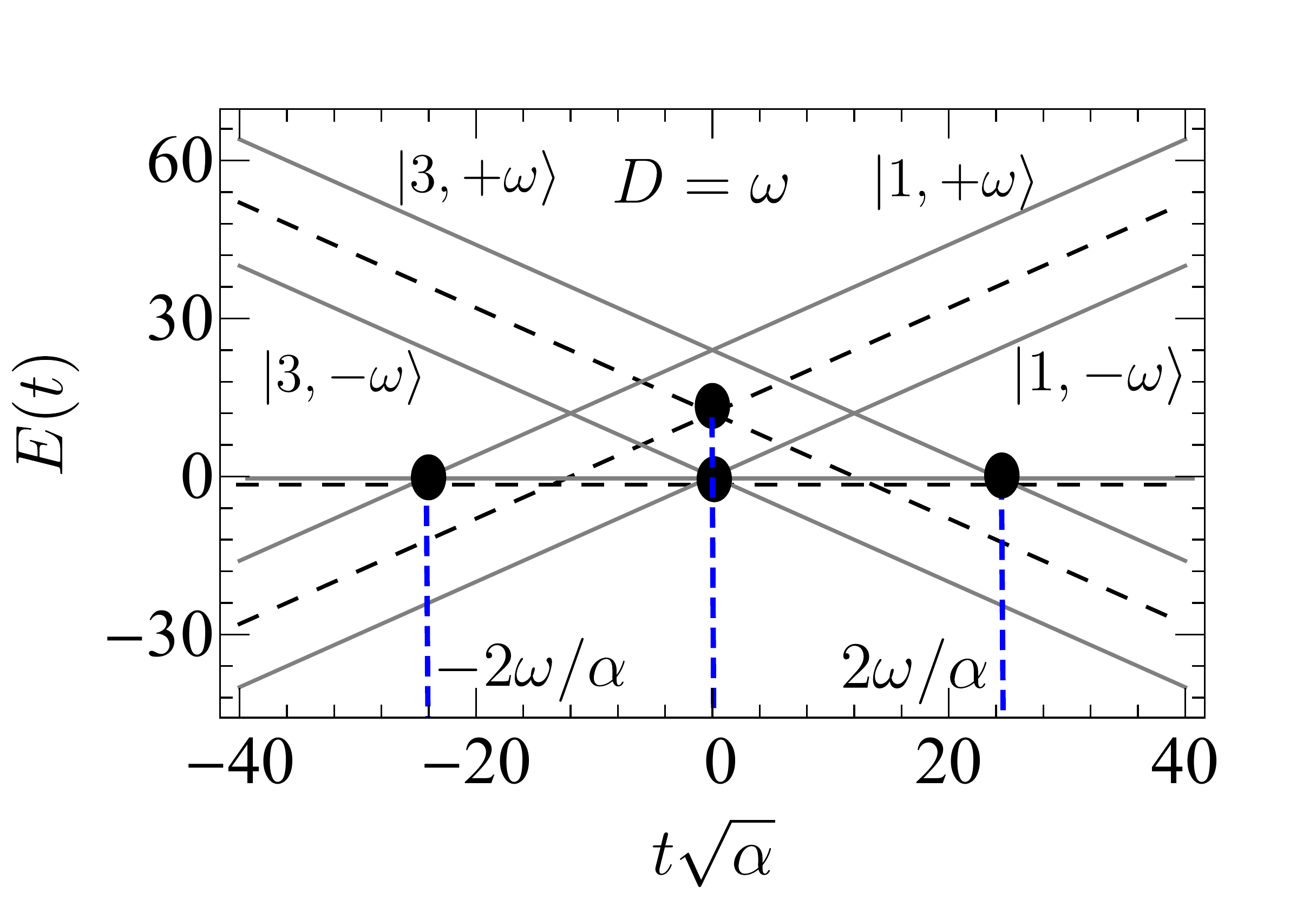}\hspace{0.2cm}
		\includegraphics[width=8.cm, height=6cm]{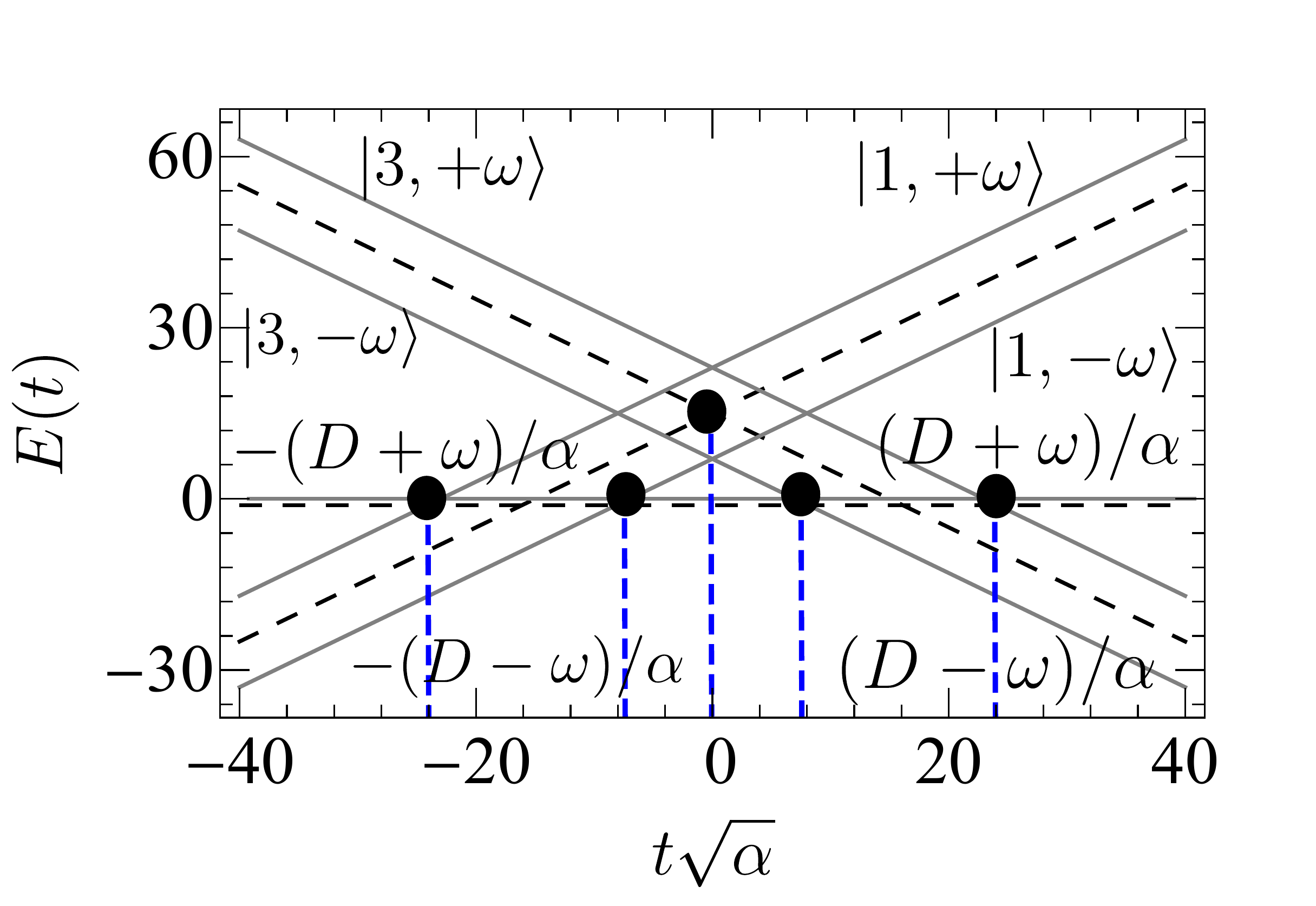}\vspace{-0.7cm}
		\includegraphics[width=9cm, height=6cm]{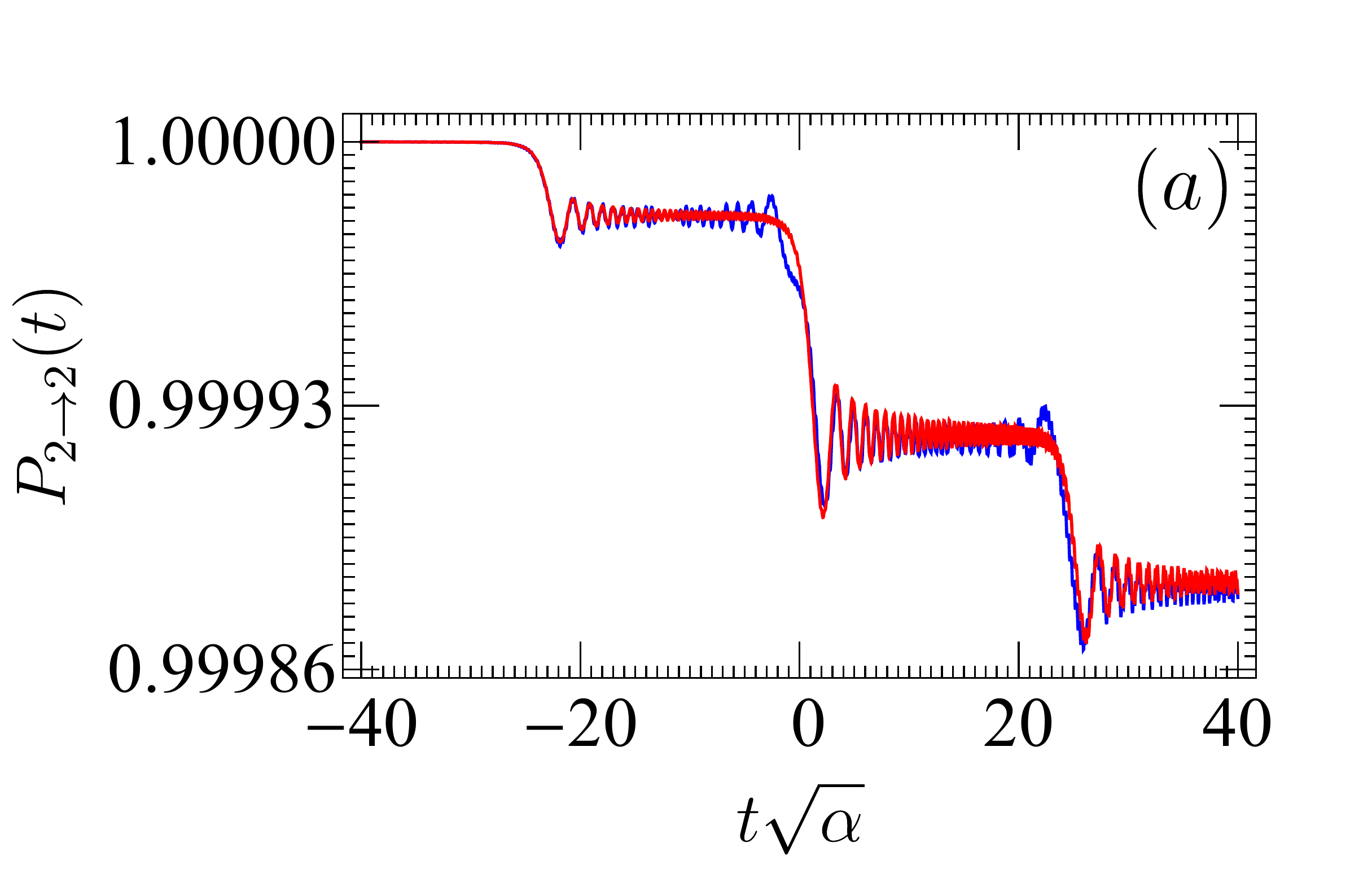}\hspace{-0.7cm}
		\includegraphics[width=9cm, height=6cm]{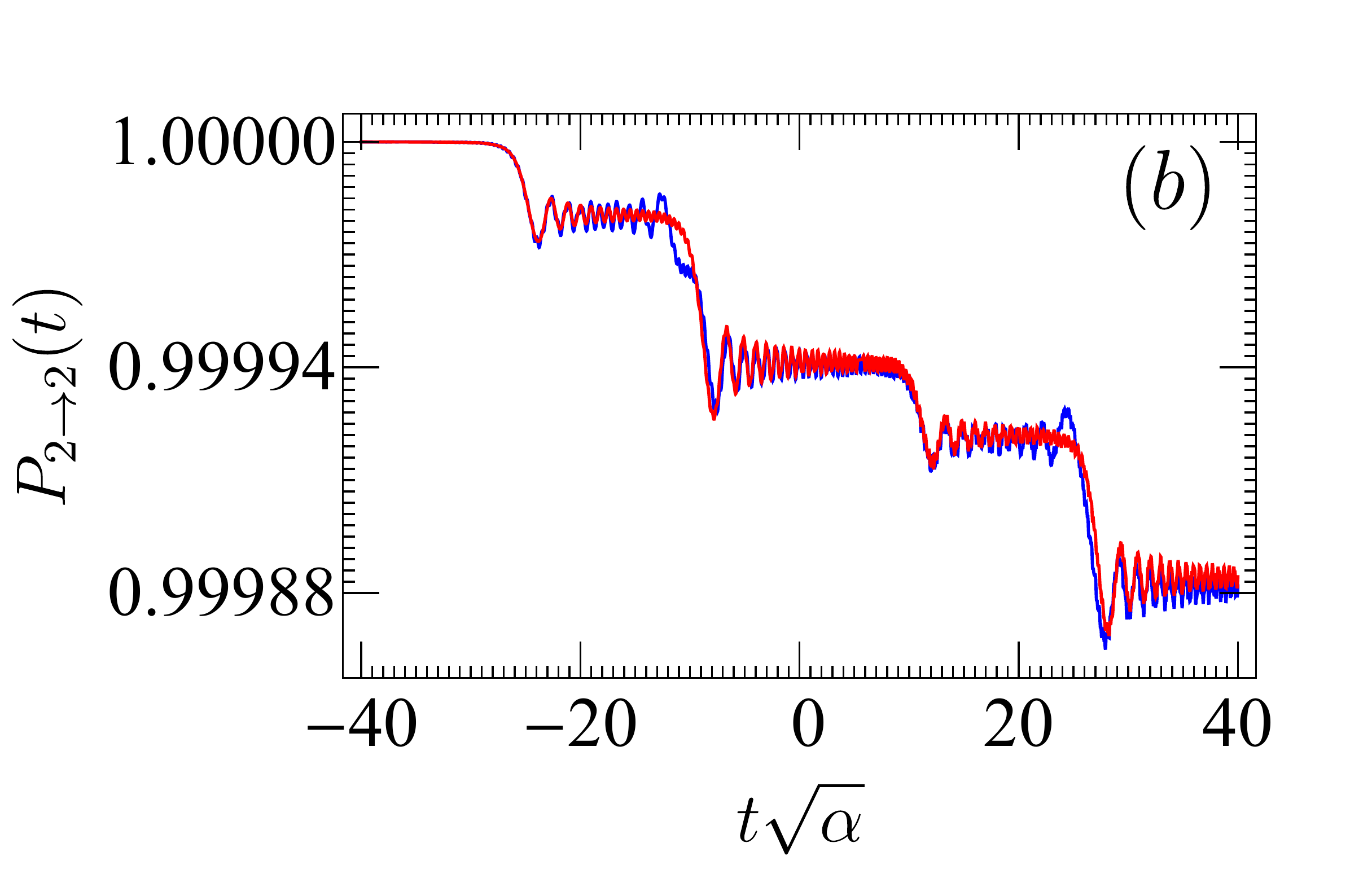}
		\vspace{-0.4cm}
		\caption{(Color Online) {\bf Upper panel:} Same as Fig.\ref{Figure4}.
			{\bf Lower panel:} Correspondence between the $SU(3)$ LZSM model with a transverse periodic drive and the $SU(5)$ model Eq.(\ref{equ15}) with time-independent couplings. Fig.\ref{Figure4k}$(a)$ (three-step): $D/\sqrt{\alpha}=12$, $\omega/\sqrt{\alpha}=12$, $\lambda_{1,2}^{\bf p_{a}}=\lambda_{\bar{1},2}^{\bf p_{a}}=0.00176\sqrt{2\alpha}$, $\lambda_{2,\bar{3}}^{\bf p_{b}}=\lambda_{2,3}^{\bf p_{b}}=0.0025\sqrt{2\alpha}$. Fig.\ref{Figure4k}$(b)$(four-step): $D/\sqrt{\alpha}=18$, $\omega/\sqrt{\alpha}=8$, $\lambda_{1,2}^{\bf p_{a}}=\lambda_{2,\bar{3}}^{\bf p_{b}}=0.00176\sqrt{2\alpha}$, $\lambda_{\bar{1},2}^{\bf p_{a}}=\lambda_{2,3}^{\bf p_{b}}=0.0025\sqrt{2\alpha}$. 
			For all plots $A/\sqrt{\alpha}=0.005$. Blue solid lines are analytical results in Eq.(\ref{equ14}) and red solid lines are numerical solutions of the model Eq.(\ref{equ15}). } \label{Figure4k}
	\end{figure}
	
\end{widetext}

 Remark, by canceling/erasing the first and last rows as well as the first and last columns in Eq.(\ref{equ15}) the remaining Hamiltonian is nothing but the $SU(3)$ LZSM model\cite{Kiselev} with anisotropy $D-\omega$ and is equivalent to the result of the rotative wave approximation (RWA) directly applied to Eq.(\ref{equ1}). These last observations indicate that we are clearly beyond the RWA which will fail in explaining the results of the semi-classical treatment. If one does same (canceled/erased) with the second and fourth rows/columns, the resulting $SU(3)$ LZSM Hamiltonian has anisotropy $D+\omega$. Remarkably, when $D=\omega=0$ the degeneracy of the state $\ket{1}$ and $\ket{3}$ is lifted. The model (\ref{equ15}) reduces to a spin-$1$ $SU(2)$ LZSM model. The energy diagrams associated with the model Eq.(\ref{equ15}) are found in Figs. \ref{Figure4} and \ref{Figure4k} (upper panels). They demonstrate that for an initial preparation of the system far from the left of the crossing point $t=-(D+\omega)/\alpha$ in the state $|2\rangle$, it passes through a maximum of four successive crossings (with $\delta t>t_{\rm relax}$) before  going far from the right of the crossing point $t=(D+\omega)/\alpha$. At each of the crossing, the system undergoes a non-adiabatic LZSM transition. This explains the maximum of four steps observed in Fig.\ref{Figure3}$(b)$. When $D=\omega$, (see left  upper panel in Fig.\ref{Figure4k}) two lines cross at the point $t=0$. The number of crossings along the direction of $|2\rangle$ reduces to three. From the left to the right, the system undergoes three successive LZSM transitions (with $\delta t>t_{\rm relax}$) and this explains the three-step pattern behavior observed in Fig.\ref{Figure3}$(a)$. It is clear from these diagrams (see left  upper panel in Fig. \ref{Figure4}) that for small $D/\sqrt{\alpha}$ or small $\omega/\sqrt{\alpha}$ as compared to $1$, only two crossings are created (see Fig.\ref{Figure2}$(a)$) justifying why we see two-step pattern in these regimes. Indeed, when we breakdown the $SU(3)$ symmetry by canceling the easy-axis anisotropy ($D=0$), three lines cross at a single point and the model (\ref{equ15}) exhibits two crossings (see Fig.\ref{Figure4} left upper panel). The two-step pattern observed in this case (see Figs.\ref{Figure2} and \ref{Figure4}) results from interactions between paths in two separated spin-1 $SU(2)$ LZSM scenarios. We have two coupled spin-1 $SU(2)$ non-adiabatic LZSM transitions. The same remarks hold for the three- and four- step patterns; indeed, the model Eq.(\ref{equ15}) recalls that each LZSM transition corresponds to a passage through a point of minimum energy (crossing) between at least two energy levels. The crossing can either be a splitter or a mixer. Thus, the three- and four- step patterns are equivalent to interplay (consecutive passage) through three and four coupled spin-1/2 (two crossing lines) separated in time (such that $\delta t>t_{\rm relax}$) $SU(2)$ non-adiabatic LZSM processes respectively. Hence, our procedure of constructing  a five-level model with constant couplings which is equivalent to a three-level model with transverse periodic couplings corresponds to a decomposition of the ThLS' dynamics into all its spin-(1/2 or 1) $SU(2)$ LZSM components. The  corresponding LZSM interference patterns are reported in Fig.\ref{Figure5} in the strong drive limit $\omega/\sqrt{\alpha}\gg1$.  The patterns clearly highlight the effects of the transverse drive. The patterns observed in Ref.\onlinecite{Kiselev} with constant couplings are here "doubled" as the frequency of the drive is large.

To explain the coexistence between beats and steps, let us note from the right upper panel in Fig.\ref{Figure4} that this is  clearly a consequence of the fact that the inter distance between the crossings $t=-(D+\omega)/\alpha$ and $t=-(D-\omega)/\alpha$  on one hand [time domain I] and $t=(D-\omega)/\alpha$ and $t=(D+\omega)/\alpha$ on the other hand [time domain II] is small compared to $t_{\rm relax}$ and the distance between the regions I and II is large as compared to $t_{\rm relax}$. The dwell time in the  regions I is such that $\delta t< t_{\rm relax}$ (beats form) after passing this region, the system relaxes enough before arriving at II (steps form) where again $\delta t< t_{\rm relax}$ (beats form). 

For a satisfactory correspondence between the semi-classical model (\ref{equ1}) and the effective quantum model (\ref{equ15}), one must compulsorily solve equation (\ref{equ15a}). Making $g_{a}$ and $g_{b}$ the subject of that equation, we can infer $g_{a}=A/4\sqrt{n_{a}}$ and $g_{b}=A/4\sqrt{n_{b}}$ as possible solutions whereby, setting $\phi=0$ leads us to 
\begin{eqnarray} \label{equ16} 
\lambda_{\kappa,\kappa'}^{\bf p_{a}/p_{b}}=\frac{A}{\sqrt{4n_{a/b}}}.
\end{eqnarray}
Note that the model Eq.(\ref{equ15} ) only reproduces the results for $P_{2\to2}(t)$ and fails for the other cases (see Appendix \ref{App2} for further explanations). 
We have observed that to reproduce the two-step pattern, one should tune the optical pumps such that $n_{a,b}=2$ while for beat-two-step pattern $n_{a,b}=1$ (see Fig.\ref{Figure4}). For three- and four- steps, special combinations are required (see  Fig.\ref{Figure4}). For three-step, the pump ${\bf p_{a}}$ should inject two photons ($n_{a}=2$) while the pump ${\bf p_{b}}$ supplies with $n_{b}=1$ photon. For four steps, the pump ${\bf p_{a}}$ injects two photons for the tunneling $|2\rangle\leftrightarrow|1,\omega\rangle$, one is absorbed and the remaining is released to $|1,-\omega\rangle$ such that for the coupling $\lambda_{1,2}^{\bf p_{a}}$ between the state $|1,\omega\rangle$ and $|2\rangle$ one requires $n_{a}=2$. The state $|1,\omega\rangle$ absorbs one and rejects the other to $|1,-\omega\rangle$ such that the coupling $\lambda_{\bar{1},2}^{\bf p_{a}}$ between $|1,-\omega\rangle$ and $|2\rangle$ also takes one photon. The pump  ${\bf p_{b}}$ operates in a similar manner. The same scenario holds with $1$ replaced by $3$. This last process is similar to photon-assisted tunneling in a linearly coupled triple quantum dot\cite{Gallego2015} and similar sequential LZSM transitions were observed in optical lattices with ultracold atoms\cite{Blochs} and in an ensemble of interacting BEC two-level atoms  interacting with cavity modes\cite{Huang}.
We can therefore assert that the number of steps corresponds to the number of photons absorbed. 

Note that, when $\omega/\sqrt{\alpha}\ll1$ or approaches zero, the inter-level distance considerably reduces between the hybrid states $|1,\omega\rangle$ and $|1,-\omega\rangle$ on one hand, and the  $|3,-\omega\rangle$ and $|3,\omega\rangle$ on the other hand (the degeneracy is lifted). The system rather performs two sequential LZSM transitions. The five-level model (\ref{equ15}) reduces to the three-level model discussed in Ref.\onlinecite{Kiselev}. In the extreme opposite case $\omega/\sqrt{\alpha}\gg1$, the degeneracy accentuates allowing the system to relax enough in between two crossings and to perform four consecutive LZSM transitions. 

 The treatment adopted to construct (\ref{equ15}) is comparable to a reverse engineering procedure. We move from the solutions (numerical and analytical) to a problem. The model quite well reproduce the tendencies observed earlier ranging from one-step to four-step patterns through beat-two-step pattern (see Figs.\ref{Figure4} and \ref{Figure4k}). With precision of representation, the model (\ref{equ15}) hides an $SU(5)$ symmetry. The fact that we used $SU(2)$ coherent states to create a correspondence between a pseudo $SU(5)$ model that ensures a correspondence  with the $SU(3)$ model Eq.(\ref{equ1}) suggests a local isomorphism $SU(5)\approx SU(3)\times SU(2)\times U(1)$ where $U(1)$ is the circle group of all complex numbers of modulus 1.
 
We can now answer the fundamental question: what is the role played by steps? As one might have thought, technically, steps or the LZSM mechanism helps in characterizing the complex dynamics of a system bathing in its environment (boson bath) or coupled to a time-dependent field, a transverse periodic drive in the circumstances. Indeed, the number of steps indicates the number of crossings traversed by the system in the course of time (or another control parameter) in the non-adiabatic  limit. The crossings are resonances points where two energy levels come close. Then, steps are good indicators for counting the number of particles in a system. The moment when the probability function sharply drops gives an estimate of the crossing time (these messages are provided by the interferometric functions $F(...)$ and $G(...)$ in Eqs.(\ref{equ14a}) and (\ref{equ14b})). Thus, one can invent a plausible and simpler scenario reflecting  the complex dynamics of the system+bath by constructing an {\it effective} LZSM model (a few dimensional higher than the original one with time-independent couplings) which nicely reproduces the intrinsic dynamics of the system. In our case, we move from an $SU(3)$ model with a transverse periodic drive and constructed the corresponding ''pseudo'' $SU(5)$ model with time-independent couplings and remarkably reproduce all the tendencies observed in the $SU(3)$ picture with time-dependent couplings. As a direct consequence to this construction, one may envisage a long time asymptotic solution to the $SU(3)$ through its equivalent counterpart $SU(5)$ by classical multiplication of probabilities\cite{Sun, KenmoeFuture}. The good news is that for finite times, solutions to the  $SU(5)$ LZSM problem in the non-adiabatic limit are known in advance for $P_{2\to2}(t)$ and are given by Eq.(\ref{equ14}).

\subsection{Further extensions}\label{Sec3d}

\begin{widetext}
 	
 	\begin{figure}[]
 		\centering
 		\vspace{-0.5cm}
 		\includegraphics[width=8.5cm, height=6.cm]{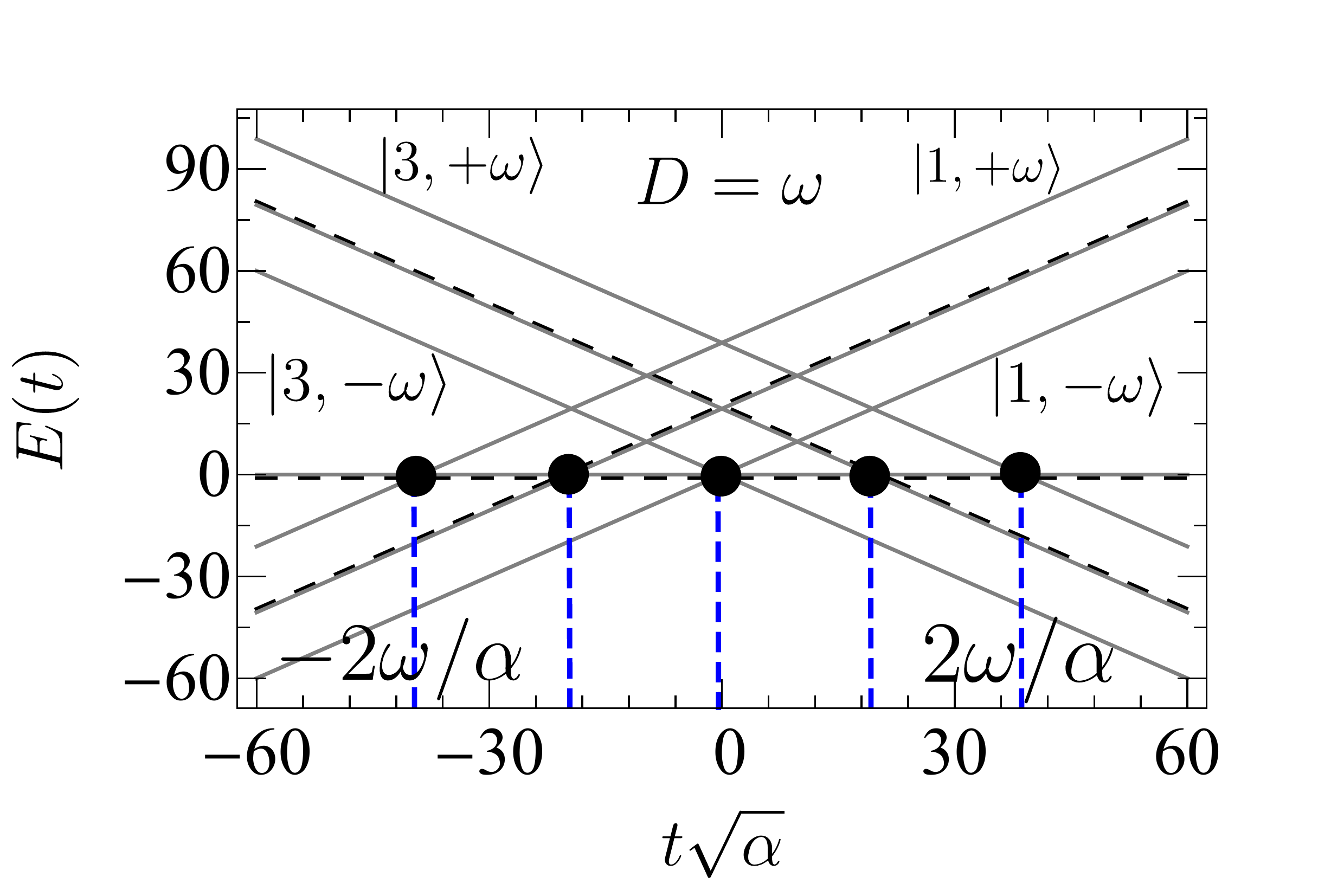}\hspace{-0.7cm}
 		\includegraphics[width=8.5cm, height=6.cm]{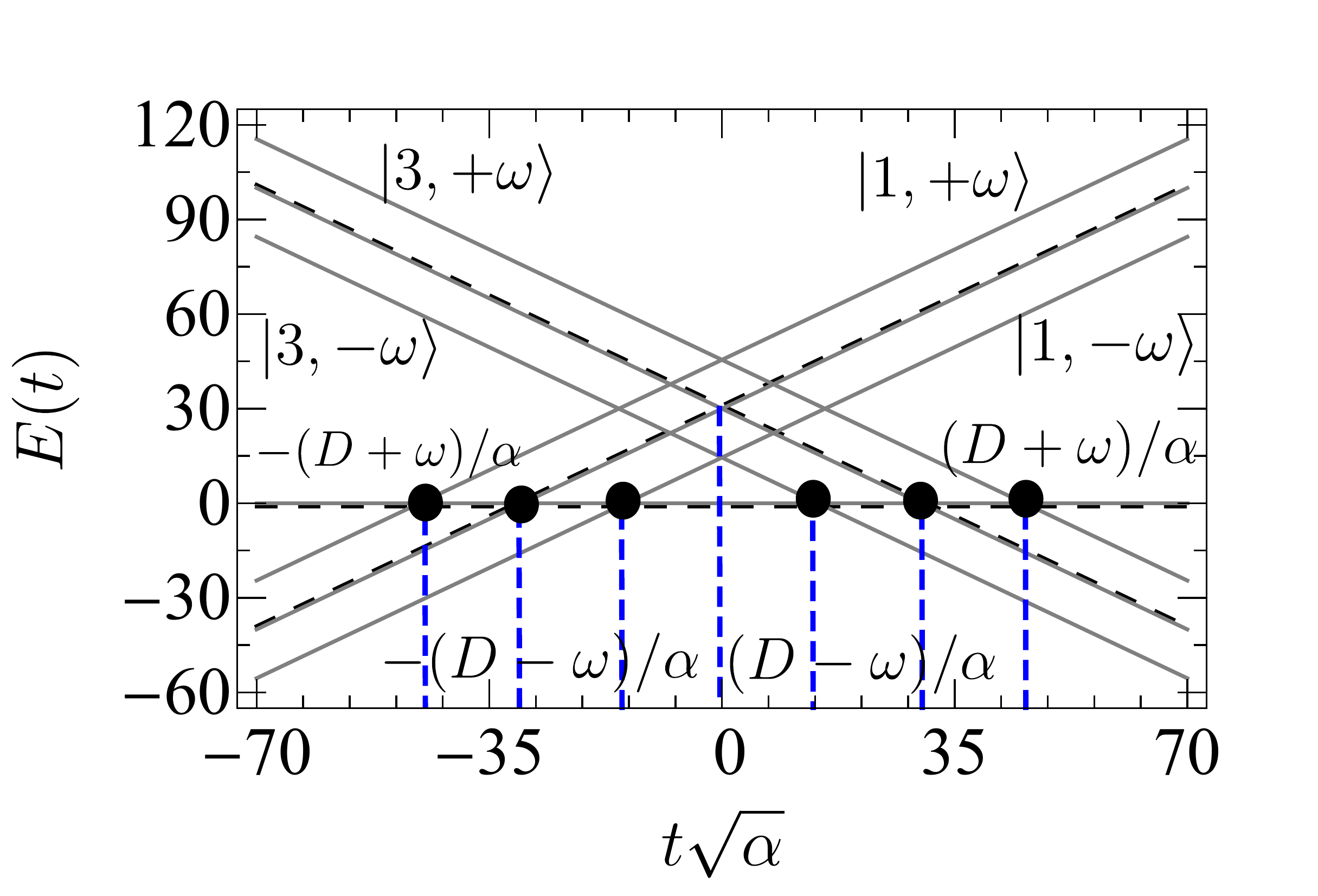}\\\vspace{-0.6cm}
 		\includegraphics[width=8.5cm, height=6.cm]{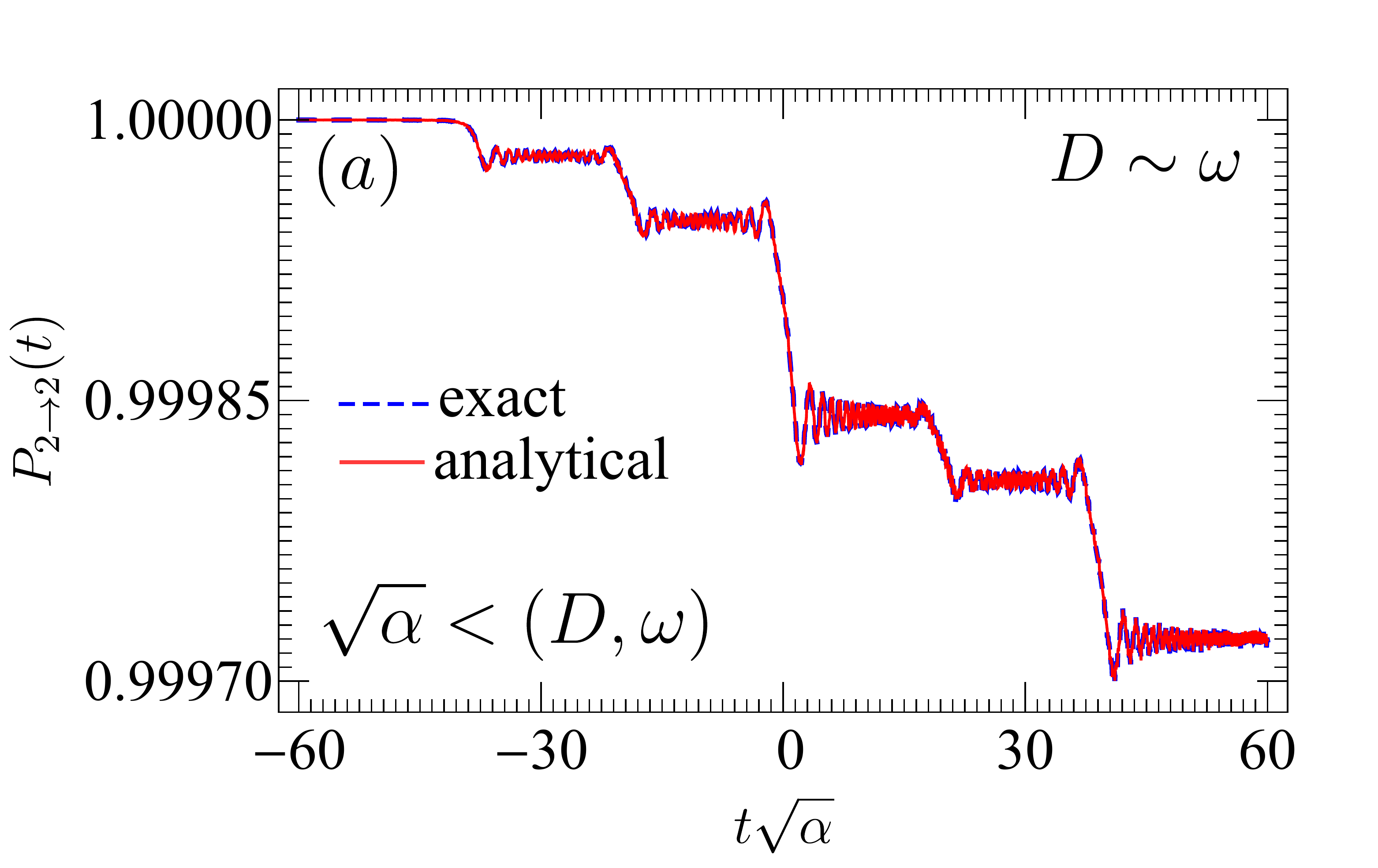}\hspace{-0.6cm}
 		\includegraphics[width=8.5cm, height=6.cm]{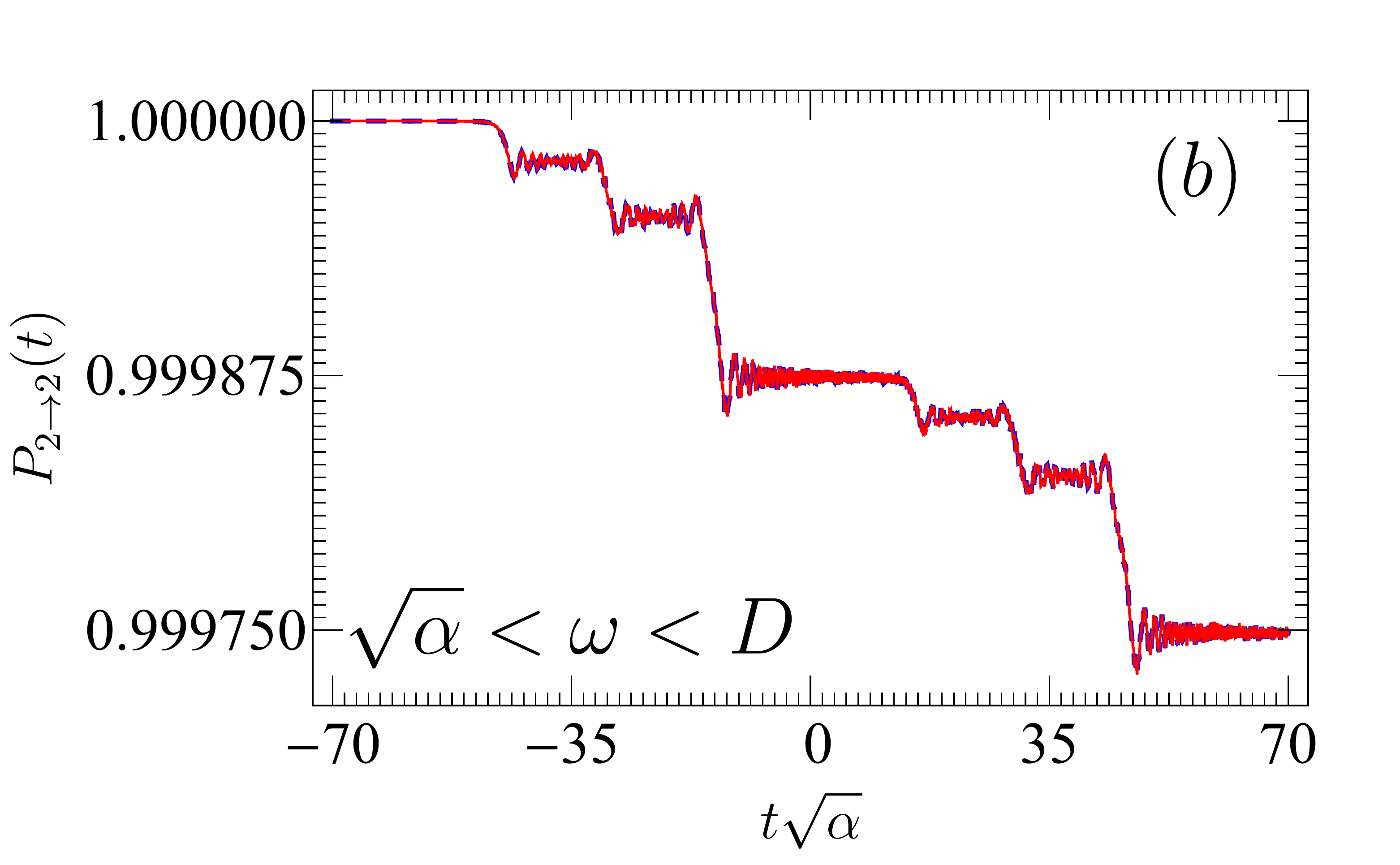}
 		\vspace{-0.5cm}
 		\caption{(Color Online)  {\bf Upper panel:} Energy diagrams for five- and six- step patterns plotted for the case when the periodic drive is added to the constant coupling in the original $SU(3)$ LZSM model. They may be used to build the $(6\times6)$ LZSM Hamiltonian with constant couplings that reflect the same patterns. Gray dashed lines correspond to diabatic energies in the absence of the periodic signal while black solid lines show their splitting in the presence of the periodic signal (Stark effects). Black balls indicate crossings. {\bf Lower panel:} For panel $(a)$ $A/\sqrt{\alpha}= 0.005$, $D/\sqrt{\alpha}=19.4163$, $\omega/\sqrt{\alpha}=19.4163$, $\Delta/\sqrt{\alpha}= 0.00167$ and for panel $(b)$ $A/\sqrt{\alpha}= 0.005$, $D/\sqrt{\alpha}= 30$, $\omega/\sqrt{\alpha}=15.5$, $\Delta/\sqrt{\alpha}=0.00167$. It should be noted that the number used for calculations are scrupulously chosen up to decimals. Any change in one of these numbers has a drastic incidence on the structure (not the number) of steps.  Beats and steps also coexist in this case  when $\omega\sim\sqrt{\alpha}$ and $D>10\sqrt{\alpha}$.} \label{Figure6}
 	\end{figure} 
 	
 \end{widetext}

We have observed so far that interactions with a transverse drive creates harmonics interactions between the three levels of the system somehow causing a  dynamical Stark effect.  For a periodic modulation of the coupling between level positions, beats and steps patterns can coexist and that the number of steps increases from $1$ to a maximum of $4$. 
 Now, let us add the periodic field $f(t)=A\cos[\omega t+\phi]$ to the tunnel matrix element (constant coupling) $\Delta$ and  probe the sensitivity of the $SU(3)$ LZSM interferometer. Henceforth, transitions between Zeemann levels in the interferometer are electromagnetically assisted as
\begin{eqnarray}\label{r1}
f(t)\to\Delta+f(t).
\end{eqnarray}
 In this paradigm, we define the additional level crossing parameter $\eta=\Delta^{2}/\alpha$. Numerical investigations reveal that the number of steps corresponding to sequential LZSM transitions increases (see Fig.\ref{Figure6}). For large $D/\sqrt{\alpha}$ and $\omega/\sqrt{\alpha}$ of the same order of magnitude, we observed five steps whilst under the same conditions with $D/\sqrt{\alpha}>\omega/\sqrt{\alpha}$ we achieve six steps. Analytical expressions that describe these oscillations in the populations of levels in the non-adiabatic limits $(\delta,\eta)\ll1$ are of the same structure as Eq.(\ref{equ9})  with 
\begin{eqnarray}\label{r2}
p_{\pm}(t)\to p_{\pm}(t)+q_{\pm}(t)+r_{\pm}(t),
\end{eqnarray}
where
\begin{eqnarray}\label{r3} 
r_{\pm}(t)=\pi\eta F\Big(t\pm\frac{D}{\alpha},t\pm\frac{D}{\alpha}\Big),
\end{eqnarray}
and
\begin{widetext}
		
	\begin{eqnarray}\label{r4} 
	\nonumber q_{\pm}(t)=2\pi\sqrt{\delta\eta}\Big[F\Big(t\pm\frac{D}{\alpha},t\pm\frac{D\mp\omega}{\alpha}\Big)\cos\chi^{\mp}+F\Big(t\pm\frac{D}{\alpha},t\pm\frac{D\pm\omega}{\alpha}\Big)\cos\xi^{\pm}\\
	-G\Big(t\pm\frac{D}{\alpha},t\pm\frac{D\mp\omega}{\alpha}\Big)\sin\chi^{\mp}-G\Big(t\pm\frac{D}{\alpha},t\pm\frac{D\pm\omega}{\alpha}\Big)\sin\xi^{\pm}\Big],
	\end{eqnarray}

\end{widetext}
with $\chi^{\mp}=D^{2}/2\alpha-\Psi^{\mp}$ and $\xi^{\mp}=D^{2}/2\alpha+\varphi^{\mp}$ (it appears that $\phi=0$ leads to  $\chi^{\mp}=\xi^{\mp}$) where
$
\Psi^{\pm}=\phi+(D\pm\omega)^{2}/2\alpha$, and $\varphi^{\pm}=\phi-(D\pm\omega)^{2}/2\alpha.
$
They are simultaneously plotted on Fig.\ref{Figure6} (red solid lines) with numerical results and a remarkable agreement can be noted. It is clear from here that setting $\Delta=0$ i.e. $\eta=0$ brings us automatically back to Eq.(\ref{equ14}). Also, $A=0$ i.e. $\delta=0$ reduces our results to Ref.\onlinecite{Kiselev}. Note that when $\Delta>A$, the tunnel matrix element $\Delta$ prevails on the periodic drive $f(t)$  i.e. $\Delta+f(t)\sim\Delta$ (given that $-A\le f(t)\le A$). To clearly see this, one should note that $r_{\pm}(t)$ in Eq.(\ref{r3}) is the contribution from the magnetic field and is the only term which does not contain the finger prints ($A,\omega,\phi$) of the periodic drive (and it is not affected by interference terms). When $\Delta>A$, then $\eta>\delta$ and in the limit $(\eta,\delta)\ll1$, the dominant contribution in Eq.(\ref{r2}) comes from the magnetic field and the effect of the periodic drive is canceled. The splitting of diabatic states is not effective, one exclusively observes two steps and the study globally returns to Ref.\onlinecite{Kiselev}. This explains why in the ongoing discussion, we consider the case $\Delta<A$.

The analytic results in Eq.(\ref{r2}) stand as a good platform for comparing the $SU(3)$ LZSM interferometers; one with coupling $\Delta$ between levels on one hand and another with  $\Delta+A\cos[\omega t+\phi]$ on the other hand. The second case offers more controllable parameters and may be useful experimentally to achieve high-precision measurement (possibly up to the desirable Heisenberg limit posed by the Heisenberg uncertainty principle) which cannot be reached by the first one or to glean more about qutrit for quantum information processing. In order to emphasize on the role played by the steps in the physics of LZSM transitions, let us remark that the six- and five- step patterns observed in Fig.\ref{Figure6} suggest a priori that the {\it effective} LZSM model (model with time-independent couplings) which could best reproduce these patterns is constructed in a similar fashion as we did earlier for four-step and should be of ''pseudo'' $SU(6)$ symmetry. This is written but not shown, the corresponding energy diagrams are  plotted for five- and six- step patterns (see Fig.\ref{Figure6} upper panels).  
 
 Let us now summarize what we have learned so far about the effects of inter-level renormalization on the $SU(3)$ LZSM interferometry. When the coupling $f(t)=\Delta$ is constant, diabatic states do not split, the ThLS undergoes a maximum of two steps\cite{Kiselev}. When $f(t)=A\cos(\omega t+\phi)$ periodically changes, diabatic states split into two and one observes a maximum of four consecutive steps. Finally,  when $f(t)=\Delta+A\cos(\omega t+\phi)$, diabatic states split into three and the ThLS achieves up to six steps. It seems as if the more the number of components in $f(t)$ increases, the more the number of steps increases as well.  These observations remind us that we are tackling a Zeeman's effect (splitting due to magnetic field) accompanying a Stark's effect (splitting due to electric field). A fundamental and natural question pops up in mind. Are each of the signals composing $f(t)$ responsible for a splitting? In order to give a reliable answer, we generalize $f(t)$ as  
 \begin{eqnarray}\label{equ3.16}
 f(t)=\sum_{n=0}^{N}A_{n}\cos(\omega_{n} t+\phi_{n}),
\end{eqnarray}
i.e. as a polychromatic wave or in other words a train of $N$ monochromatric waves of amplitude $A_{n}$, frequency $\omega_{n}$ and phase shift $\phi_{n}$.  Representing the coupling in (\ref{equ3.16}) as a superposition of higher harmonics extends our study to a wide range of experiments given that the amplitude $A_{n}$, the frequencies $\omega_{n}$ and the phase $\phi_{n}$ can rather be well controlled for flexible design of the pulse. Entanglement can also be created\cite{Blattmann} when for instance $\omega_{n}=n\omega$, $\phi_{n}=n\omega t_{0}$ ($t_{0}$ being the initial time) and $A_{n}=A/n$ (spike signal) with $N=5$. 

\begin{widetext}
	
	\begin{figure}[] 
		\vspace{-0.4cm}	
		\includegraphics[width=8.5cm, height=6cm]{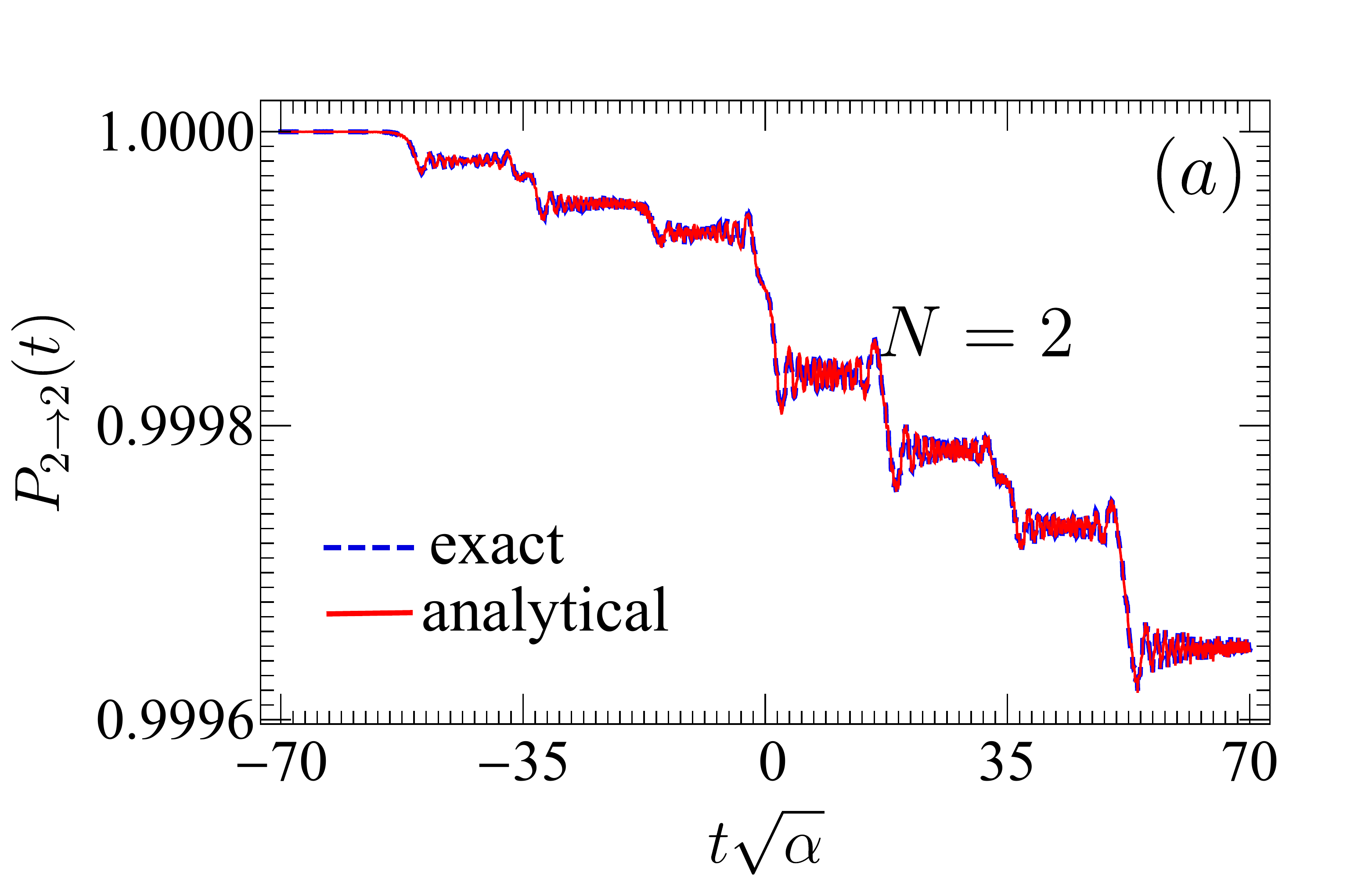}\hspace{-0.2cm}\includegraphics[width=8.5cm, height=6cm]{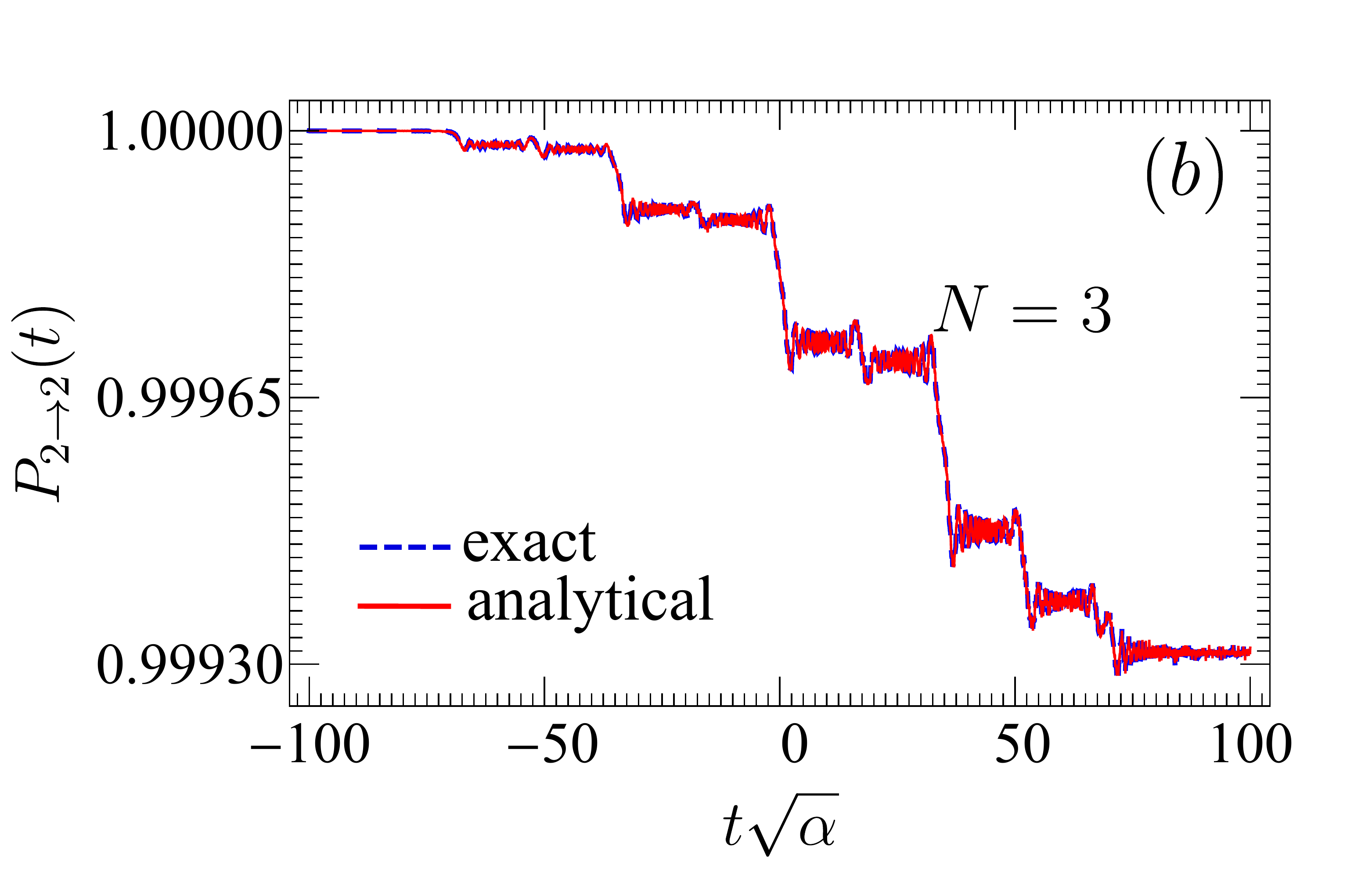}\vspace{-0.6cm}\\
		\includegraphics[width=8.5cm, height=6cm]{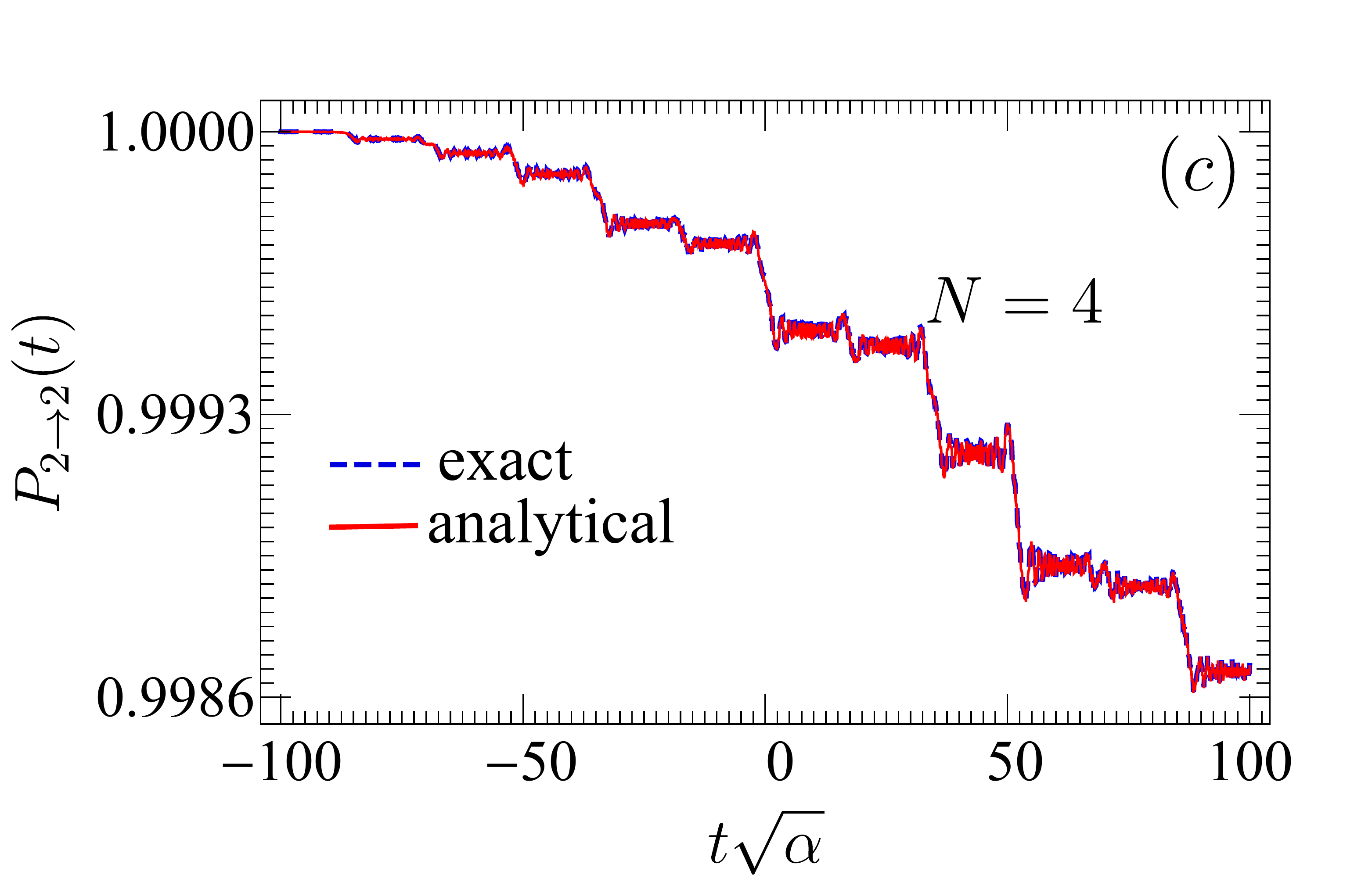}\hspace{-0.2cm}\includegraphics[width=8.5cm, height=6cm]{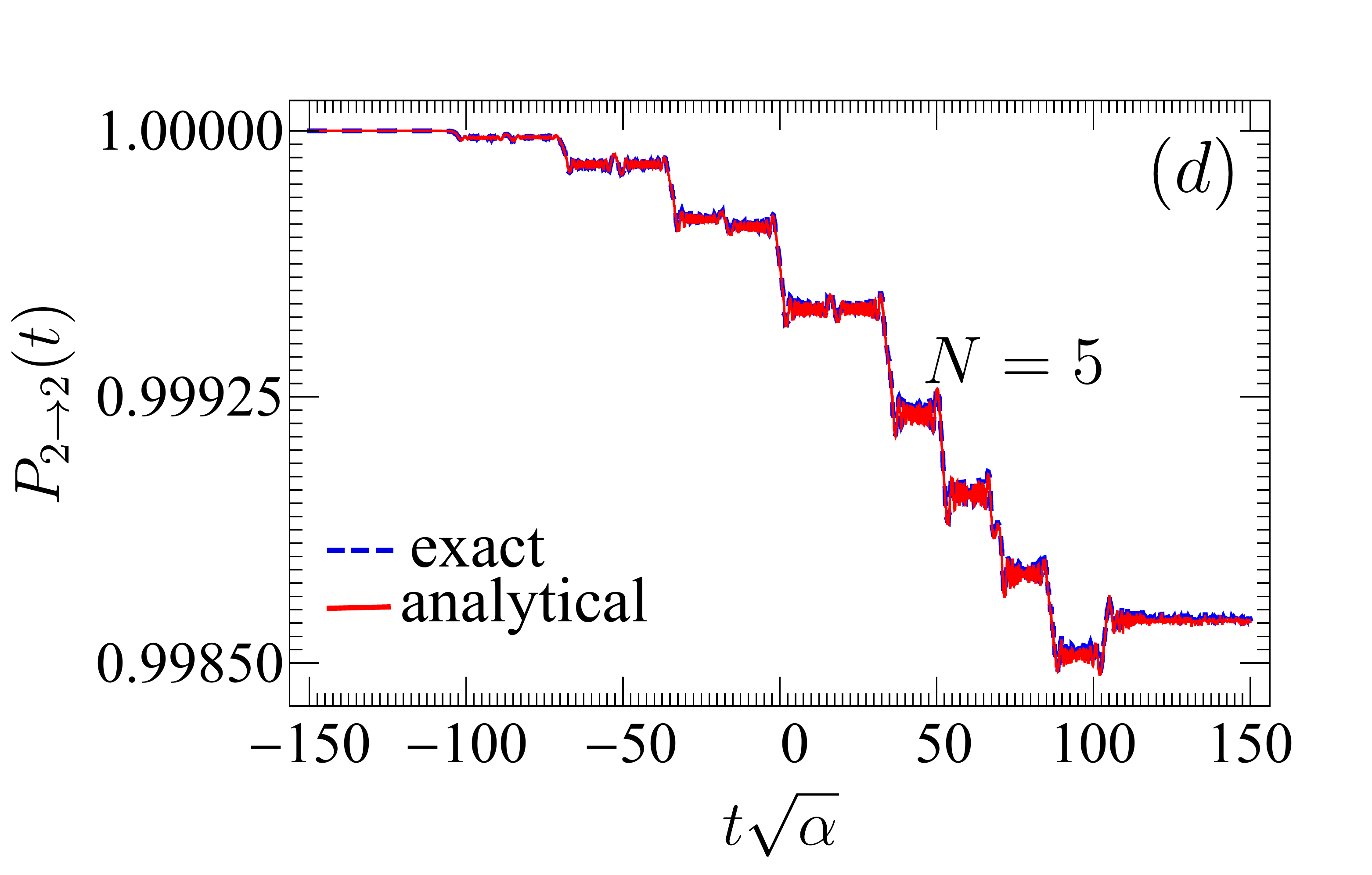}
		\vspace{-0.4cm}
		\caption{(Color Online) Cascaded LZSM transitions due to the polychromatic transverse drive Eq.(\ref{equ3.16}). Blue dashed lines are exact numerical results and red solid lines are analytical results Eq.(\ref{equ3.17}). For calculations, we have considered $A_{0}=0$ (i.e. $\delta_{00}=0$), $A_{n}=A$ (with $n>0$ such that $\delta_{nm}=A^2/4\alpha$ with $A/\sqrt{\alpha}=0.005$ for arbitrary $m>0$), $\phi_{n}=0$, $\omega_{n}=n\omega$,  $D/\sqrt{\alpha}=\omega/\sqrt{\alpha}=(2M+1)\pi/2$ with $M=5$ and the time in the unit $1/\sqrt{\alpha}$.} \label{Figure5Q}
	\end{figure} 
	
\end{widetext}

Now, solutions to Eqs.(\ref{equ4})-(\ref{equ6}) considering (\ref{equ3.16}) after a long and tedious algebra yield  
\begin{widetext}
	
\begin{eqnarray}\label{equ3.17}
\nonumber p_{\pm}(t)=\sum_{n=0}^{N}\sum_{m=0}^{N}\pi \delta _{n m}\Big(\cos\left[\Psi _n^{\mp }-\Psi _m^{\mp }\right]F\Big(t\pm \frac{D\mp \omega _n}{\alpha },t\pm \frac{D\mp \omega _m}{\alpha}\Big)+\cos\left[
\Psi_n^{\mp}+\varphi_m^{\pm}\right]F\Big(t\pm \frac{D\mp \omega_n}{\alpha},t\pm\frac{D\pm \omega _m}{\alpha}\Big)\\\nonumber+\cos[\varphi_n^{\pm}+\Psi _m^{\mp}]F\Big(t\pm\frac{D\pm\omega _n}{\alpha },t\pm\frac{D\mp\omega_m}{\alpha}\Big)+\cos[\varphi_n^{\pm }-\varphi_m^{\pm }]F\Big(t\pm \frac{D\pm \omega_n}{\alpha},t\pm \frac{D\pm \omega _m}{\alpha}\Big)\\\nonumber-\sin[\Psi_n^{\mp}-\Psi_m^{\mp }]G\Big(t\pm\frac{D\mp \omega _n}{\alpha},t\pm \frac{D\mp \omega_m}{\alpha}\Big)+\sin[\varphi _n^{\pm }+\Psi_m^{\mp }]G\Big(t\pm\frac{D\pm\omega_n}{\alpha},t\pm \frac{D\mp\omega_m}{\alpha }\Big)\\-\sin[\Psi_{n}^{\mp}+\varphi _{m}^{\pm}] G\Big(t\pm \frac{D\mp \omega _n}{\alpha },t\pm \frac{D\pm \omega _m}{\alpha}\Big)+\sin[\varphi_n^{\pm}-\varphi_m^{\pm }]G\Big(t\pm\frac{D\pm \omega_n}{\alpha},t\pm \frac{D\pm \omega_m}{\alpha}\Big)\Big),\hspace{0.9cm}
\end{eqnarray}

\end{widetext}
where $\delta_{nm}=A_{n}A_{m}/4\alpha$$, 
\Psi^{\pm}_{n}=\phi_{n}+(D\pm\omega_{n})^{2}/2\alpha$, and $\varphi^{\pm}_{n}=\phi_{n}-(D\pm\omega_{n})^{2}/2\alpha$. In the presence of a polychromatic transverse drive containing $N$ monochromatic signals, the diabatic energies $\alpha t+D$ and $-\alpha t+D$ split into $2N+1$ sublevels $\alpha t+(D\pm\omega_{n})$ and $-\alpha t+(D\pm\omega_{n})$ respectively while the middle diabatic state $|2\rangle$ remains unaltered. This give rises to several level crossings denoted as $t_{n}^{(sign)}=sign(D\pm\omega_{n})/\alpha$ where $sign=\pm$ (the $+$ refers to the states issued from the splitting of $\alpha t+D$ while $-$ refers to those of $-\alpha t+D$).  We have observed cascaded LZSM transitions with this model (see Fig.\ref{Figure5Q}) and we can assert that each of the signals in $f(t)$ is responsible for a splitting of diabatic states in the original $SU(3)$ LZSM model. Eq.(\ref{equ3.17}) is a generalization of our previous results and has also been tested numerically and remarkably holds for $\delta_{nm}\ll1$ and arbitrary $D$, $\omega_{n}$ and $\phi_{n}$ (see Fig.\ref{Figure5Q}). This result is more expressive and instructive than the others and opens a deeper perspective for investigating periodic drive of three-level systems with polychromatic signals. Indeed, Eq.(\ref{equ3.17}) shows that repeated passages depend on the phase acquired in between crossings. The cosine and sine of phase sums and differences are interference terms. It stands out that $p_{\pm}(t)$ is the sum of all contributions (constructive and destructive) from all the $N$ signals in $f(t)$. The phases $\Psi^{\mp}_{n}$ and $\varphi^{\pm}_{n}$ are accumulated by the system when it traverses the $n$th crossing points $t_{\Psi, n}^{(\mp)}=\mp(D\mp \omega_n)/\alpha$ and $t_{\varphi,n}^{(\pm)}=\mp(D\pm \omega_n)/\alpha$ respectively ($\Psi$ and $\varphi$ are used here as subscripts to refer to the phases accumulated). Two waves coming from the $n$th and $m$th signals emerging from the same family of sub-levels and interfering constructively, contribute as $\cos[\chi_{n}^{(i)}-\chi_{m}^{(i)}]F(t\pm t_{\chi,n}^{(i)}, t\pm t_{\chi, m}^{(i)})$ where   $\chi=\Psi,\varphi$ and the phases $\Psi_{n}^{(i)}=\phi_{n}+\alpha (t_{\Psi,n}^{(i)})^{2}/2$ and $\varphi_{n}^{(i)}=\phi_{n}-\alpha (t_{\varphi,n}^{(i)})^{2}/2$ are picked up at  crossings $t_{\chi,n}^{(i)}$ by linear sweep. Then, the phase difference
\begin{eqnarray}\label{equ3.18}
\chi_{n}^{(i)}-\chi_{m}^{(j)}=sign\int_{t_{\chi,n}^{(i)}}^{t_{\chi, m}^{(j)}}\alpha t'dt',
\end{eqnarray}
is accumulated between the crossings $t_{\chi,n}^{(i)}$ and $t_{\chi,m}^{(j)}$. Note that $sign=+$
corresponds to the $\Psi$-phase acquired when the ThLS traverses the level crossings due to the sub-levels of $\alpha t+D$ while $sign=-$ in the case of $\varphi$-phase is achieved in the regions of the sub-levels of $-\alpha t+D$. When the waves interfere destructively, their contributions read $\sin[\chi_{n}^{(i)}-\chi_{m}^{(i)}]G(t\pm t_{\chi,n}^{(i)}, t\pm t_{\chi, m}^{(i)})$. When now they are from different family and interfere constructively $\cos[\chi_{n}^{(i)}+\bar{\chi}_{m}^{(j)}]F(t\pm t_{\chi,n}^{(i)}, t\pm t_{\bar{\chi}, m}^{(j)})$ and in case of destructive interferences $\sin[\chi_{n}^{(i)}+\bar{\chi}_{m}^{(j)}]G(t\pm t_{\chi,n}^{(i)}, t\pm t_{\bar{\chi}, m}^{(j)})$.
 The final population in $\ket{2}$ after interactions is the sum of contributions (constructive and destructive) from all the crossings due to all signals composing $f(t)$.
 
 For $D/\sqrt{\alpha}=\omega/\sqrt{\alpha}=(2M+1)\pi/2$ and $\phi_{M}=2\pi M$ (with $M=0,1,2,3,...$), we have observed the  cascaded LZSM  transitions depicted on Fig.\ref{Figure5Q} for $N=2,3,4,5$. We have restricted ourselves to the case $M=5$. It clearly appears that the number of steps increases with $N$ validating our previous assertions and confirming the accuracy of our analytical results Eq.(\ref{equ3.18}) given that they and the numerical exact solutions are barely discernible. These results are desirable and relevant for periodic drive of ThLSs.
 
\section{Adiabatic evolution}\label{Sec4}
By slowly changing the constant sweep velocity $\alpha$ and turning the transverse drive such that $A^{2}/\alpha\gg 1$, the Hamiltonian slowly changes throughout the course of time and the ThLS follows adiabatic trajectories. This behavior is described in the adiabatic basis achieved by rotating the system from diabatic to adiabatic basis. Thus, defining an orthogonal rotation matrix $\mathbf{W}(t)$ with elements $w_{n\kappa}(t)$, the elements of $\boldsymbol{\mathrm{\rho}}(t)$ in Eq.(\ref{equ3}) are rotated as 
\begin{eqnarray}\label{equa1}
\rho_{nm}(t)=\sum_{\kappa,\ell=1}^{3}w_{n\kappa}(t)\varrho_{\kappa\ell}(t)w_{m\ell}^{*}(t),
\end{eqnarray}
where $\varrho_{\kappa\ell}(t)$ is the density matrix in adiabatic basis. Techniques for constructing  $w_{n\kappa}(t)$ are presented in Ref.\onlinecite{KenmoePRB} and one verifies that they have the following properties
\begin{equation} \label{equa1a} 
\sum_{\kappa=1}^{3}w_{\kappa i}(t)w_{\kappa j}(t)=\delta_{ij}, \quad \sum_{\ell,\kappa=1}^{3}w_{\ell \kappa}^{2}(t)=3.
\end{equation}
 Now the indices $1$, $2$ and $3$ respectively match $+$, $0$ and $-$ such that $\varrho_{++}(t)$, $\varrho_{00}(t)$ and $\varrho_{--}(t)$ are diagonal elements representing the population measured from the adiabatic basis. Coherence factors are $\varrho_{\kappa\ell}(t)=\varrho_{\ell\kappa}^{*}(t)$. Assuming that the system starts off in the state $|\kappa'\rangle$, then, the initial condition (\ref{equ3a}) writes in adiabatic basis as
\begin{eqnarray}\label{equa1b}
\varrho_{\kappa\ell}(t_{0})=w_{\kappa'\kappa}(t_{0})w_{\kappa'\ell}(t_{0}).
\end{eqnarray}
Plugging Eq.(\ref{equa1}) into Eq.(\ref{equ3}), it can be shown that $\varrho_{\kappa\ell}(t)$ obey
 \begin{eqnarray}\label{equa2}
\nonumber i\dot{\varrho}_{\kappa\ell}(t)=\Big(E_{\kappa}(t)-E_{\ell}(t)\Big)\varrho_{\kappa\ell}(t)\\-
i\sum_{n=1}^{3}\Big(\nu_{\kappa n}(t)\varrho_{n\ell}(t)+\varrho_{\kappa n}(t)\nu_{\ell n}(t)\Big),
\end{eqnarray}
where $\nu_{\kappa\ell}(t)$ evaluates the strength of non-adiabatic couplings between adiabatic states with index $\kappa$ and $\ell$:
\begin{equation} \label{equa3} 
\nu_{\kappa\ell}(t)=-\frac{\langle \varphi_{\kappa}(t)|(\partial_{t}\mathcal{H})|\varphi_{\ell}(t)\rangle}{E_{\kappa}(t)-E_{\ell}(t)},\quad \kappa\neq\ell.
\end{equation}  
Here $\ket{\varphi_{\ell}(t)}$ are eigenstates of $\mathcal{H}(t)$ (see Ref.\onlinecite{KenmoePRB}). As we do not want adiabatic states to talk at all (we want the ThLS to remain out of the triangle), we demand that the energy difference between adiabatic states becomes larger than the non-adiabatic couplings i.e. $\nu_{\kappa\ell}(t)\ll |E_{\kappa}(t)-E_{\ell}(t)|$. Thus, setting $\nu_{\kappa\ell}(t)=0$ (strong adiabatic evolution) guarantees that the energies are well separated in time. Therefore, Eq.(\ref{equa2}) easily integrates  and $\varrho_{\kappa\ell}(t)=e^{-i\Lambda_{\kappa\ell}(t,t_{0})}\varrho_{\kappa\ell}(t_{0})$. Here, 
\begin{equation} \label{equa4} 
\Lambda_{\kappa\ell}(t,t_{0})=\int_{t_{0}}^{t}\Big(E_{\kappa}(t')-E_{\ell}(t')\Big)dt',
\end{equation}
is the surface in between the states with energies $E_{\kappa}(t)$ and $E_{\ell}(t)$. It may be associated with the dynamical phase acquired by the system during adiabatic evolutions. It can be shown that population in the same adiabatic state remains constant throughout the sweeping and driving processes $\varrho_{\kappa\kappa}(t)=\varrho_{\kappa\kappa}(t_{0})$.

Now, a key question arises. Due to the transverse signal $f(t)$, how separated are adiabatic states such that one can guarantee no-transition between them? The LZSM protocol somehow violates this requirement. Indeed, the original $SU(3)$ as well as $SU(2)$ (with constant interaction between levels) showed us that at far removed times $t=\pm\infty$ diabatic and adiabatic trajectories are identical. Thus, non-adiabatic transitions (system in the same  diabatic state) result in transitions between adiabatic states. In the present case, this occurs several times (see Fig.\ref{Figure1}(a)) as the coupling between levels changes periodically. Thus, the largest separation between adiabatic states is achieved  when $A^{2}/\alpha\gg 1$.
 
Upon determining the condition for evolution out of the triangle and returning to the original diabatic basis, one finds that  
\begin{equation} \label{equa5} 
P_{\kappa'\to\kappa}(t_{0},t)=\sum_{n,j=1}^{3}p_{nj}^{\kappa'}(t_{0})p_{nj}^{\kappa}(t)\cos\Lambda_{nj}(t,t_{0}),
\end{equation}
where $p_{nj}^{\kappa}(t)=w_{\kappa n}(t)w_{\kappa j}(t)$. The presence of the cosine term here is an evidence of interferences. Using the first property in (\ref{equa1a}), one proves that $\sum_{\kappa=1}^{3}p_{nj}^{\kappa}(t)=\delta_{nj}$. Based on that, we prove that the total probability is a conserved quantity $\sum_{\kappa=1}^{3}P_{\kappa'\to\kappa}(t_{0},t)=1$. The formula given by Eq.(\ref{equa5}) is strictly equivalent to Eq.(3.32) in Ref.\onlinecite{KenmoePRB}.

We have however to stress whether the $3$ in the summand (\ref{equa5}) can be extended to an arbitrary number of level $\mathcal{N}>3$. The naive answer is affirmative and the stress immediately translocates to $\Lambda_{nj}(t,t_{0})$ and how to construct $w_{\kappa n}(t)$ analytically. We are afraid that even equipped with current symbolic calculators such as Mathematica and Maple this formula is hard to extend beyond $\mathcal{N}=3$. On the other hand, if the entries of the Hamiltonian in matrix form are numbers, this task can be easily executed by either of these programming languages. Our concerns remain about generalization of (\ref{equa5}) for $\mathcal{N}$-level systems.

\section{Experimental relevance}\label{Sec5}
In the present section, we highlight the results of the paper pinpointing their experimental relevance for a few examples. As already mentioned in the introductory part, quantum triangles are visible in the energy diagram of versatile set-ups. However, due to the complexity of the $SU(3)$ symmetry which accompanies these set-ups, and the non-linearity due to atom-atom interactions in BJJ\cite{Lee, Bouk}, optical lattices\cite{Blochs}, BEC\cite{Huang, Huang2012}, theoretical descriptions have so far been considered as huge challenges. We wish to demonstrate that the discrimination that grants more interests to experimental treatments in spite of analytical ones  may be alleviated.  

\subsection{Bose-Einstein Condensates}\label{Sec5.a} 
 Bose-Einstein condensates (BEC) are vital for quantum interferometry due to their long coherence time, high controllability and the property of atoms to possess a common phase. A prerequisite to produce BEC  is the ability to manipulate and control atoms (occupying hyperfine levels at microscales) at ultra low temperature close to the absolute zero where their degrees of freedom are frozen and hallmark quantum effects prevail\cite{Blochs, Huang}.  Remarkable efforts and unprecedented progresses have been made in this regard to offer the possibility to prepare macroscopic quantum systems by trapping or cooling down microscopic systems (single-particle systems). It is currently possible to trap ultra cold atoms and guide them using atom chips devices\cite{Shumm} issued from the technology of microfabrications. Upon completing preparations, macroscopic quantum entanglement is created and intriguing exotic many-body effects may be observed among which are interaction blockade\cite{Blochs}, coherent spatial splitting\cite{Shumm}, internal-state Rabi oscillations\cite{Treut} etc and are described in the schemes of the mean-field theory. Due to entanglement, the total state of the macroscopic system  is no longer expressed as a linear superposition of its microscopic states. The crucial objective is to achieve a macroscopic quantum coherence desirable to produce high-precise interferometers.  Indeed, it has been demonstrated that single-particle states interferometry accuracy can only reach the standard quantum limit while the multi-particle states interferometry significantly approaches the Heisenberg limit\cite{Shumm}. These observations explain why quantum coherence is desired in quantum interferometry with many-body systems. The quantum coherence is accurately achieved by applying a magnetic field and/or an electric field. If for instance the magnetic field linearly varies in time and changes its sign at a resonance point, we have a many-body LZSM tunneling. The minimal model of Bose-Hubbard type which describes the  many-body effects in the two-mode approximation should account for atom-atom interactions that are here embedded in the parameter $D$. It is in general possible to map the spin operators onto suitable combinations of the modes of the many-body system such that the resulting Hamiltonian ultimately achieves the form (\ref{equ1}). For instance\cite{Huang, Huang2012},
\begin{subeqnarray} \label{equ5.1} 
& S^{x}=\frac{1}{2}\Big(a^{\dagger}_{\sigma}b_{\sigma'}+b^{\dagger}_{\sigma'}a_{\sigma}\Big),\\
& S^{y}=\frac{1}{2i}\Big(a^{\dagger}_{\sigma}b_{\sigma'}-b^{\dagger}_{\sigma'}a_{\sigma}\Big),\\
& S^{z}=\frac{1}{2}\Big(a^{\dagger}_{\sigma}a_{\sigma}-b^{\dagger}_{\sigma'}b_{\sigma'}\Big),
\end{subeqnarray}
where $\sigma$ designates the set of the system's degrees of freedoms. 
 However, an invariant denoted as $n=a^{\dagger}_{\sigma}a_{\sigma}+b^{\dagger}_{\sigma'}b_{\sigma'}$ and representing the total number of bosons, is associated with (\ref{equ5.1}). Considering the operator $\mathbf{n}=n/2$, then the invariance in $n$ suggests that $[\mathbf{n},S^{\nu}]=[n,S^{\nu}]=0$  (with $\nu=x,y,z$), $(S^{z}+\mathbf{n})=n_{\sigma}$ and $(\mathbf{n}-S^{z})=n_{\sigma'}$ with $n_{\sigma}$ and $n_{\sigma'}$ representing the number of bosons in mode $|\sigma\rangle$ and $|\sigma'\rangle$ respectively. A second invariant $\mathcal{K}=(S^{x})^2+(S^{y})^2+(S^{z})^2=\mathbf{n}(\mathbf{n}+1)$ (the casimir operator) goes along with (\ref{equ5.1}) such that $[\mathcal{K}, S^{\nu}]=0$.  If in addition to the linearly varying magnetic field the condensate is trapped in an optical cavity, we have a many-body LZSM tunneling in a condensate-cavity system. Following the same strategy, one can in a semi-classical limit achieve the model (\ref{equ1}) through the substitutions (\ref{equ5.1}).

In order to gain more insights into the relevance of our results  for both experimental and  theoretical analyses, let us compare them with other results. In Ref.\onlinecite{Blochs}, the authors considered ultra cold atoms in an optical superlattice in which lattice sites are converted into double-well potentials. They observed that for four atoms in the wells, the energy diagram of the system against the bias depicts a triangular geometry. This clearly explains why they also observed two- (with three atoms) and three- (with four atoms) step patterns reminiscent to LZSM transitions in the population dynamics of wells by variation of the bias potential between negative and positive values for large onsite interactions between two atoms in the wells. This study falls into the scope of this paper. Indeed, for a model of double wells described by the two-mode Bose-Hubbard model for which $a_{\sigma}=a_{L}$ (creation of atoms in the left well) and $b_{\sigma'}=a_{R}$ (creation of atoms in the right well) the theoretical model Hamiltonian used in that reference through the transformations (\ref{equ5.1}) acquires the form 
\begin{eqnarray} \label{equ5.2} 
 \mathcal{H}_{1}(t)=-\sum_{\nu=x,z}\mathbf{B}_{\nu}(t)S^{\nu}+D(S^{z})^{2}+D\mathbf{n}(\mathbf{n}-1).
\end{eqnarray}
Here, $\mathbf{B}_{x}(t)$ is the tunneling matrix element, $\mathbf{B}_{z}(t)$ the potential bias along the double well axis and $D$ the onsite interaction energy between two atoms. The model (\ref{equ5.2}) appears in various experiments involving double-well potentials\cite{Blochs, Blochs2007}. It is for instance used in Ref.\onlinecite{Blochs2007} for time-resolved observations of the correlated tunneling of two interacting ultra cold atoms through a barrier in
a double-well potential (second order atom tunneling).  The last term $\mathbf{n}(\mathbf{n}-1)=\mathcal{K}-2\mathbf{n}$ is clearly an irrelevant term and can be disregarded by shifting the zero-point energy of the Hamiltonian. Indeed, the Hamiltonian $\mathcal{H}_{1}(t)$ preserves the total number of particles $[\mathbf{n},\mathcal{H}_{1}(t)]=0$ and $[\mathcal{K},\mathcal{H}_{1}(t)]=0$. The resulting model also describes atoms in  Bose Josephson Junctions\cite{Huang2012}. By linearly varying $\mathbf{B}_{z}$ and periodically changing $\mathbf{B}_{x}$ in an experiment similar to the ones in Refs.[\onlinecite{Blochs, Blochs2007}], our results may be relevant to describe the LZSM transitions observed and to prepare BECs.

As yet another remarkable work which supports the feasibility of our results, we mention the one 
in Ref.\onlinecite{Huang}. The authors considered an ensemble of interacting condensed spin-$1/2$ Bose atoms  optically coupled to a single-mode cavity. The authors reported an observation of four asymmetric LZSM transitions for a linear variation of the detuning from negative to positive values in the non-adiabatic regime and large atom-atom interactions. It was clearly established that interplay between large values of atom-atom interactions (i.e. the parameter $D$ here), the detuning (in the fast drive limit) and the number of photons  in the cavity via the atom-cavity coupling ($\lambda_{\kappa,\kappa'}^{{\bf P_{a}/P_{b}}}$) is responsible for the asymmetric sequential LZSM transitions. Their numerical results are found to be in  remarkable qualitative agreement with ours. However, finite time solutions important to describe the temporal evolution of a system were not derived. It should be noted that our three-level system can always be reduced to a two-level system by adiabatically eliminating one of the states with extremal spin projections. Thus, our analytical results may be relevant to explore in-depth the numerical results in Ref.\onlinecite{Huang}.

\subsection{Quantum Information Processing}\label{Sec5.b}
The technology of quantum information processing (QIP) sits in the concept of the quantum bit\cite{Patton} (qubit). It is universally admitted that qubit is ubiquitous in nature and holds promise to be a good candidate for developing and  implementing quantum technologies. The qubit has been investigated in various realistic situations and several of its features have been uncovered. However, due to the impossibility to completely isolate the qubit from unwanted external nuisances, the actual quantum computers operate with a limited number of qubits. In addition, interplay between two qubits involves entanglement which leads to intriguing quantum phenomena such as decoherence\cite{KenmoePRB}. These key points raise the question of coherent control of the dynamics of two entangled qubits. We demonstrate in what follows that the model (\ref{equ1}) is suitable to fulfill this requirement vital for the coveted quantum computer. We show that (\ref{equ1}) is remarkably equivalent to the Kibble-Zurek model which has proven to be intimately related to the LZSM model and this has enhanced  the possibility of use for problems of control\cite{Raymond, Jose}.
The starting point of our description is the Kibble-Zurek model\cite{Raymond, Jose} 
\begin{eqnarray} \label{equ5.3} 
\mathcal{H}_{2}(t)=\sum_{\nu=x,z}\mathbf{B}_{\nu}(t)(\boldsymbol{\mathrm{\sigma}}_{\nu}^{(1)}+\boldsymbol{\mathrm{\sigma}}_{\nu}^{(2)})+J\boldsymbol{\mathrm{\sigma}}_{z}^{(1)}\boldsymbol{\mathrm{\sigma}}_{z}^{(2)}, 
\end{eqnarray}
where $\boldsymbol{\mathrm{\sigma}}_{\nu}^{(j)}$ ($j=1,2$) denotes the $\nu$-component of the Pauli matrix associated with the $j$th qubit. The first two terms describe the actions of the transverse and longitudinal components of a magnetic field on the two qubits while the third term depicts the entanglement. One can prove that the coupling $J$ between the qubits may be relevantly associated with the parameter $D$ of our study. To this end, one can define the spin-1/2 collective operator $S^{z}=(\boldsymbol{\mathrm{\sigma}}_{z}^{(1)}+\boldsymbol{\mathrm{\sigma}}_{z}^{(2)})/2$ and show that $(S^{z})^{2}=(1+\boldsymbol{\mathrm{\sigma}}_{z}^{(1)}\boldsymbol{\mathrm{\sigma}}_{z}^{(2)})/2$. Thus, up to an  irrelevant additional term we have $J\sim D/2$. The model (\ref{equ5.3}) is used in Ref.\onlinecite{Makhlin} to control two qubits in Josephson junctions.

Considering representations of a two-spin system, the space of the Hamiltonian (\ref{equ5.3}) can be subdivided into two subspaces; one for spin $S=1$ containing the triplets $\ket{\uparrow\uparrow}$, $\ket{\downarrow\downarrow}$ and the Bell's state $\ket{+}=(\ket{\uparrow\downarrow}+\ket{\downarrow\uparrow\rangle})/\sqrt{2}$ (maximally entangled state) and another subspace for $S=0$ and set by the singlet $\ket{-}=(\ket{\uparrow\downarrow}-\ket{\downarrow\uparrow\rangle})/\sqrt{2}$. For the same reasons as in Refs.[\onlinecite{Raymond, Jose}], we discard the singlet $\ket{-}$ and consider the remaining three triplets. After considering the collective spin operators $S^{x}=(\boldsymbol{\mathrm{\sigma}}_{x}^{(1)}+\boldsymbol{\mathrm{\sigma}}_{x}^{(2)})/2$ and $S^{z}=(\boldsymbol{\mathrm{\sigma}}_{z}^{(1)}+\boldsymbol{\mathrm{\sigma}}_{z}^{(2)})/2$ and neglecting the irrelevant Abelian term, the leading Hamiltonian acquires the form (\ref{equ1}) with a spin $S=1$. For a linear variation of $\mathbf{B}_{z}(t)$, it was shown in Refs.[\onlinecite{Raymond, Jose}] that the energy diagram of a two-qubit quantum simulator maps a triangular geometry. Thus, the theory elaborated in this paper is a solid test-bed for creating entangled states and controlling the temporal evolution of two entangled qubits by applying a longitudinal linearly changing magnetic field and a transverse periodically varying electric field. The interest for QIP lies in the major fact that the system described by (\ref{equ5.3}) is more robust against external nuisances and encodes more information as compared to a single qubit\cite{KenmoePRB}.

\section{Conclusions}\label{Sec6}

We have revisited the $SU(3)$ LZSM interferometry originally set up with constant tunnel matrix elements and adopted three realistic versions that may be useful for high-precision measurements; in spectroscopy analysis to glean more information about this set up  or used for implementing quantum technologies. In the first case (i), the tunnel matrix element periodically changes while in the second case (ii) a static part is added to it and finally, in the case (iii) it is tailored as a superposition of higher harmonics. We have observed that in (i) the renormalization splits each of the original paths into two, consequently doubling the number of patterns. Indeed, for constant couplings, the maximum number of steps is two while four steps are achieved for periodic couplings. The interference patterns observed in the former case is doubled in the latter case. We have detected a new regime in which beats and steps coexist. This pattern appears when the frequency of the transverse signal matches the frequency of non-adiabatic oscillations and the uni-axial anisotropy is large. Three steps are equally observed. This pattern occurs when the uniaxial anisotropy matches the driving frequency of the signal with both being of large order of magnitude. In case (ii), the periodic drive is rather added to the tunnel matrix element in the original model. The original paths are split into three leading thus to observation of six steps. We have argued that the appearance of steps is related to the number of individual monochromatic signals in the main signal. This assertion is defended in case (iii) as we consider superposition of higher harmonics. In each of the cases discussed in this paper, numerical tests are implemented for verification of the range of validity of analytical results. Remarkable agreements are observed between analytical and numerical results that are barely discernible. The role played by steps in the physics of LZSM transitions is clarified by rewriting the model in terms of quantized fields as a five-level model with pseudo $SU(5)$ symmetry. It is established that a multi-photon process is enough to describe oscillations  appearing in the population dynamics. We have compared our results with known theoretical and experimental results with a special attention granted to many-body LZSM tunneling and QIP. We therefore believe that we have opened up a new route for exploring interferometry in many-body atom-boson systems for high-precision measurements and quantum information processing.
  
\section*{Acknowledgments}

MBK is grateful to AIMS-Ghana for the warm hospitality where this project was entirely written and to E. F. Kazeem and F. Ngoran for careful reading of the manuscript and linguistic suggestions. We equally acknowledge valuable discussions with R. M. Koch and A. N. Kammogne.

\appendix

\section{Complete Transition Matrix}\label{App1}
Solving the integral-differential equations (\ref{equ4})-(\ref{equ6}) we got the transition matrix
\begin{eqnarray} \label{equ9} 
\nonumber\mathbf{P}_{\kappa'\to \kappa}(t)\approx \left[\begin{array}{ccc} {1-p_{+}}(t) & {p_{+}}(t) & {0} \\ {p_{+}}(t) & {1-p_{+}(t)-p_{-}(t)} & {p_{-}}(t) \\ {0} & {p_{-}}(t) & {1-p_{-}}(t) \end{array}\right].\\
\end{eqnarray}
where we first defined
\begin{eqnarray}\label{equ10}
p_{\pm}(t)=2\pi\delta\Big[F^{\pm}_{cc}(t)+F^{\pm}_{cs}(t)\Big],
\end{eqnarray}
and where we introduced the two additional functions
$F^{\pm}_{cc}(t)=[G^{\pm}_{cc}(t)]^{2}$ and $F^{\pm}_{cs}(t)=[G^{\pm}_{cs}(t)]^{2}$. Here 
\begin{eqnarray}\label{equ11}
\nonumber G^{\pm}_{cc}(t)=\dfrac{1}{2}\Big[\mathcal{C}\Big(t\pm\dfrac{D\mp\omega}{\alpha}\Big)\cos\Psi^{\mp}+\mathcal{C}\Big(t\pm\dfrac{D\pm\omega}{\alpha}\Big)\cos\varphi^{\pm}\\\nonumber+\mathcal{S}\Big(t\pm\dfrac{D\mp\omega}{\alpha}\Big)\sin\Psi^{\mp}-\mathcal{S}\Big(t\pm\dfrac{D\pm\omega}{\alpha}\Big)\sin\varphi^{\pm}\Big],\\
\end{eqnarray}
and
\begin{eqnarray}\label{equ12}
\nonumber G^{\pm}_{cs}(t)=\dfrac{1}{2}\Big[\mathcal{S}\Big(t\pm\dfrac{D\mp\omega}{\alpha}\Big)\cos\Psi^{\mp}+\mathcal{S}\Big(t\pm\dfrac{D\pm\omega}{\alpha}\Big)\cos\varphi^{\pm}\\\nonumber-\mathcal{C}\Big(t\pm\dfrac{D\mp\omega}{\alpha}\Big)\sin\Psi^{\mp}+\mathcal{C}\Big(t\pm\dfrac{D\pm\omega}{\alpha}\Big)\sin\varphi^{\pm}\Big],\\
\end{eqnarray}
with
\begin{subeqnarray}\label{equ13a}
	& \mathcal{C}(x(t))=C\Big(\sqrt{\dfrac{\alpha}{\pi}}x(t)\Big)-C\Big(\sqrt{\dfrac{\alpha}{\pi}}x(t_{0})\Big)\hspace{-0.4cm}\quad, \\
	& \mathcal{S}(x(t))=S\Big(\sqrt{\dfrac{\alpha}{\pi}}x(t)\Big)-S\Big(\sqrt{\dfrac{\alpha}{\pi}}x(t_{0})\Big),
\end{subeqnarray}
where $C(...)$ and $S(...)$ are Fresnel integrals\cite{Abramo}. It is worth noting that in this scenario, our solutions hold for all possible initial time $t_{0}=t_{\rm initial}$ and final time $t=t_{\rm final}$ allowing adequate manipulation of the ThLS  in the non-adiabatic regime.  

In order to clearly see the interference processes occurring in the triangle by splitting and recombination of wave functions at vertices,  we introduced the functions in Eqs. (\ref{equ14a}) and (\ref{equ14b}) as
$
F(x,y)=[\mathcal{C}(x)\mathcal{C}(y)+\mathcal{S}(x)\mathcal{S}(y)]/2,
$
and
$
G(x,y)=[\mathcal{C}(x)\mathcal{S}(y)-\mathcal{S}(x)\mathcal{C}(y)]/2,
$
defined such that $F(x,y)=F(y,x)$ (symmetric by argument permutation) and $G(x,y)=-G(y,x)$ (anti-symmetric). These functions also have the following properties:  $F^{2}(x,y)+G^{2}(x,y)=F(x,x)F(y,y)$ and $(F(x,y)\pm G(x,y))^{2}=F(x,x)F(y,y)\pm 2F(x,y)G(x,y)$ and one of them is widely known in the physics of LZSM transitions\cite{Kiselev, KenmoePRB}, $F(t,t)\equiv F(t)$ with
\begin{eqnarray}\label{equ13b}
F(t)=\frac{1}{2}\Big[\Big(\frac{1}{2}+C\Big(\sqrt{\frac{\alpha}{\pi}}t\Big)\Big)^{2}+\Big(\frac{1}{2}+S\Big(\sqrt{\frac{\alpha}{\pi}}t\Big)\Big)^{2}\Big].
\end{eqnarray}
This function is encountered in the theory of light diffraction, where it describes the intensity of light passing through a semi-infinite plane
bounded by a sharp straight edge with $t$ standing for the lateral distance of the edge from the point of observation.\cite{KenmoePRB, Born}

\section{The model (\ref{equ15})}\label{App2}
In this appendix, we present additional materials for the derivation of the model Eq.(\ref{equ15}) and integrate it for some specific cases. Let us recall that the quantized $SU(3)$ model reads (see subsection \ref{Sec3c})
\begin{eqnarray}\label{equation1}
\mathcal{H}(t)=\alpha tS^{z}+\mathcal{H}_{\rm bath}+\mathcal{H}_{\rm ThLS-bath}+D(S^{z})^{2},
\end{eqnarray}
where
\begin{eqnarray}\label{equation2}
\mathcal{H}_{\rm bath}=\omega(\hat{a}^{\dagger}\hat{a}-\hat{b}^{\dagger}\hat{b}),
\end{eqnarray}
and
\begin{eqnarray}\label{equation3}
\mathcal{H}_{\rm ThLS-bath}=[ g_{a}(\hat{a}^{\dagger}+\hat{a})+g_{b}(\hat{b}^{\dagger}+\hat{b})]S^{x}.
\end{eqnarray}
Let $\ket{m,\{n_a,n_b\}}=\ket{m}\otimes\ket{\{n_a,n_b\}}$ denote the eigenstates of $\mathcal{H}(t)-\mathcal{H}_{\rm ThLS-bath}$ [The $\ket{m}$ are eigenstates of $\alpha tS^{z}+D(S^{z})^{2}$ and $\ket{\{n_a,n_b\}}$ are those of $\mathcal{H}_{\rm bath}$]. They generate a vector space $\mathcal{F}$ equipped with the orthogonality relation $\bra{\{n_{a}',n_{b}'\},m'}{m,\{n_{a},n_{b}\}}\rangle=\delta_{m,m'}\delta_{\{n_{a},n_{b}\},\{n_{a}',n_{b}'\}}$ and the closure relation $\sum_{m, n_{a},n_{b}}\ket{m,\{n_{a},n_{b}\}}\bra{\{n_{a},n_{b}\},m}=\hat{\mathbf{1}}$ (unit matrix). 
Here,
\begin{eqnarray}\label{equation4}
\delta_{\{n_{a},n_{b}\},\{n_{a}',n_{b}'\}}=\delta_{n_{a},n_{a}'}\delta_{n_{b},n_{b}'},
\end{eqnarray}
where $\delta_{a,b}$ is the Kronecker delta of $a$ and $b$. It takes the value $1$ when $a=b$ and $0$ otherwise.
The creation and annihilation operators of bosons act on $\mathcal{F}$ according to
\begin{eqnarray}\label{equation5}
\nonumber &\hat{a}\ket{m,\{n_a,n_b\}}=\sqrt{n_a}\ket{m,\{n_a-1,n_b\}},\\
\nonumber &\hat{b}\ket{m,\{n_a,n_b\}}=\sqrt{n_b}\ket{m,\{n_a,n_b-1\}},\\
\nonumber &\hat{a}^{\dagger}\ket{m,\{n_a,n_b\}}=\sqrt{n_a+1}\ket{m,\{n_a+1,n_b\}},\\  \nonumber &\hat{b}^{\dagger}\ket{m,\{n_a,n_b\}}=\sqrt{n_b+1}\ket{m,\{n_a,n_b+1\}},\\  \nonumber &\hat{a}^{\dagger}\hat{a}\ket{m,\{n_a,n_b\}}=n_a\ket{m,\{n_a,n_b\}},\\ &\hat{b}^{\dagger}\hat{b}\ket{m,\{n_a,n_b\}}=n_b\ket{m,\{n_a,n_b\}}.
 \end{eqnarray}
Considering these actions, the matrix elements of (\ref{equation1}) are obtained as
\begin{eqnarray}\label{equation6}
\hspace{-0.7cm}\nonumber\bra{\{n_{a}',n_{b}'\},m'}\mathcal{H}(t)\ket{m,\{n_a,n_b\}}=\Big[\alpha tm+Dm^{2}+\omega(n_a-n_b)\Big]\times\\\hspace{-0.7cm}\nonumber\delta_{m,m'}\delta_{\{n_{a},n_{b}\},\{n_{a}',n_{b}'\}}
+\bra{\{n_{a}',n_{b}'\},m'}\mathcal{H}_{\rm ThLS-bath}\ket{m,\{n_a,n_b\}},\\
\end{eqnarray}
where
\begin{eqnarray}\label{equation8}
\nonumber\bra{\{n_{a}',n_{b}'\},m'}\mathcal{H}_{\rm  ThLS-bath}\ket{m,\{n_a,n_b\}}=\\\Delta_{ab}^{+}+\Delta_{ba}^{+}+\Delta_{ba}^{-}+\Delta_{ab}^{-},
\end{eqnarray}
and where
\begin{eqnarray}\label{equation9}
\nonumber\Delta_{\gamma\gamma'}^{+}=\frac{g_{\gamma}}{2}\sqrt{n_{\gamma}+1}\Big(\sqrt{2-m(m+1)}\delta_{m',m+1}\\+\sqrt{2-m(m-1)}\delta_{m',m-1}\Big)\delta_{n_{\gamma}',n_{\gamma}+1}\delta_{n_{\gamma'}',n_{\gamma'}},
\end{eqnarray}
and
\begin{eqnarray}\label{equation10}
\nonumber\Delta_{\gamma\gamma'}^{-}=\frac{g_{\gamma}}{2}\sqrt{n_{\gamma}}\Big(\sqrt{2-m(m+1)}\delta_{m',m+1}\\+\sqrt{2-m(m-1)}\delta_{m',m-1}\Big)\delta_{n_{\gamma}',n_{\gamma}-1}\delta_{n_{\gamma'}',n_{\gamma'}}.
\end{eqnarray}
Obtaining the off diagonal components of the quantized model, we have considered the following action of the transverse component for the spin-$1$ vector onto the spin part of $\mathcal{F}$
\begin{eqnarray}\label{equation11}
\nonumber S^{x}\ket{m}=\frac{1}{2}\Big(\sqrt{2-m(m+1)}\ket{m+1}\\+\sqrt{2-m(m-1)}\ket{m-1}\Big).
\end{eqnarray}
To reproduce all the results of the semi-classical analysis including $P_{2\to2}(t)$ and all other transition probabilities, we  consider a four-photon tunneling process just as we did in the main text. In each of the $a,b$ modes, we assume a maximum of two photons; $n_{a/b}=(1,2)$ being the maximum number of excitations. This restriction reduces  the space $\mathcal{F}$  to a subspace $\mathcal{F}_{sub}$ i.e.
\begin{eqnarray}\label{equation12}
\mathcal{F}=\Big[\cdots \mathcal{F}_{sub}\cdots\Big].
\end{eqnarray}
This subspace is designed such that the actions of creation and annihilation operators on a state $\ket{m,\{n_a,n_b\}}$ which does not belong to $\mathcal{F}_{sub}$ is set to zero. This does not necessary hold for spin operators simply because the  states $\ket{m,\{n_a,n_b\}}$ are determined by $n_{a/b}$. The subspace $\mathcal{F}_{sub}$ in the four-photon restriction explicitly writes
\begin{eqnarray}\label{equation13a}
\nonumber\mathcal{F}_{sub}=\Big[\ket{1,\omega},\ket{1},\ket{1,-\omega},\ket{2,\omega},\ket{2},\\\hspace{-8cm}\ket{2,-\omega},\ket{3},\ket{3,-\omega},\ket{3,\omega}\Big].
\end{eqnarray}
Here, we have adopted the following representations for the sake of convenience in notations $\ket{m,\{n_a,n_b\}}\equiv \ket{m,(n_a-n_b)\omega}$ or explicitly
\begin{eqnarray}\label{equation14}
\nonumber	&\ket{1,\{2,1\}}=\ket{1,(2-1)\omega}=\ket{1,\omega},\\
\nonumber	&\ket{1,\{2,2\}}=\ket{1,\{1,1\}}=\ket{1,0}=\ket{1},\\
\nonumber	&\ket{1,\{1,2\}}=\ket{1,(1-2)\omega}=\ket{1,-\omega},\\
\nonumber	&\ket{0,\{2,1\}}=\ket{0,(2-1)\omega}=\ket{2,\omega},\\
\nonumber	&\ket{0,\{2,2\}}=\ket{0,\{1,1\}}=\ket{0,0}=\ket{2},\\
\nonumber	&\ket{0,\{1,2\}}=\ket{0,(1-2)\omega}=\ket{2,-\omega},\\
\nonumber	&\ket{-1,\{1,2\}}=\ket{-1,(1-2)\omega}=\ket{3,-\omega},\\
\nonumber	&\ket{-1,\{2,2\}}=\ket{-1,\{1,1\}}=\ket{-1,0}=\ket{3},\\
	&\ket{-1,\{2,1\}}=\ket{-1,(2-1)\omega}=\ket{3,\omega}.
\end{eqnarray}
Remark, no selection rule is applied. All diabatic states are allowed to contain the same number of photon of each mode. Denoting as $\lambda_{i,j}=\langle i|\mathcal{H}_{\rm  ThLS-bath}|j\rangle$ (with $i$ and $j$ the position of the states in $\mathcal{F}_{sub}$), in the basis (\ref{equation13a}), the matrix elements (\ref{equation6}) yields the Hamiltonian in matrix form 

\begin{widetext}
	
	\begin{eqnarray} \label{equation13b} 
	\nonumber
	\mathcal{H}(t)=
	\left[
	{\begin{array}{*{20}c}
		\alpha t+(D+\omega) &  0 & 0 & 0 & \lambda_{1,5}/\sqrt{2} & 0  & 0 &  0 & 0\\
		0  & \alpha t+D &  0 &  \lambda_{2,4}/\sqrt{2} &  0  &  \lambda_{2,6}/\sqrt{2} & 0  & 0  &  0\\
		0 &   0  & \alpha t+(D-\omega) & 0  & \lambda_{3,5}/\sqrt{2} & 0 & 0  &  0  & 0\\
		0  &  \lambda_{4,2}/\sqrt{2} & 0 & \omega & 0 & 0  &  0  &  \lambda_{4,8}/\sqrt{2}  &  0\\
		\lambda_{5,1}/\sqrt{2} &  0  & \lambda_{5,3}/\sqrt{2}  & 0 & 0  & 0 & \lambda_{5,7}/\sqrt{2} & 0 & \lambda_{5,9}/\sqrt{2} \\
		0  & \lambda_{6,2}/\sqrt{2} & 0 & 0 & 0 & -\omega & 0 &  \lambda_{6,8}/\sqrt{2}  & 0\\
		0 & 0 & 0 & 0 & \lambda_{7,5}/\sqrt{2} & 0 & -\alpha t+(D-\omega) & 0  & 0\\
		0 & 0 &  0 &  \lambda_{8,4}/\sqrt{2}  & 0 & \lambda_{8,6}/\sqrt{2} & 0 &-\alpha t+D & 0\\
		0 & 0 &  0 &  0 & \lambda_{9,5}/\sqrt{2}  & 0 & 0  & 0 &-\alpha t+(D+\omega)\\
		\end{array} } \right]. \\
	\end{eqnarray} 
	
\end{widetext}
 Here, we have numbered the couplings according to their positions in the matrix. They can always be associated with the states in (\ref{equation13a})  and with pumps. The Hamiltonian (\ref{equation13b}) reproduces $P_{2\to2}(t)$ as (\ref{equ15}) does, and also takes care of all other possible results of the semi-classical analysis provided suitable choices of the coupling terms. Why have we considered (\ref{equ15})  in the main text instead of (\ref{equation13b})? Indeed, we realized that the model (\ref{equation13b})  can be factorized into two effective models (one of which is (\ref{equ15})) that can be dissociated and treated separately. In order to prove this assertion, let us realize that if $\lambda_{i,j}=\lambda$ with $i,j=1,...,9$, then (\ref{equation13b})  describes two interacting spin-$1$ systems  locally subject each to a $z$-oriented magnetic field and coupled through $XX$ exchanges. Thus,
 \begin{eqnarray} \label{equation13bb} 
\mathcal{H}(t)=\alpha tS^{z}_1+\omega S^{z}_2+JS^{x}_1S^{x}_2+DS^{z}_1S^{z}_1,
\end{eqnarray} 
 with the exchange term $J=\sqrt{2}\lambda$.  
 Therefore, one can rotate the Hamiltonian (\ref{equation13b}) with the help of the unitary and hermitian rotation matrix\cite{Grimaudo}
  \begin{widetext}
 \begin{eqnarray}\label{equation13c}
 U=\left(
 \begin{array}{ccccccccc}
 0 & 0 & 0 & 0 & 1 & 0 & 0 & 0 & 0 \\
 1 & 0 & 0 & 0 & 0 & 0 & 0 & 0 & 0 \\
 0 & 0 & 0 & 0 & 0 & 1 & 0 & 0 & 0 \\
 0 & 1 & 0 & 0 & 0 & 0 & 0 & 0 & 0 \\
 0 & 0 & 0 & 0 & 0 & 0 & 1 & 0 & 0 \\
 0 & 0 & 1 & 0 & 0 & 0 & 0 & 0 & 0 \\
 0 & 0 & 0 & 0 & 0 & 0 & 0 & 1 & 0 \\
 0 & 0 & 0 & 1 & 0 & 0 & 0 & 0 & 0 \\
 0 & 0 & 0 & 0 & 0 & 0 & 0 & 0 & 1 \\
 \end{array}
 \right),
 \end{eqnarray} 
which transforms  (\ref{equation13b}) into the block form
 \begin{eqnarray} \label{equation13d} 
 \mathbf{H}(t)=U^{T}\mathcal{H}(t)U=
 \left(
 {\begin{array}{*{20}c}
 	\mathbf{H}_{up}(t)  & \mathbf{O}_{4\times5}(t) \\
 	\mathbf{O}_{5\times4}(t) & \mathbf{H}_{down}(t)\\
 	\end{array} } \right),
 \end{eqnarray} 
where

 	\begin{eqnarray} \label{equation13e} 
 	\mathbf{H}_{up}(t)=
 	\left(
 	{\begin{array}{*{20}c}
 		\alpha t+D  & \lambda _{2,4}/\sqrt{2} & \lambda _{2,6}/\sqrt{2} & 0 \\
 		\lambda _{4,2}/\sqrt{2} & \omega  & 0 & \lambda _{4,8}/\sqrt{2} \\
 		\lambda _{6,2}/\sqrt{2} & 0 & -\omega  & \lambda _{6,8}/\sqrt{2}\\
 		0 & \lambda _{8,4}/\sqrt{2} & \lambda _{8,6}/\sqrt{2} & -\alpha t+D\\
 		\end{array} } \right),
 	\end{eqnarray} 
 	and
 	\begin{eqnarray}\label{equation13f}
 	\mathbf{H}_{down}(t)=\left(
 	\begin{array}{ccccccccc}
 	\alpha t+(D +\omega)  & 0 & \lambda _{1,5}/\sqrt{2} & 0 & 0 \\
 	0 &\alpha t+(D -\omega)  & \lambda _{3,5}/\sqrt{2} & 0 & 0 \\
 	\lambda_{5,1}/\sqrt{2} & \lambda_{5,3}/\sqrt{2} & 0 & \lambda _{5,7}/\sqrt{2} & \lambda_{5,9}/\sqrt{2}\\
 	0 & 0 & \lambda _{7,5}/\sqrt{2} &-\alpha t +(D-\omega)  & 0 \\
 	0 & 0 & \lambda_{9,5}/\sqrt{2} & 0 &-\alpha t+(D+\omega)  \\
 	\end{array}
 	\right).
 	\end{eqnarray}
 \end{widetext}
 The $\mathbf{O}_{n\times m}(t)$ is an $(n\times m)$  matrix [$n$ rows and $m$ columns] with zeros as entries. From Eq.(\ref{equation13f}) one can recognize the model (\ref{equ15}). Let us now demonstrate that $\mathbf{H}_{up}(t)$ and  $\mathbf{H}_{down}(t)$ are completely decoupled from each other and may be analyzed separately. As a starting point, let us note that   they respectively sit in 
\begin{eqnarray}\label{equation13h}
\mathcal{F}_{sub}^{up}=[\ket{1}, \ket{2,\omega}, \ket{2,-\omega}, \ket{3}],
\end{eqnarray}
and
\begin{eqnarray}\label{equation13g}
\mathcal{F}_{sub}^{down}=\Big[\ket{1,\omega},\ket{1,-\omega},\ket{2},\ket{3,-\omega},\ket{3,\omega}\Big].
\end{eqnarray}
As a consequence, $\mathcal{F}_{sub}$ is a composition of $\mathcal{F}_{sub}^{up}$ and $\mathcal{F}_{sub}^{down}$. Then, in the block form, if one constructs the wave function as $\ket{\Psi(t)}=[\ket{\Psi_{up}(t)},\ket{\Psi_{down}(t)}]^{T}$ the dynamics described by the models (\ref{equation13e}) and (\ref{equation13f}) are completely decoupled from each other and can be treated separately. In other words, one obtains the total wave function $\ket{\Psi(t)}$ after solving $i\partial_{t}\ket{\Psi_{up}(t)}=\mathbf{H}_{up}(t)\ket{\Psi_{up}(t)}$ and $i\partial_{t}\ket{\Psi_{down}(t)}=\mathbf{H}_{down}(t)\ket{\Psi_{down}(t)}$ separately. This explains why as we were only interested  in $P_{2\to2}(t)$ we could only and safely focused on $\mathbf{H}_{down}(t)$ i.e. (\ref{equ15}) and dropped down $\mathbf{H}_{up}(t)$. The same strategy we used for $P_{2\to2}(t)$ can be used with (\ref{equation13e}) to describe other transition probabilities. For instance, $\mathbf{H}_{up}(t)$ can be used to describe the results of the semi-classical analysis for $P_{2\to1}(t)$ and $P_{2\to3}(t)$. As yet another consequence of this separation, there is no transition between the states in (\ref{equation13h}) and those in (\ref{equation13g}).

Let us now integrate the models $\mathbf{H}_{up}(t)$ and $\mathbf{H}_{down}(t)$ (or equivalently  (\ref{equ15})) and obtain their long positive time asymptotic solutions. The method for doing this is elaborate in details in Refs.[\onlinecite{Sun, Sinitsyn2, Sinitsyn3}]. It is worth noting that $\mathbf{H}_{down}(t)$ ($\mathbf{H}_{up}(t)$) depicts different dynamics depending on the values of the static shifts  of the detuning $D+\omega$ and $D-\omega$ ($D$ for $\mathbf{H}_{up}(t)$). This suggests that different values of $D$ and $\omega$ yield different non-adiabatic trajectories and consequently different expressions for transition probabilities. For $\mathbf{H}_{down}(t)$ for instance, we have observed that when $D=0$, the model describes two consecutive spin-$1$ LZSM processes (see Fig.\ref{Figure8} left upper panel) i.e. three diabatic/adiabatic states cross/come close at $-\omega/\alpha$ and at $+\omega/\alpha$. When $D=\omega$ (see Fig.\ref{Figure8} middle upper panel), three lines cross at $t=0$ and two cross at $-2\omega/\alpha$ and  at $+2\omega/\alpha$. We have a consecutive sequence of three LZSM processes, one spin-$1/2$ followed by a spin-$1$ and ending by another spin-$1/2$ LZSM process. When $\sqrt{\alpha}\ll \omega\ll D$ (see Fig.\ref{Figure8} right upper panel) the model depicts four distinct separated-in-time crossing points where two lines cross. These are the three cases considered below for each model.

In what follows, we set all couplings equals to $\lambda\sqrt{\alpha}$. For $\mathbf{H}_{up}(t)$, we only consider initial occupations of the diabatic state $\ket{2,\omega}$ at $t=-\infty$. For $\mathbf{H}_{down}(t)$, we concentrate on $\ket{2}$ at $t=-\infty$. In both cases, we leave other preparations to Ref.\onlinecite{KenmoeFuture}. The work in Ref.\onlinecite{Sinitsyn3} teaches us how to compute transition probabilities in multilevel LZSM systems such as the ones described by $\mathbf{H}_{up}(t)$ and $\mathbf{H}_{down}(t)$. We will not review the technique in full detail but review some of its basic principles that are relevant to our task.  When two diabatic energy  levels $i$ and $j$ with slopes $\alpha_i$ and $\alpha_j=-\alpha_i$ cross at a resonance point $t_0$, if the coupling $\Delta_{ij}$ between them is nonzero (spin-$1/2$ LZSM problem), then the probability amplitude for remaining in the same diabatic state after traversing the crossing is given by $c_{i\to i}=\sqrt{p_{i\to j}}e^{i\varphi}$. Here, $p_{i\to j}=e^{-2\pi\Delta_{ij}^2/|\alpha_i-\alpha_j|}$ is the celebrated LZSM formula\cite{lan, zen, stu, Majorana} and $\varphi$ a phase picked up by the system during the linear sweep. The probability amplitude for changing diabatic state is given by $c_{i\to j}=\sqrt{1-p_{i\to j}}e^{i\varphi'}$. If a third diabatic energy level let say $k$ with slope zero crosses the previous ones at the same point in the direction of positive times (spin-$1$ LZSM problem), then the probability amplitude to remain in that state after passing the unique crossing is given by $c_{k\to k}=(2p_{i\to j}-1)e^{i\varphi''}$ while $c_{k\to j}=\sqrt{2(p_{i\to j}-p_{i\to j}^2)}e^{i\varphi'''}$  is that for changing diabatic state\cite{Kenmoe2013}. Importantly,  when the spin-$1$ LZSM process is preceded and/or followed by another process, the coupling in the last two formulas is divided by a factor of $\sqrt{2}$. Considering these rules, and Ref.\onlinecite{Sinitsyn3}, analytical solutions to $\mathbf{H}_{up}(t)$ and $\mathbf{H}_{down}(t)$ are obtained.

\begin{widetext}
	
	\begin{figure}[]
		\centering
		\vspace{-0.5cm}
		\includegraphics[width=6.cm, height=6cm]{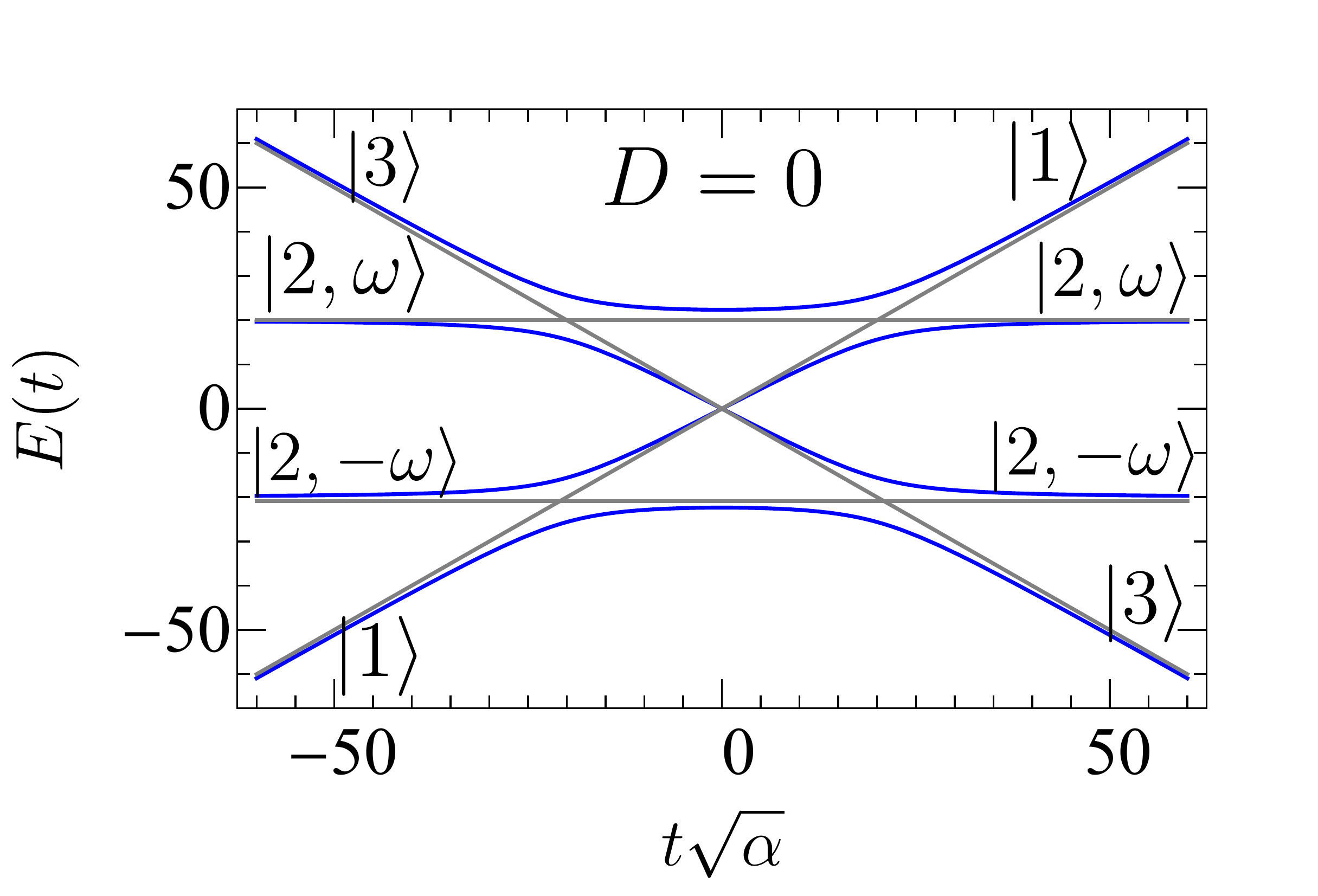}\hspace{-0.5cm}
		\includegraphics[width=6.cm, height=6cm]{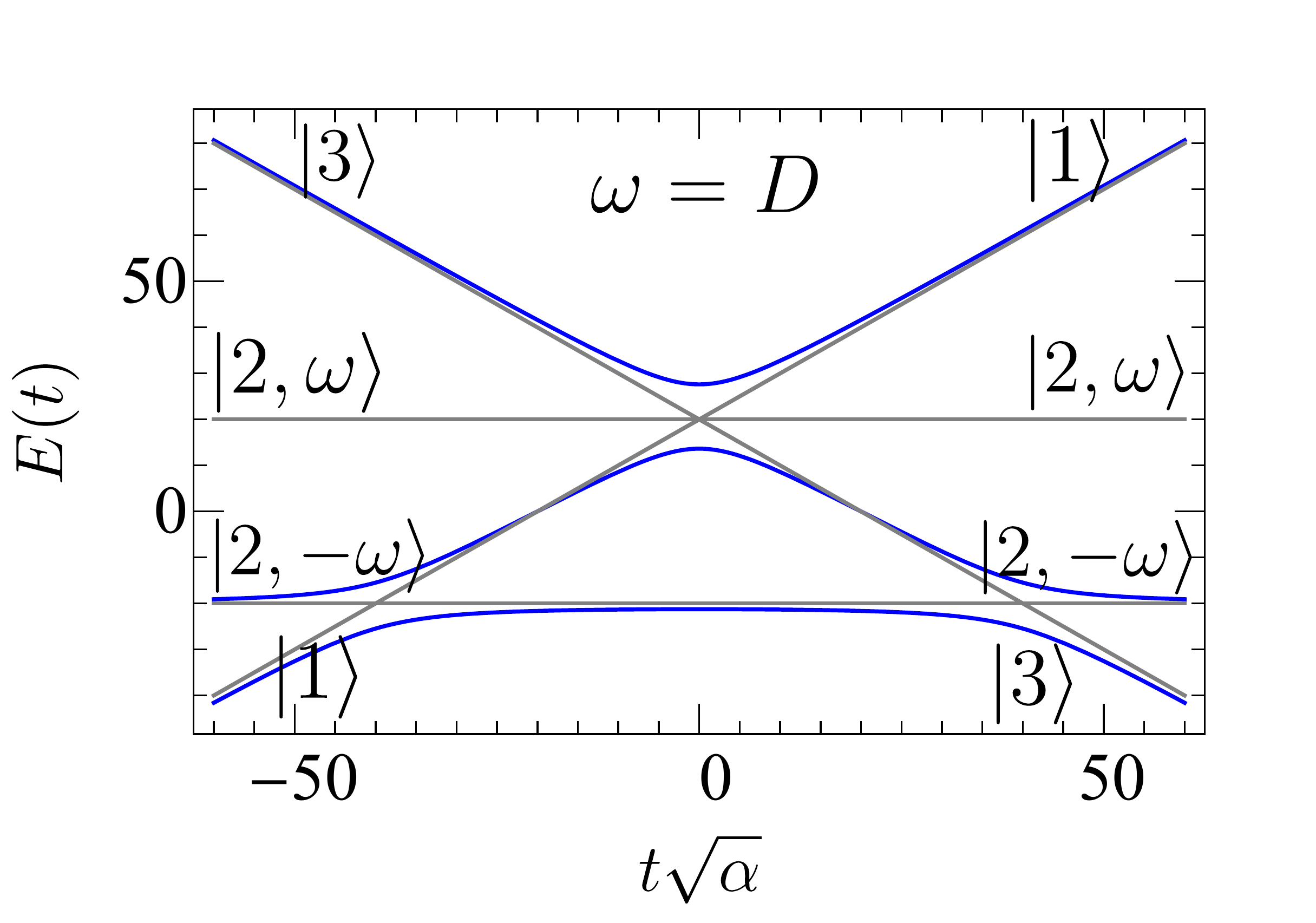}\hspace{-0.5cm}
		\includegraphics[width=6.cm, height=6cm]{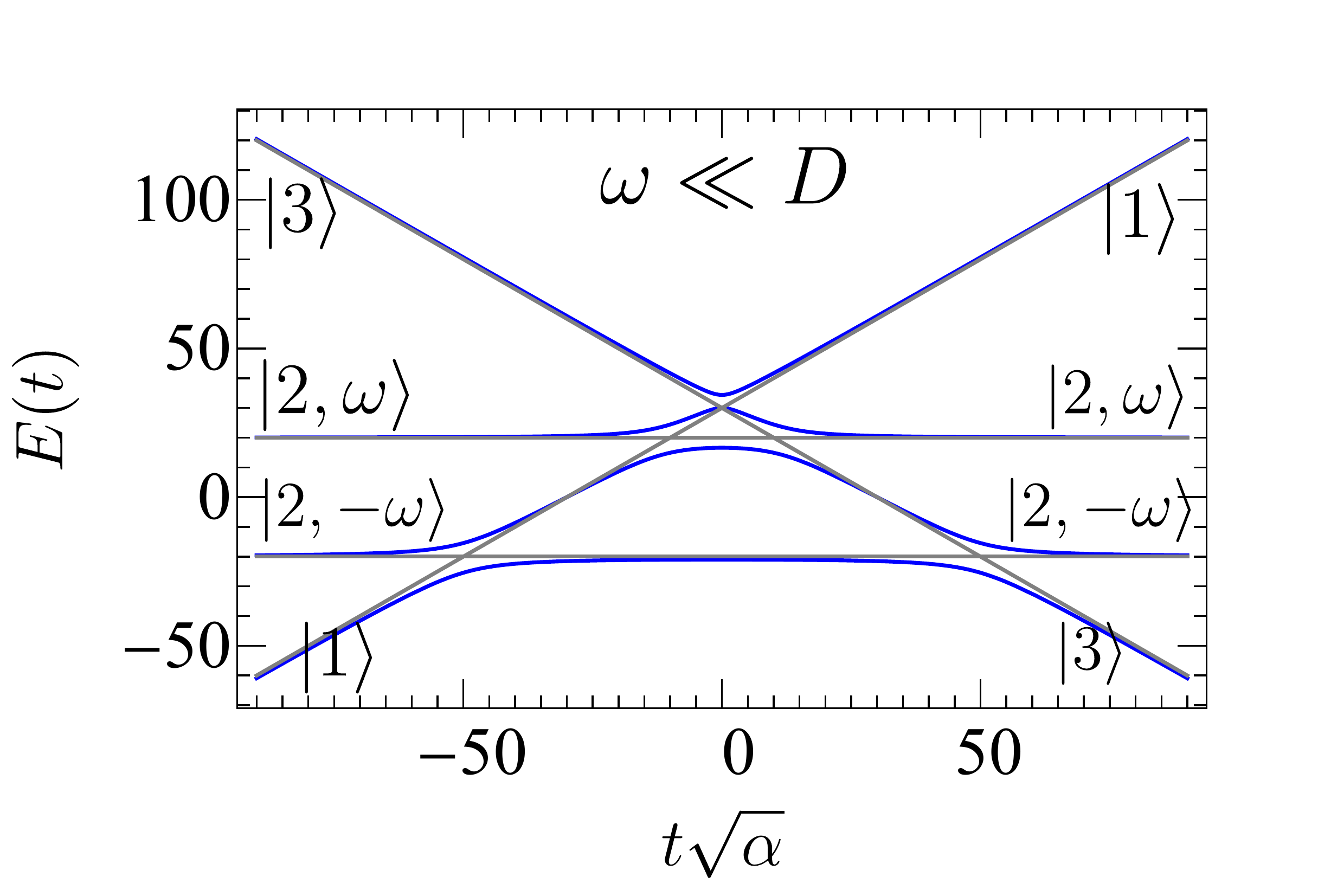}\\\vspace{-0.7cm}
		\includegraphics[width=6.cm, height=6cm]{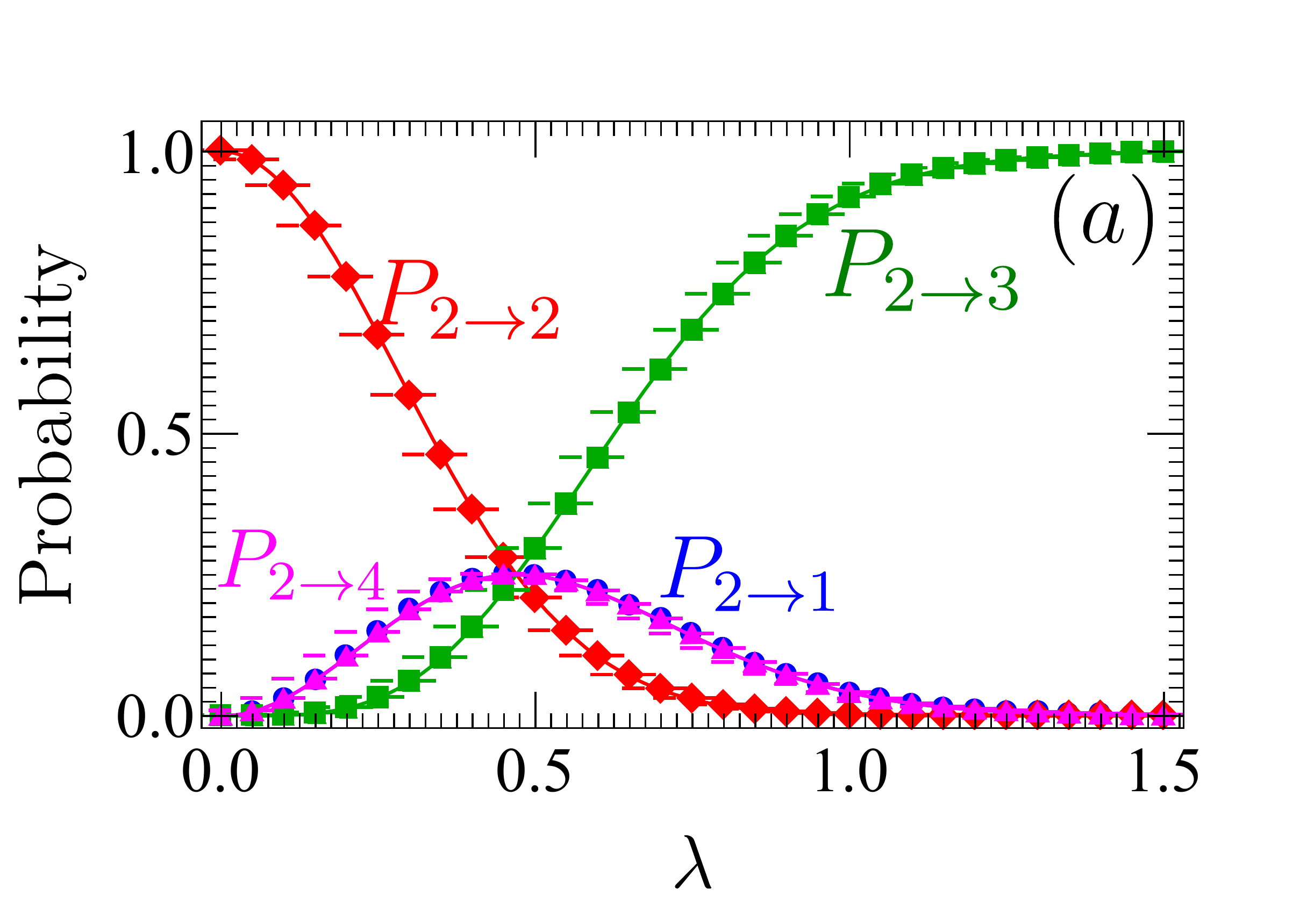}\hspace{-0.5cm}
		\includegraphics[width=6.cm, height=6cm]{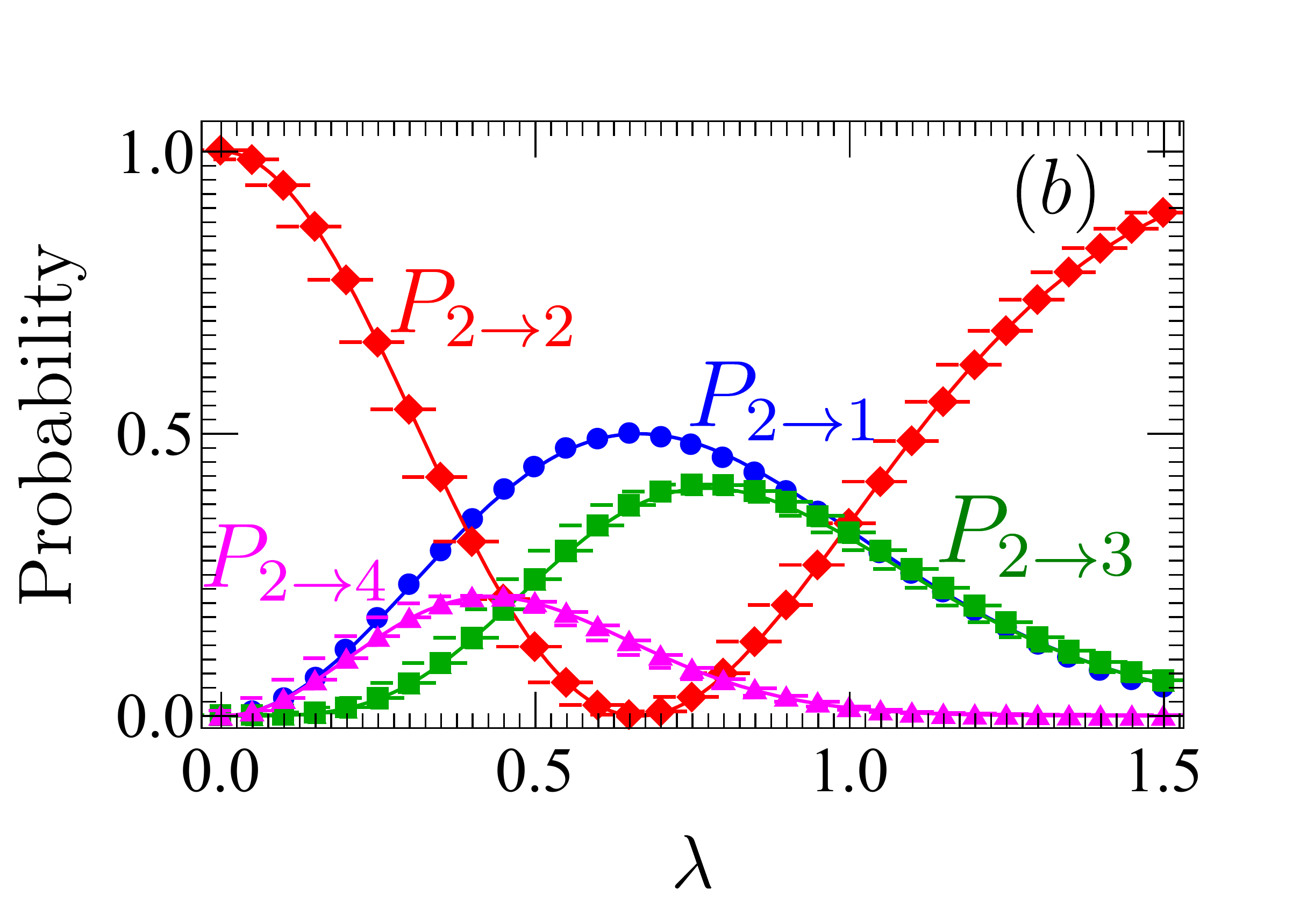}\hspace{-0.5cm}
		\includegraphics[width=6.cm, height=6cm]{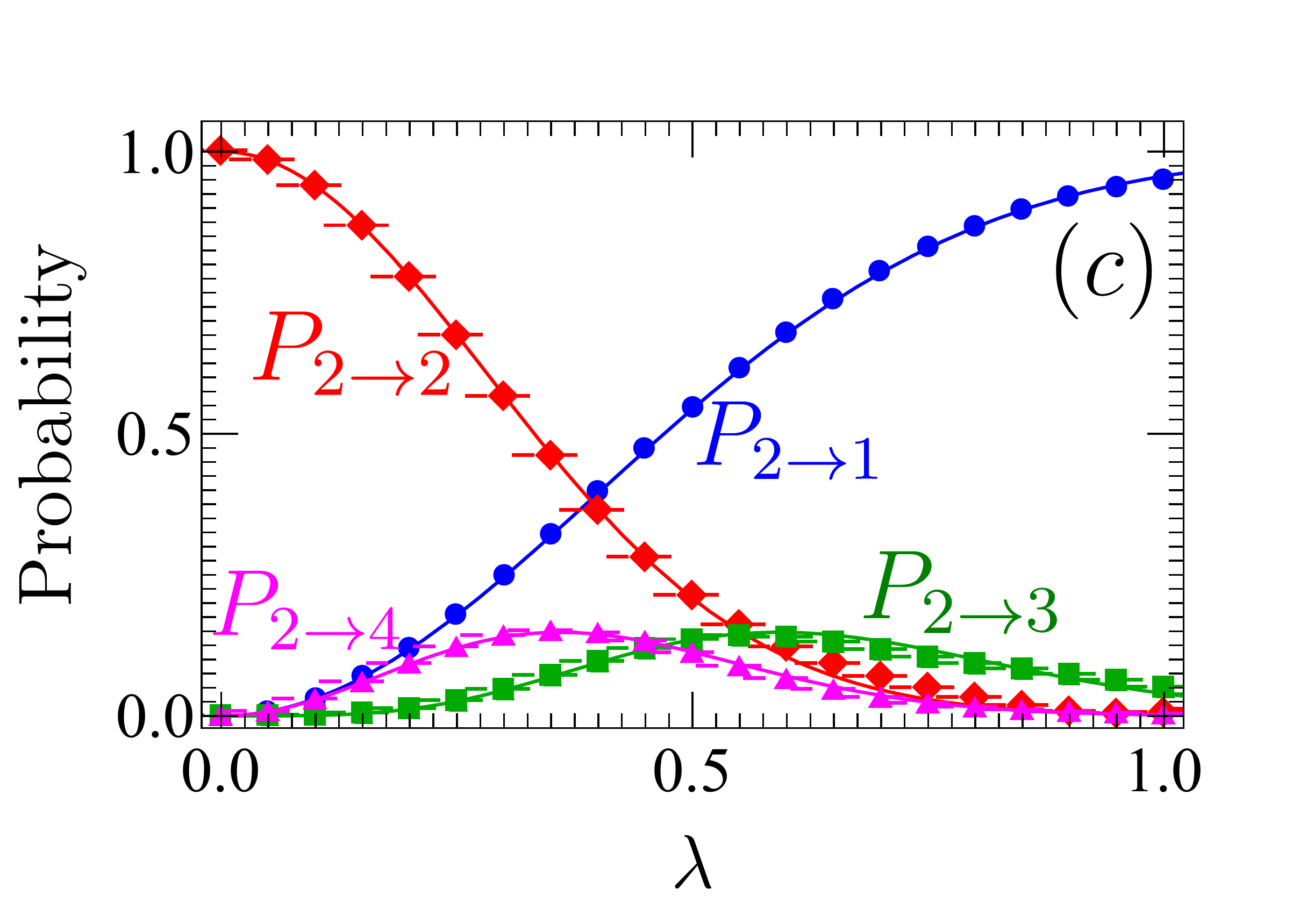}
		\vspace{-0.4cm}
		\caption{(Color Online) {\bf Upper panel:} Eigenenergies (blue solid lines) and diabatic energies (gray solid lines) of (\ref{equation13e}). Left panel corresponds to the case $D=0$, middle panel $D=\omega$ and the right panel depicts the case $\omega\ll D$. For all calculations, $\lambda_{i,j}=4\sqrt{\alpha}$ with $i,j=2,4,6,8$. {\bf Lower panel:} Transition probabilities in the four-level model (\ref{equation13e}) for an initial occupation of  $\ket{2,\omega}$. Solid objects and solid lines are respectively the numerical and analytical results. We have considered $\lambda_{i,j}=\lambda\sqrt{\alpha}$ with $i,j=2,4,6,8$. Left panels:  $D/\sqrt{\alpha}=0$ and $\omega/\sqrt{\alpha}=20$. Middle panels:  $D/\sqrt{\alpha}=20$ and $\omega/\sqrt{\alpha}=20$. Right panels:  $D/\sqrt{\alpha}=20$ and $\omega/\sqrt{\alpha}=10$. The integration time runs from $t\sqrt{\alpha}=-500$ to $t\sqrt{\alpha}=500$. Here, $P_{2\to 1}$, $P_{2\to 2}$, $P_{2\to 3}$ and $P_{2\to 4}$ respectively correspond to the probabilities for the transitions $\ket{2,\omega}\to \ket{1}$, $\ket{2,\omega}\to \ket{2,\omega}$, $\ket{2,\omega}\to \ket{2,-\omega}$ and $\ket{2,\omega}\to \ket{3}$.} \label{Figure9}
	\end{figure}
	
\end{widetext}
\subsection{Solutions for  $\mathbf{H}_{up}(t)$}\label{App2.1}
For $D=0$,
\begin{eqnarray}\label{equation15a}
\nonumber	& P_{2\to1}=e^{-\frac{\pi\lambda^{2}}{\alpha}}\Big(1-e^{-\frac{\pi\lambda^{2}}{\alpha}}\Big)=P_{2\to4},\\
\nonumber	& P_{2\to2}=e^{-\frac{2\pi\lambda^{2}}{\alpha}},\\
& P_{2\to3}=\Big(1-e^{-\frac{\pi\lambda^{2}}{\alpha}}\Big)^2.
\end{eqnarray}
For $D=\omega$,
\begin{eqnarray}\label{equation15b}
\nonumber	& P_{2\to1}=2\Big(e^{-\frac{\pi\lambda^{2}}{2\alpha}}-e^{-\frac{\pi\lambda^{2}}{\alpha}}\Big),\\
\nonumber	& P_{2\to2}=\Big(2e^{-\frac{\pi\lambda^{2}}{2\alpha}}-1\Big)^2,\\
\nonumber & P_{2\to3}=2\Big(e^{-\frac{\pi\lambda^{2}}{2\alpha}}-e^{-\frac{\pi\lambda^{2}}{\alpha}}\Big)\Big(1-e^{-\frac{\pi\lambda^{2}}{\alpha}}\Big),\\
  &
 P_{2\to4}=2\Big(e^{-\frac{\pi\lambda^{2}}{2\alpha}}-e^{-\frac{\pi\lambda^{2}}{\alpha}}\Big)e^{-\frac{\pi\lambda^{2}}{\alpha}}.
\end{eqnarray}
For $\omega\ll D$,
\begin{eqnarray}\label{equation15c}
\nonumber	& P_{2\to1}=1-e^{-\frac{\pi\lambda^{2}}{\alpha}},\\
\nonumber	& P_{2\to2}=e^{-\frac{2\pi\lambda^{2}}{\alpha}},\\
\nonumber & P_{2\to3}=e^{-\frac{\pi\lambda^{2}}{\alpha}}\Big(1-e^{-\frac{\pi\lambda^{2}}{\alpha}}\Big)^2,\\
& P_{2\to4}=e^{-\frac{2\pi\lambda^{2}}{\alpha}}\Big(1-e^{-\frac{\pi\lambda^{2}}{\alpha}}\Big).
\end{eqnarray}
Here, $P_{2\to 1}$,$P_{2\to 2}$,$P_{2\to 3}$ and $P_{2\to 4}$ respectively represent the probabilities for the transitions $\ket{2,\omega}\to \ket{1}$, $\ket{2,\omega}\to \ket{2,\omega}$, $\ket{2,\omega}\to \ket{2,-\omega}$ and $\ket{2,\omega}\to \ket{3}$. It can readily be verified that in all cases $\sum_{j\in \mathcal{F}_{sub}^{up}}P_{2\to j}=1$ i.e. the total probability is conserved.  The analytical results (\ref{equation15a})-(\ref{equation15c}) are compared/tested with numerical ones (see Fig.\ref{Figure9}). A satisfactory agreement is observed between the two solutions that are barely discernible. Same can be done for other initial occupations of diabatic states.
\begin{widetext}
	
	\begin{figure}[]
		\centering
		\vspace{-0.5cm}
		\includegraphics[width=6.cm, height=6cm]{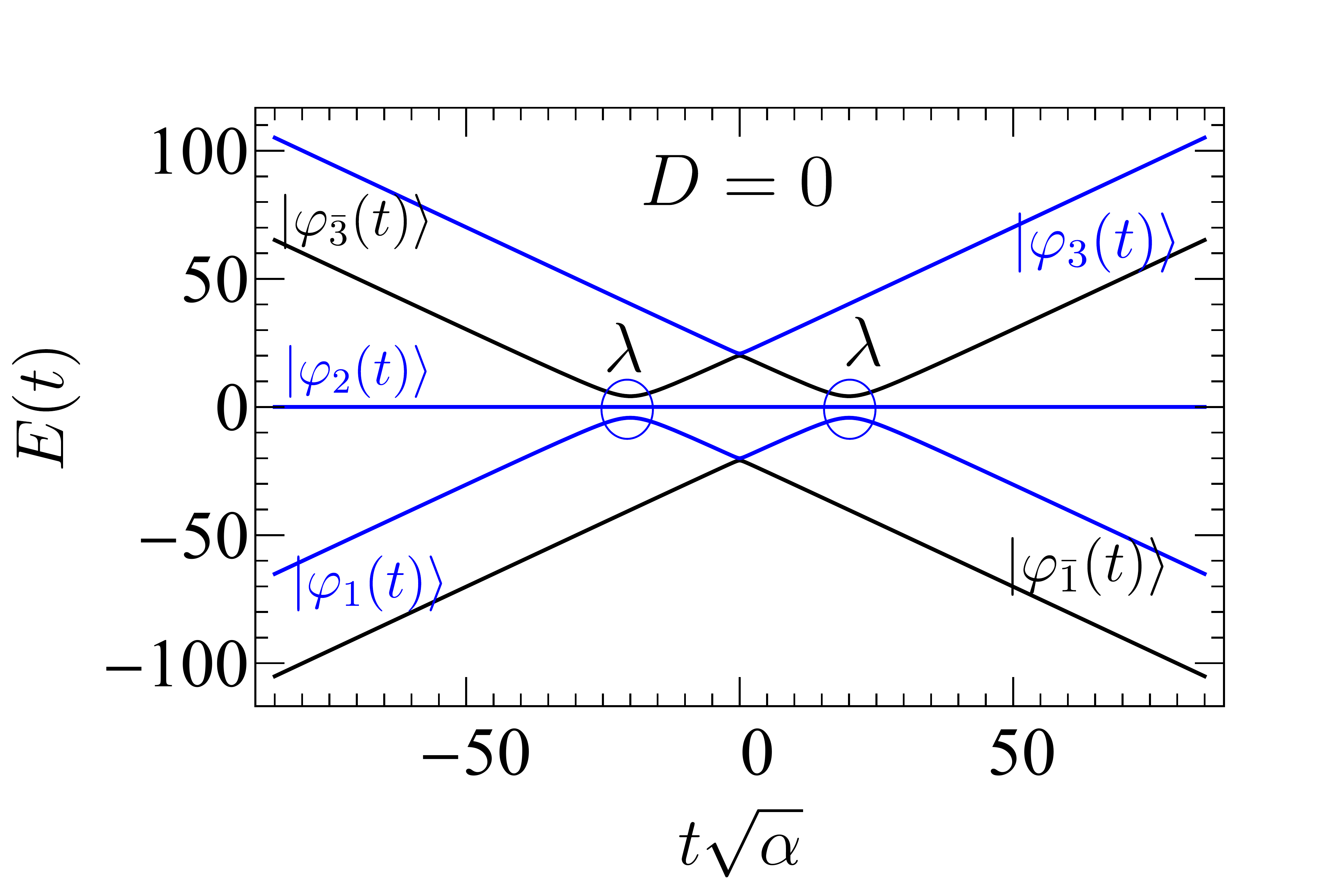}\hspace{-0.5cm}
		\includegraphics[width=6.cm, height=6cm]{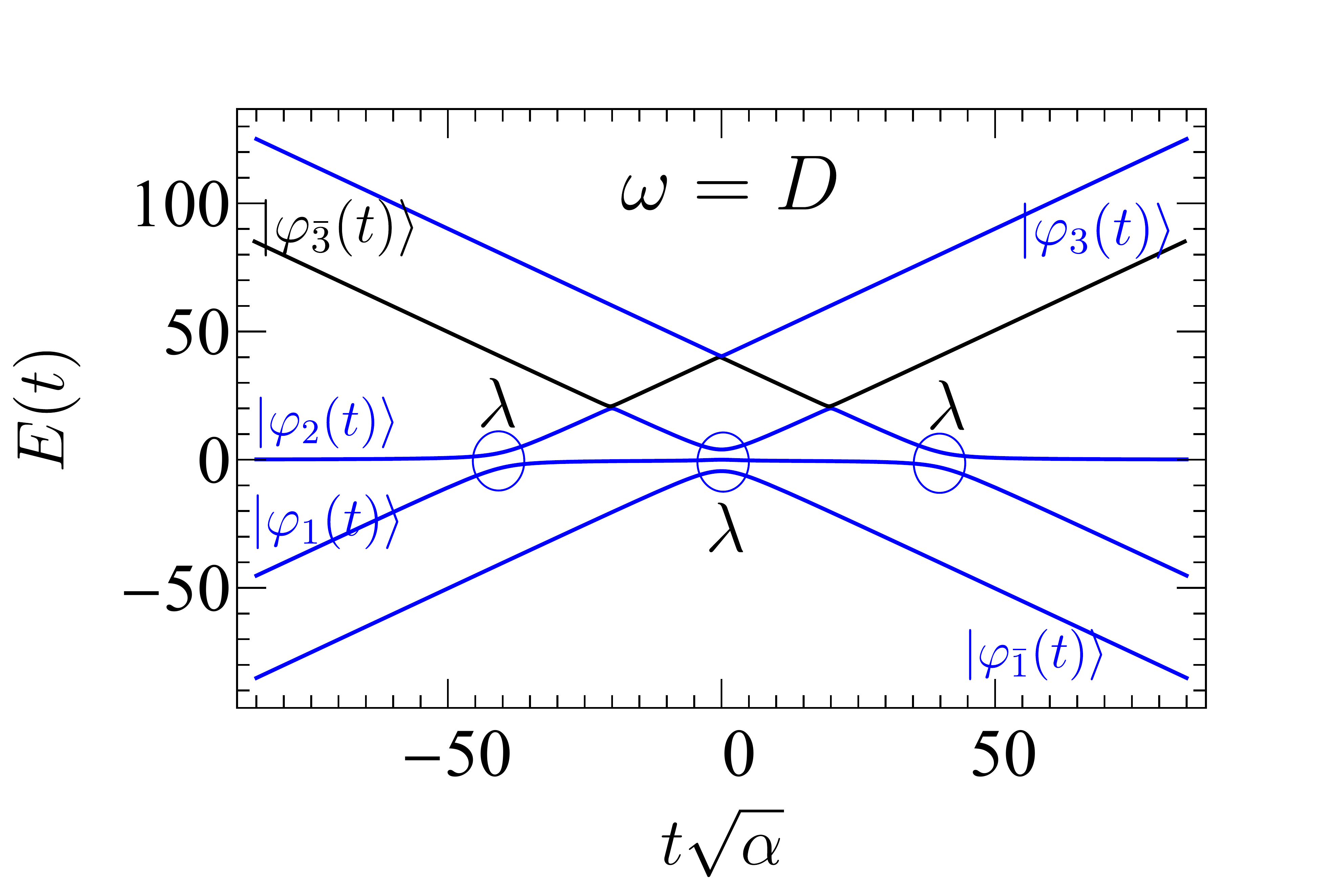}\hspace{-0.5cm}
		\includegraphics[width=6.cm, height=6cm]{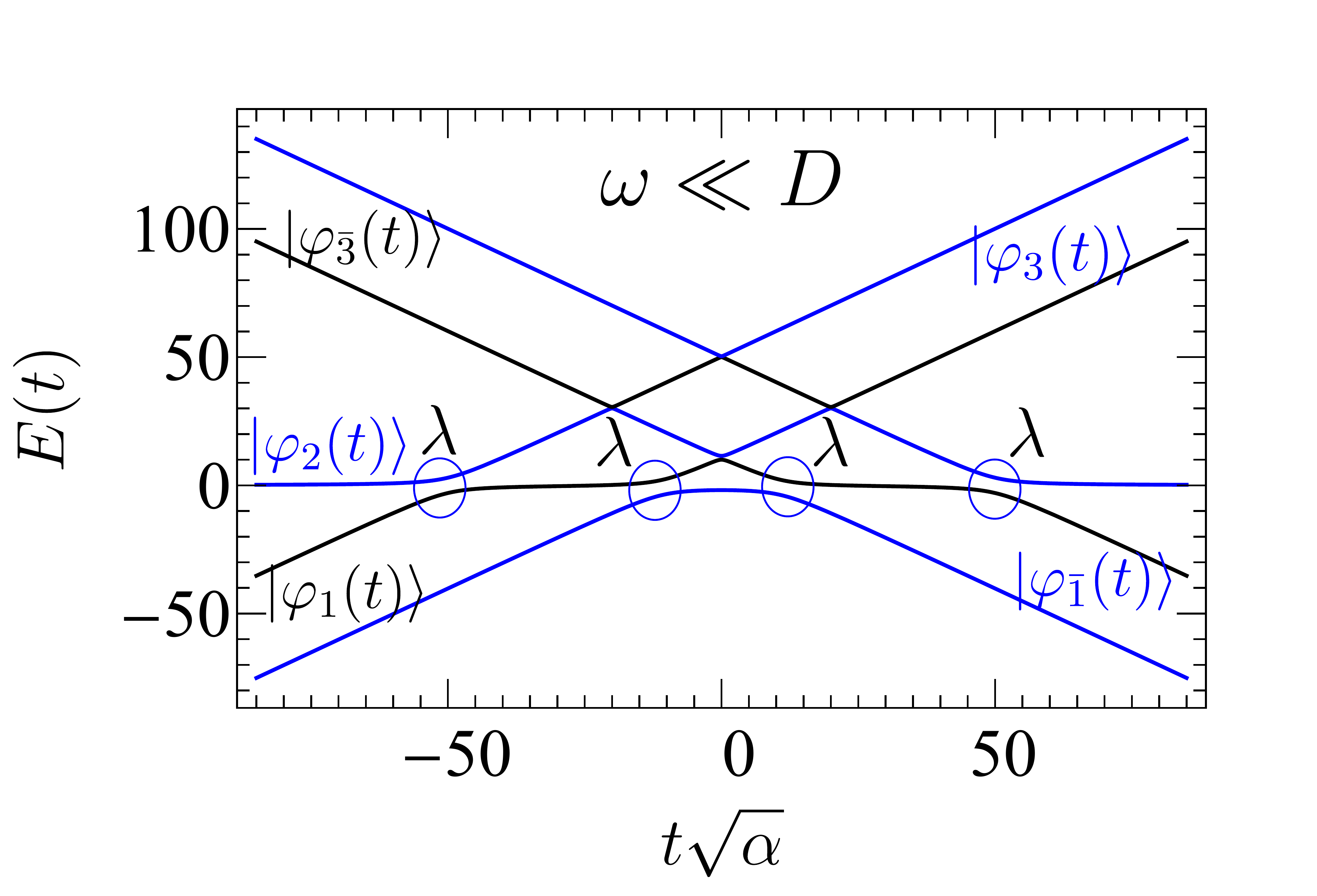}\\\vspace{-0.7cm}
		\includegraphics[width=6.cm, height=6cm]{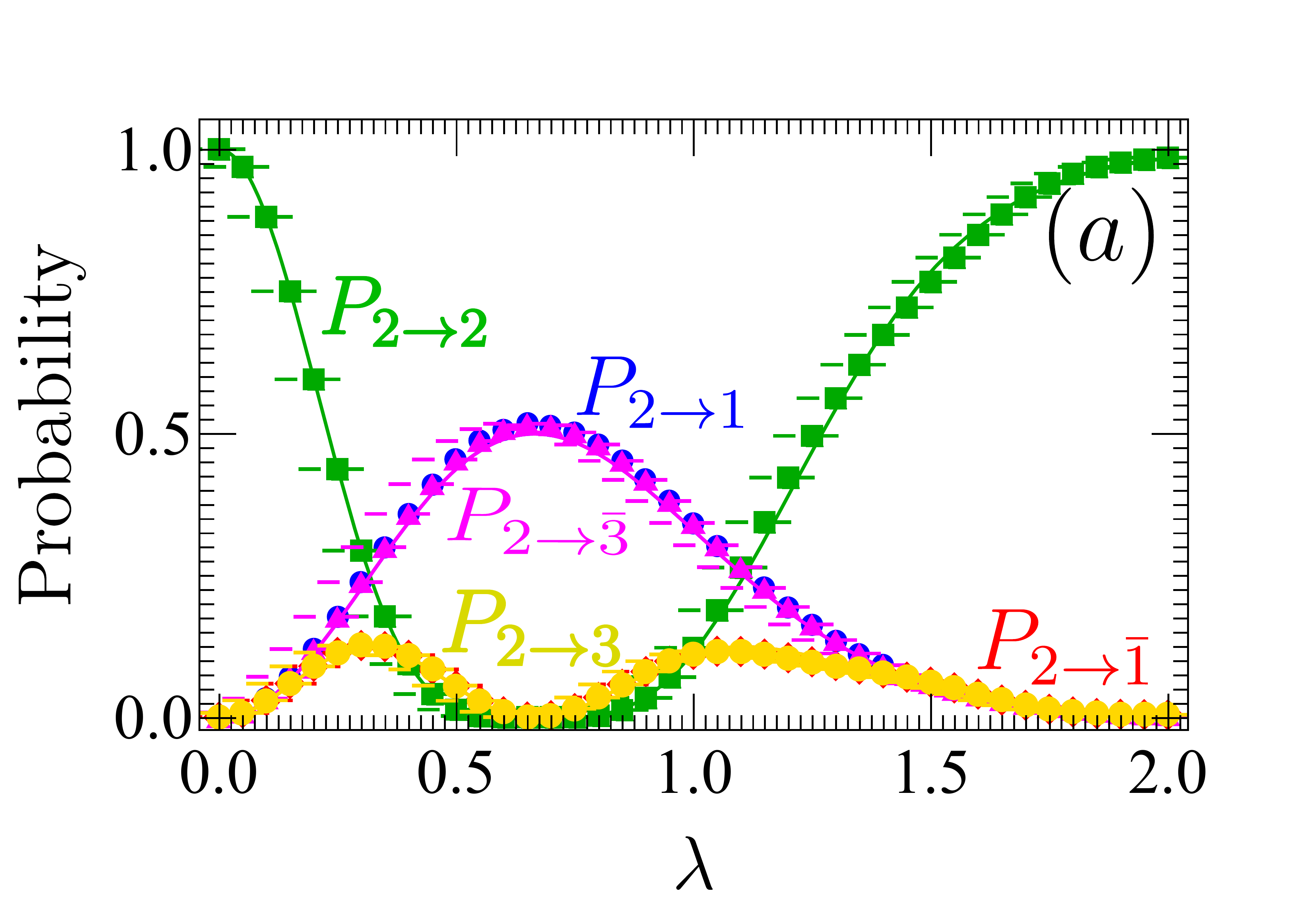}\hspace{-0.5cm}
		\includegraphics[width=6.cm, height=6cm]{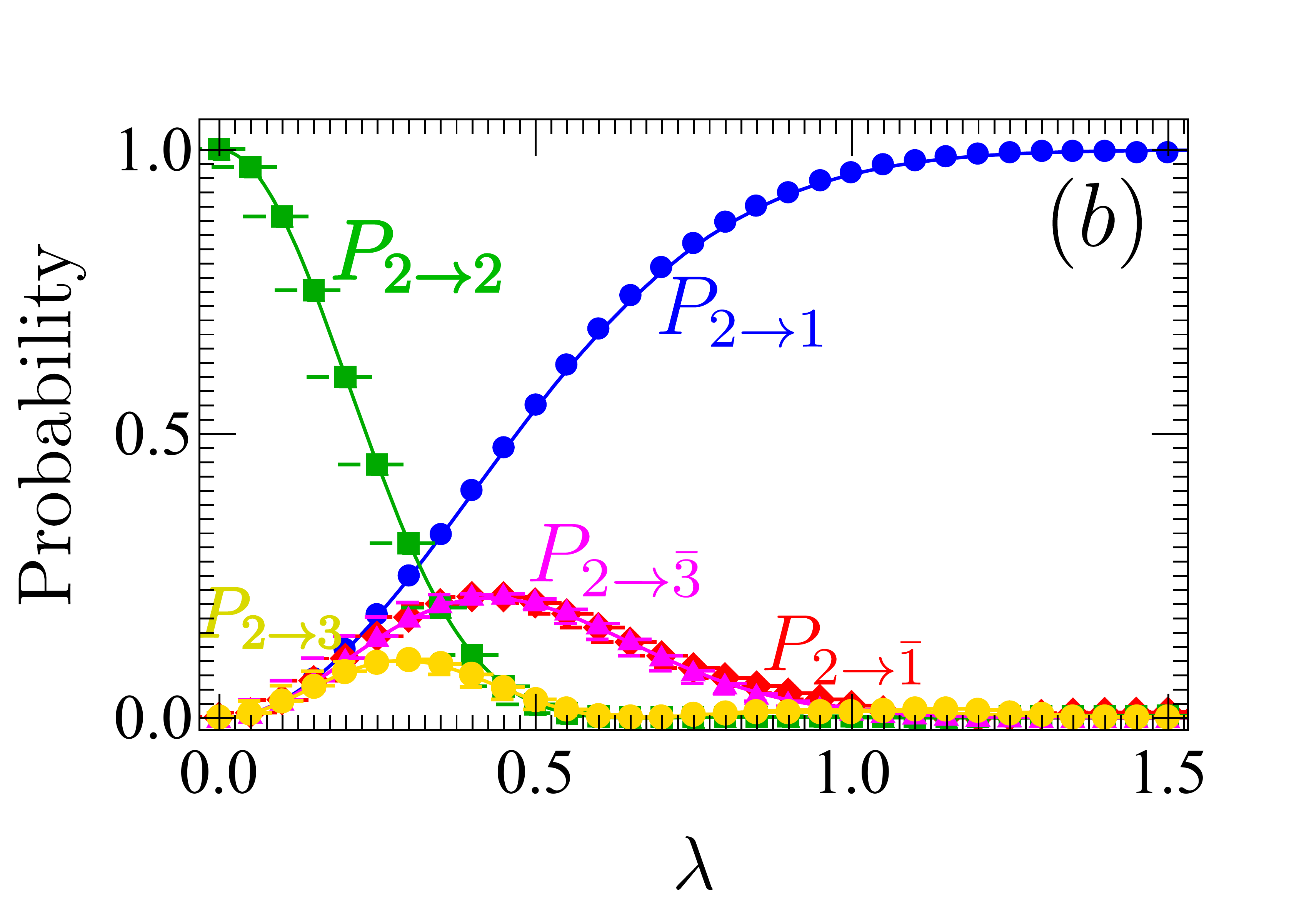}\hspace{-0.5cm}
		\includegraphics[width=6.cm, height=6cm]{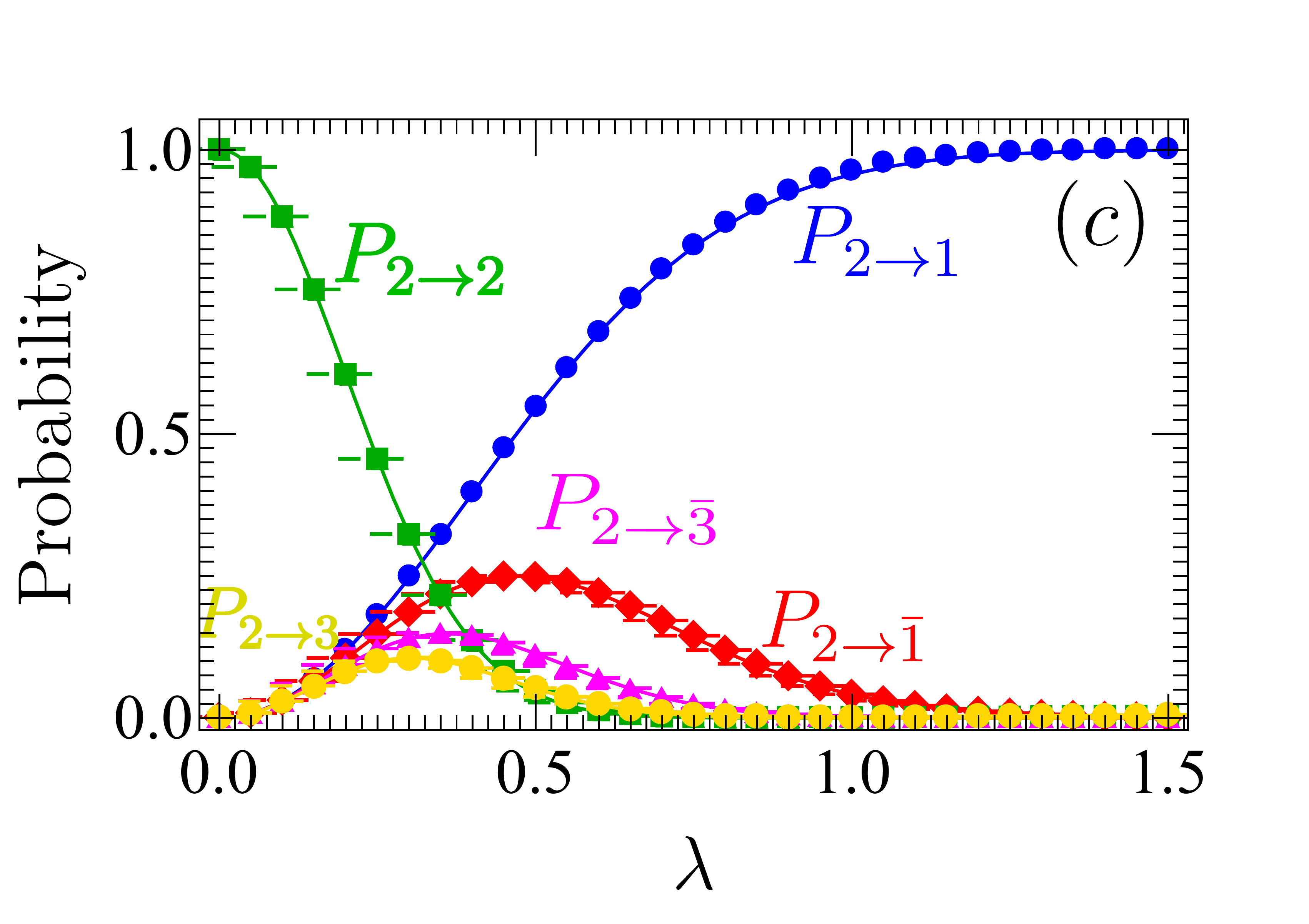}
		\vspace{-0.4cm}
		\caption{(Color Online) {\bf Upper panel:} Eigenenergies of the model Hamiltonian (\ref{equ15}). $\ket{\varphi_{j}(t)}$ (with $j=1,2,3$) and $\ket{\varphi_{\bar{j}}(t)}$ (with $j=1,3$) denote the eigenstates of (\ref{equ15}). Left panel corresponds to the case $D=0$ (two steps), middle panel $D=\omega$ (three steps) and the right panel depicts the case $\omega\ll D$ (four steps). For all calculations, $\lambda_{j,2}^{\bf P_a/P_b}=\lambda_{\bar{j},2}^{\bf P_a/P_b}=3\sqrt{\alpha}$ with $j=1,3$. Blue circles indicate avoided level crossings. {\bf Lower panel:} Transition probabilities in the five-level model (\ref{equ15}) for an initial occupation of  $\ket{2}$. Solid objects and solid lines are respectively the numerical and analytical results. We have considered $\lambda_{j,2}^{\bf P_a/P_b}=\lambda_{\bar{j},2}^{\bf P_a/P_b}=\lambda\sqrt{\alpha}$ with $j=1,3$. Left panels:  $D/\sqrt{\alpha}=0$ and $\omega/\sqrt{\alpha}=20$. Middle panels:  $D/\sqrt{\alpha}=20$ and $\omega/\sqrt{\alpha}=20$. Right panels:  $D/\sqrt{\alpha}=30$ and $\omega/\sqrt{\alpha}=20$. The integration time runs from $t\sqrt{\alpha}=-500$ to $t\sqrt{\alpha}=500$.} \label{Figure8}
	\end{figure}
	
\end{widetext}

\subsection{Solutions for  $\mathbf{H}_{down}(t)$}\label{App2.2}

For $D=0$ (two spin-$1$),
\begin{eqnarray}\label{equation15}
\nonumber	& P_{2\to1}=2\Big(e^{-\frac{\pi\lambda^{2}}{2\alpha}}-e^{-\frac{\pi\lambda^{2}}{\alpha}}\Big)=P_{2\to\bar{3}},\\
\nonumber	& P_{2\to\bar{1}}=2\Big(2e^{-\frac{\pi\lambda^{2}}{2\alpha}}-1\Big)^{2}\Big(e^{-\frac{\pi\lambda^{2}}{2\alpha}}-e^{-\frac{\pi\lambda^{2}}{\alpha}}\Big)=P_{2\to3},\\
  & P_{2\to2}=\Big(2e^{-\frac{\pi\lambda^{2}}{2\alpha}}-1\Big)^{4}.
\end{eqnarray}
For $D=\omega$ (two spin-$1/2$ and one spin-$1$),
\begin{eqnarray}\label{equation16}
\nonumber	& P_{2\to1}=1-e^{-\frac{\pi\lambda^{2}}{\alpha}},\\
\nonumber	& P_{2\to\bar{1}}=2e^{-\frac{\pi\lambda^{2}}{\alpha}}\Big(e^{-\frac{\pi\lambda^{2}}{2\alpha}}-e^{-\frac{\pi\lambda^{2}}{\alpha}}\Big)=P_{2\to\bar{3}},\\
\nonumber	& P_{2\to2}=e^{-\frac{2\pi\lambda^{2}}{\alpha}}\Big(2e^{-\frac{\pi\lambda^{2}}{2\alpha}}-1\Big)^{2},\\
& P_{2\to3}=e^{-\frac{\pi\lambda^{2}}{\alpha}}\Big(2e^{-\frac{\pi\lambda^{2}}{2\alpha}}-1\Big)^{2}\Big(1-e^{-\frac{\pi\lambda^{2}}{\alpha}}\Big).
\end{eqnarray}
For $\omega\ll D$ (four spin-$1/2$),
\begin{eqnarray}\label{equation17}
\nonumber	& P_{2\to1}=1-e^{-\frac{\pi\lambda^{2}}{\alpha}},\\
\nonumber	& P_{2\to\bar{1}}=e^{-\frac{\pi\lambda^{2}}{\alpha}}\Big(1-e^{-\frac{\pi\lambda^{2}}{\alpha}}\Big),\\
\nonumber	& P_{2\to2}=e^{-\frac{4\pi\lambda^{2}}{\alpha}},\\
\nonumber	& P_{2\to\bar{3}}=e^{-\frac{2\pi\lambda^{2}}{\alpha}}\Big(1-e^{-\frac{\pi\lambda^{2}}{\alpha}}\Big),\\
& P_{2\to3}=e^{-\frac{3\pi\lambda^{2}}{\alpha}}\Big(1-e^{-\frac{\pi\lambda^{2}}{\alpha}}\Big).
\end{eqnarray}
One can verify that in all cases $\sum_{j\in \mathcal{F}_{sub}^{down}}P_{2\to j}=1$. Here, $P_{2\to j}$ represents the probability for the transition $\ket{2}\to \ket{j,\omega}$ and $P_{2\to \bar{j}}$ that for $\ket{2}\to \ket{j,-\omega}$ (with $j=1,3$). We have compared the analytical results (\ref{equation15})-(\ref{equation17}) with numerics (see Fig.\ref{Figure8}). We have noted a good agreement between the two solutions that are barely discernible. Note that same can be done for other initial occupations of diabatic states. As indicated in the main text, the model (\ref{equ15}) reproduces the results of the semi-classical analysis only for $P_{2\to2}$. It should be noted that apart from $P_{2\to2}$ in (\ref{equation15})-(\ref{equation17}) other transition probabilities cannot be related to the result of the semi-classical analysis. Then, in the weak coupling limit $\lambda_{\kappa,\kappa'}^{\bf p_{a}/p_{b}}\ll1$, considering (\ref{equ16}), Eq.(\ref{equation17}) yields
\begin{eqnarray}\label{equation18}
	 P_{2\to2}\approx 1-\frac{\pi}{\alpha}\Big(\frac{A^2}{2n_{a}}+\frac{A^2}{2n_{b}}\Big)+\mathcal{O}\Big[\Big(\frac{A^2}{\alpha}\Big)^2\Big].
\end{eqnarray}
Let us compare this result with (\ref{equ14}). If in Eq.(\ref{equ14}), we consider $t=\infty$ and $D\omega/\alpha=(2N+1)\pi/2$ ($N=0,1,2,3,...$), we obtain $P_{2\to2}\approx 1-4\pi\delta+\mathcal{O}(\delta^2)$. It is clear from (\ref{equation18}) that the two results are equivalent for $n_{a/b}=1$ . At the end of our investigations, one may ask what happens when $D=\omega=0$?. As mentioned in the main text, when such a situation is achieved, the degeneracy of the states $\ket{1}$ and $\ket{3}$ is lifted out (the states $\ket{1,\omega}$ and $\ket{1,-\omega}$ returns to the same diabatic energy level $\alpha t$ while  $\ket{3,\omega}$ and $\ket{3,-\omega}$ returns to $-\alpha t$). The five-level model (\ref{equ15}) reduces to a three-level spin-$1$ LZSM model.  Expressions for transition probabilities in this case are found in Ref.\onlinecite{Kenmoe2013}.

\bibliography{Mybib}

\begin{thebibliography}{50}
\bibitem [{com({\natexlab{b}})}]{comment4}%
  \BibitemOpen
  \enquote {\bibinfo {title} {If after passing the first crossing the system
  does not have enough time to relax and again passes through a second
  crossing, beats appear in the its population when it retuns to its original
  diabatic state. {I}f in contrary the system relaxes enough after passing the
  first crossing and traverses a second crossing, steps are observed in the
  final population after returning to its original diabatic state.}}\
  \BibitemShut {NoStop}%
\bibitem [{com({\natexlab{c}})}]{comment2}%
  \BibitemOpen
  \enquote {\bibinfo {title} {Note that by setting ${D=0}$,
  {E}qs.(\ref{equ4})-(\ref{equ6}) are also relevant to describe noise in the
  {SU(2)} {LZSM} theory},}\ \BibitemShut {NoStop}%
\end{thebibliography}
\end{document}